\documentclass[aps,twocolumn,superscriptaddress,nofootinbib,a4paper,longbibliography]{revtex4-2}

\usepackage{times}
\usepackage{epsfig}
\usepackage{amsfonts}
\usepackage{amsmath}
\usepackage{amsthm}
\usepackage{amssymb}					
\usepackage{amsthm}
\usepackage{dsfont}
\usepackage{bm}
\usepackage{mathtools}
\usepackage{color}
\usepackage{multirow}
\usepackage[normalem]{ulem}
\newcommand{\stkout}[1]{\ifmmode\text{\sout{\ensuremath{#1}}}\else\sout{#1}\fi}
\usepackage{latexsym}
\usepackage{mathrsfs}
\usepackage{natbib}
\usepackage{verbatim}
\usepackage[T1]{fontenc}
\usepackage{float}
\usepackage{graphicx}
\usepackage[table]{xcolor}
\usepackage{physics}
\usepackage{soul}
\usepackage[labelfont=bf]{caption}

\definecolor{LightCyan}{rgb}{0.88,1,1}

\renewcommand{\thetable}{\arabic{table}}

\newtheorem{theorem}{Theorem}

\newtheorem{result}{Result} 

\newcommand{\armin}[1]{\textit{\small\textcolor{red}{Armin: #1}}}
\theoremstyle{definition}
\newtheorem{definition}[theorem]{Definition}

\newcommand{\bracket}[3]{\langle#1|#2|#3\rangle}

\usepackage{amssymb}
\usepackage{amsthm}
\usepackage{mathtools} 
\usepackage[customcolors]{hf-tikz} 
\usetikzlibrary{patterns}
\usetikzlibrary{matrix,decorations.pathreplacing}

\pgfkeys{tikz/mymatrixenv/.style={decoration={brace},every left delimiter/.style={xshift=8pt},every right delimiter/.style={xshift=-8pt}}}
\pgfkeys{tikz/mymatrix/.style={matrix of math nodes,nodes in empty cells,left delimiter={[},right delimiter={]},inner sep=1pt,outer sep=1.5pt,column sep=2pt,row sep=2pt,nodes={minimum width=20pt,minimum height=10pt,anchor=center,inner sep=0pt,outer sep=0pt}}}
\pgfkeys{tikz/mymatrixbrace/.style={decorate,thick}}

\newcommand*\mymatrixbraceleft[4][m]{
	\draw[mymatrixbrace] (#1.east|-#1-#2-1.north east) -- node[right=2pt] {#4} (#1.east|-#1-#3-1.south east);
}
\newcommand*\mymatrixbracetop[4][m]{
	\draw[mymatrixbrace] (#1.north-|#1-1-#2.north west) -- node[above=2pt] {#4} (#1.north-|#1-1-#3.north east);
}

\usepackage[colorlinks=true,linkcolor=blue,citecolor=magenta,urlcolor=blue]{hyperref}

\begin{document}
	

	\title{Detecting the dimensionality of genuine multi-particle entanglement}
	
	\author{Gabriele Cobucci}
	
	\author{Armin Tavakoli}
	\email{armin.tavakoli@teorfys.lu.se}
	\affiliation{Physics Department and NanoLund, Lund University, Box 118, 22100 Lund, Sweden.}
	
	\begin{abstract}
		Complex forms of quantum entanglement can arise in two qualitatively different ways; either between many qubits or between two particles with higher-than-qubit dimension. While  the many-qubit frontier and the high-dimension frontier both are well-established, state-of-the-art quantum technology is becoming increasingly able to create and manipulate entangled states that simltaneously feature many particles and high dimension. Here, we investigate generic states that can be considered both  genuinely high-dimensional and genuine multi-particle entangled. We consider a natural quantity that characterises this key property. To detect it, we develop three different classes of  criteria. These enable us both to probe the ultimate noise tolerance of this form of entanglement and to make detection schemes using sparse or even minimal measurement resources. The approach provides a simple way of benchmarking entanglement dimensionality in the multi-particle regime and general, platform-independent, detection methods that readily apply to experimental use.\\
		
		\textbf{Teaser:} Simple and efficient methods are introduced to detect the dimensionality of entanglement between many quantum particles.
		
	\end{abstract}
	
	\date{\today}

	\maketitle
	
	\section{Introduction}
	Entanglement is a cornerstone of quantum theory and a paradigmatic resource for modern quantum information processing. It has been a subject of intense research for decades, both in its own interest \cite{Horodecki2009,Guhne2009} and for its applications in e.g.~quantum cryptography \cite{Xu2020,Pirandola2020,Portman2022}, quantum-enhanced metrology \cite{Giovannetti2011,Degen2017,Pezze2018}, quantum computation \cite{Arute2019,Zhong2020} and fundamental tests of quantum theory \cite{Pan2012,Brunner2014}. Therefore, detecting, quantifying and characterising interesting forms of entanglement is a central question for quantum theory in general and quantum information science in particular. 
	
	A central goal is to generate fully controlled entangled states featuring increasingly many particles. This is  a key requirement for quantum computing advantages. In its basic form, a quantum state comprised on $n$ particles is said to be entangled if it cannot be created via coordinated local operations on each of the particles. However, many times, the most appropriate form of entanglement goes beyond this elementary notion. For example, knowing that $n$ qubits are entangled does not reveal whether it is a feature of all the $n$ qubits, or whether just a pair of them are entangled \cite{Seevinck2001}. To ensure that the entanglement is truly $n$-partite, it is standard to consider a stronger notion, called genuine multipartite entanglement (GME), which ensures that  the state cannot be generated by entangling only some of the qubits. Much theoretical \cite{Guhne2009} and experimental \cite{Bourennane_2004,Haffner2005,Lu2007,Gao2010,Yao2012,Wang2016,Zhong2018,Saggio2019,Friis2018} research has been focused on detecting and realising GME.  GME states of up to 14, 32 and 51 qubits have been achieved with photonics \cite{Thomas2022}, trapped ions \cite{Moses2023} and superconducting circuits \cite{Cao2023}, respectively.

	A second frontier for entanglement are systems of two particles with more than two internal levels. Such high-dimensional entanglement leads to improved rates and enhanced noise- and loss-properties in quantum communication \cite{Cerf2002,Sheridan2010,Ecker2019}, it makes possible teleporation of multiple degrees of freedom \cite{Wang2015} and it leads to stronger quantum correlation phenomena \cite{Collins2002,Skrzypczyk2018,Tavakoli2021}.  In analogy with the multi-particle case, knowing that two high-dimensional particles are entangled does not reveal whether it is truly a feature of all the levels. Verifying the entanglement dimension, called the Schmidt number, means to certify that the state cannot be generated using fewer levels \cite{Terhal_2000}. High-dimensional entanglement has received much attention  \cite{Friis2019,Erhard2020}. A variety of theoretical criteria are known \cite{Sanpera2001,Hulpke2004,Shahandeh2013,Weilenmann2020,Wyderka2023,Morelli2023,Tavakoli2024} and experiments can now reach far into double-digit Schmidt numbers \cite{Dada2011,Krenn2014,PhysRevLett.118.110501,Bavaresco2018,Herrera2020,Designolle2021,Goel2024}.

	A major challenge is to combine the two above frontiers and generate entangled states with many particles and high dimension. This is typically approached by concepts that are based on bipartite entanglement and Schmidt numbers \cite{Huber_2013}. This is interesting not only because entanglement generation and distribution is a  fundamental primitive for quantum information science, but also because it allows to combine the quantum advantages associated with both regimes. Such states are also interesting for high-dimensional quantum computing \cite{Chi2022,Ringbauer2022}, quantum nonlocality \cite{Tang2013,Augusiak2019} and various other tasks \cite{Fitzi2001,Cabello2002,Wang2016}.  Only recently have experiments reported on  realisations of controlled high-dimensional multi-particle states in free space and fiber photonics \cite{Malik2016,Erhard2018,Imany2019,Xing2023}, integrated optics \cite{Reimer2019,Bao2023} and superconducting systems \cite{Cervera2022}. It currently remains hard to generate such states at low noise. 

	\begin{figure}[t!]
		\centering
		\includegraphics[width=\columnwidth]{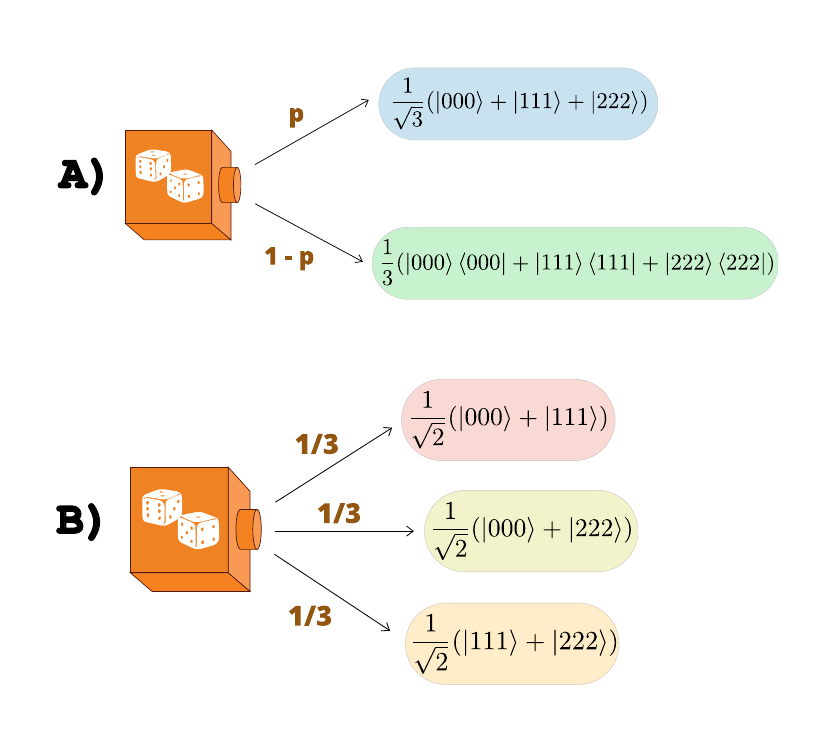}
		\caption{\textbf{Simulation example.} The source in (A) flips a biased coin ($\{p,1-p\}$) and outputs either the three-qutrit GHZ state or a classically correlated state. The mixed state is always genuinely three-particle entangled, but when $p\leq 1/2$  it can nevertheless be simulated with just two-dimensional entanglement using the alternative source in (B).}
		\label{Fig_simulation}
	\end{figure}

	In view of this progress, it is increasingly important to characterise relevant notions of high-dimensional multi-particle entanglement and to develop methods for benchmarking these properties in the lab. For this purpose, both two-particle entanglement dimensionality and GME are unsatisfactory concepts. The former cannot be generalised to multi-particle states \cite{Peres1995} and the latter does not reveal knowledge of the dimensionality, i.e.~a GME state of  $n$-particles with dimension $d$ is not necessarily a true $d$-dimensional phenomenon. Indeed, it is possible that the entanglement, albeit spread over all $n$ particles, can be simulated by using entangled states with fewer than $d$ levels. 
	
	As an illustrative example, consider a source that with probability $p$ emits a three-qutrit  Greenberger-Horne-Zeilinger (GHZ)  state $\frac{1}{\sqrt{3}}\left(\ket{000}+\ket{111}+\ket{222}\right)$ and with probability $1-p$  emits the state  $\frac{1}{3}\left(\ketbra{000}+\ketbra{111}+\ketbra{222}\right)$; see Fig.1(A). The former state is  genuinely high-dimensional and genuinely three-particle entangled but the latter is entirely classical; it is just the state of flipping three correlated three-sided coins. Using the methods of \cite{Bourennane_2004}, one can show that the mixture of these two states, which we denote $\tau_p$, is GME whenever  $p>0$. However, we shall now construct a simulation that uses \textit{only} entanglement between qubits to reproduce $\tau_p$. Specifically, let us randomly prepare one of the three different qubit GHZ states: $\ket{\psi_1}=\frac{1}{\sqrt{2}}\left(\ket{000}+\ket{111}\right)$, $\ket{\psi_2}=\frac{1}{\sqrt{2}}\left(\ket{000}+\ket{222}\right)$ and $\ket{\psi_3}=\frac{1}{\sqrt{2}}\left(\ket{111}+\ket{222}\right)$, as in Fig.1(B). One can verify that the average state, namely $\frac{1}{3}(\psi_1+\psi_2+\psi_3)$ in fact is identical to $\tau_p$ when $p=\frac{1}{2}$. Hence, for $0<p\leq \frac{1}{2}$, the original source has physical dimension three and is GME, but its genuine dimension is actually two.

	Here, we study the dimensionality of multi-particle entangled states based on ideas that naturally extend the standard concepts of GME and Schmidt number. The approach is based on a quantity that we call the GME-dimension; see also \cite{Spengler2013}. It admits a simple interpretation in terms of the smallest possible Schmidt number needed over all possible ways of bisecting the $n$ particles in order to simulate the state.  In its limiting cases, namely just having two particles or just two dimensions, this picture reduces to the Schmidt number and GME respectively; see Figure 2. The states associated with GME-dimensions ranging from $1$ up to the physical dimension, $d$, therefore constitute a hierarchy of increasingly high-dimensional forms of GME.

	\begin{figure}[t!]
		\centering
		\includegraphics[width=\columnwidth]{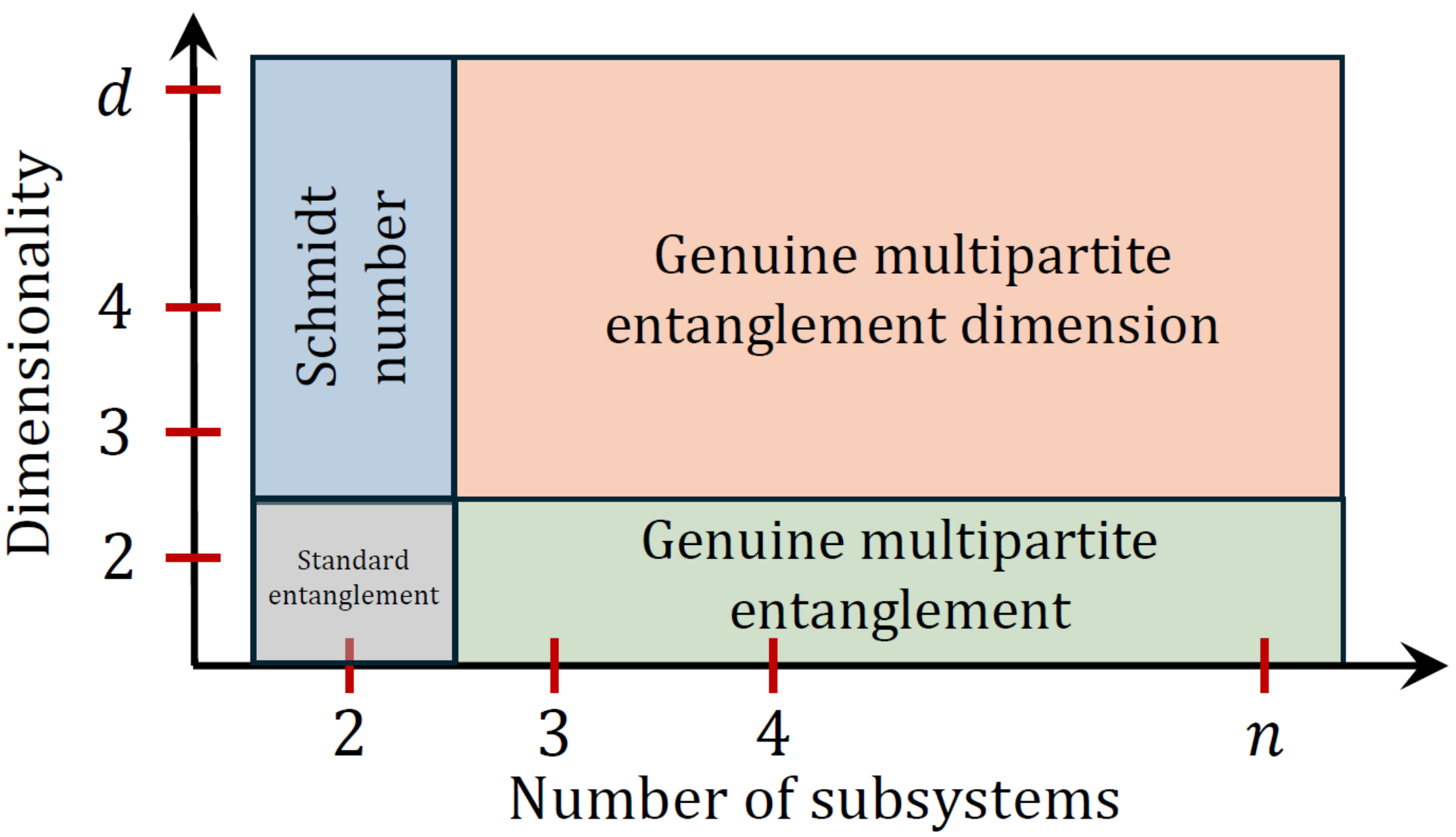}
		\caption{\textbf{Entanglement concepts.} The Schmidt number determines the genuine dimensionality of two-particle entanglement. GME determines whether all $n$ subsystems are entangled. The GME-dimension generalises both these concepts to systems with arbitrary dimensionality and number of subsystems.}
		\label{Fig_concepts}
	\end{figure}	
	

	The key question is how to construct versatile criteria for detecting the GME-dimension of arbitrary, initially uncharacterised, states. We develop  three different classes of detection criteria. Firstly, we consider measuring the fidelity with a given target state and show how this implies a bound on the GME-dimension. This  generalises the standard fidelity witness method for GME \cite{Bourennane_2004}, which is commonly used to benchmark many-qubit experiments (see e.g.~\cite{Haffner2005,Lu2007,Gao2010,Wang2016,Thomas2022}). This method has the advantage of being versatile since it applies to arbitrary pure target states and admits a simple characterisation. Secondly, we develop resource-minimalistic detection criteria, tailored for experimental limitations. This is motivated by the fact that high-dimensional multi-particle entanglement sources typically are complex, leading to limited count rates. Also, measuring the fidelity can require a lot of measurement settings, especially as  $d$ increases. Together, this can quickly become a substantial obstacle for experiments. Therefore,  our method allows the fidelity to be estimated using only the minimal number of measurements, namely two. We explicitly construct such minimal criteria for the two most broadly considered high-dimensional multi-particle states, namely  GHZ and cluster states.  Thirdly, we employ convex programming methods to determine conditions for a state's GME-dimension.  We show that this leads to considerably stronger GME-dimension criteria than what is possible with fidelity witnesses. We also show that for experiments involving a few particles and a few levels, this method can offer particularly noise-robust detection criteria using only a small number of measurement settings.

	\subsection{Dimensionality of genuine multipartite entanglement}
	The entanglement structure of quantum systems with $n>2$ subsystems is richer than in the two-particle (bipartite) case. A pure state, $\ket{\psi}$, is called biseparable if it is possible to partition the $n$ particles into two non-overlapping sets $\{S,\bar{S}\}$, such that the state factors over this bipartition. That is, there exists states such that $\ket{\psi}_{1\ldots n}=\ket{\phi}_S\otimes \ket{\varphi}_{\bar{S}}$. This is straightforwardly extended to mixed states. A mixed is biseparable if it can be generated by classically mixing pure biseparable states, i.e.~$\rho=\sum_{S|\bar{S}} \sum_j q_{S|\bar{S}}^{(j)} \ketbra{\phi_j}_S\otimes \ketbra{\varphi_j}_{\bar{S}}$, where $q_{S|\bar{S}}^{(j)}\geq 0$ and $\sum_{S|\bar{S},j}q_{S|\bar{S}}^{(j)}=1$. The index $S|\bar{S}$ runs over all choices of $\{S,\bar{S}\}$, the number of which is $2^{n-1}-1$. If a state is not biseparable, it called genuine multipartite entangled (GME). Thus, GME states admit a simple interpretation: they are the states impossible to generate using classical randomness and arbitrary $n$-partite states, each separable w.r.t some bipartition of the particle set.	
	
	In order to further develop the notion of GME so that it also addresses the dimensionality of the entanglement, we must suitably replace the factorisability condition of each pure state in the ensemble decomposition. A more appropriate constraint to capture the high-dimensional nature of the entanglement is to instead impose that the bipartite entanglement dimension, namely the  Schmidt rank, across $\{S,\bar{S}\}$, is bounded from above. The Schmidt rank is the rank of the reduced density matrix, which is necessarily equal to one for a product state but larger for pure entangled states (it generalizes to the Schmidt number for mixed states \cite{Terhal_2000}). That is, for a generic $n$-partite and $d$-dimensional density matrix $\rho$, we consider decompositions of the form
	\begin{equation}\label{decomp}
		\rho=\sum_{S|\bar{S}} q_{S|\bar{S}} \sigma_{S|\bar{S}}
	\end{equation}
	for an arbitrary probability distribution $\{q_{S|\bar{S}}\}$ and states $\sigma_{S|\bar{S}}$ with local dimensions $d^{|S|}$ and $d^{|\bar{S}|}$, whose Schmidt number is no more than $r$. The latter limitation means that for each $\{S,\bar{S}\}$ there exists a pure-state decomposition $\sigma_{S|\bar{S}}=\sum_i p_{i,S|\bar{S}} \ketbra{\psi_{i, S|\bar{S}}}$ where the largest Schmidt rank over all pure states $\{\ket{\psi_{i,S|\bar{S}}}\}_i$ does not exceed $r$.  In contrast, if no decomposition of the form \eqref{decomp} exists, the simulation of $\rho$ via classical randomness and bipartite quantum states then requires that at least one of the ensemble  states $\sigma_{S|\bar{S}}$ has a Schmidt number exceeding $r$. This naturally leads us to our main quantity of interest, namely the largest $r$ found over all  the bipartitions, in the least ``dimension-expensive'' ensemble realisations of $\rho$. We call this the GME-dimension.

	\begin{definition}[GME-dimension]
		\label{def_GMED}
		\textit{For an arbitrary $n$-partite state $\rho$, its GME-dimension is}
		\begin{equation}
			\begin{aligned}
				\mathcal{D}_\text{GME}(\rho) = \min_{\lbrace q_{S|\bar{S}}\rbrace, \lbrace \sigma_{S|\bar{S}} \rbrace} \left\lbrace r_{\text{max}}: \, \rho = \sum_{ S|\bar{S}} q_{S|\bar{S}}\,\sigma_{S|\bar{S}}  \right. \\ \left. \text{and }\, r_{\text{max}} = \max_{\lbrace S|\bar{S}\rbrace}\, r_{S|\bar{S}} \right\rbrace,
			\end{aligned}
		\end{equation}
		\textit{where $r_{S|\bar{S}}$ is the Schmidt number of $\sigma_{S|\bar{S}}$ across  $\{S,\bar{S}\}$.}
	\end{definition}
	Notably, setting $n=2$ reduces the GME-dimension to the Schmidt number, as there is only one possible bipartiton. Furthermore, setting $\mathcal{D}_\text{GME}=1$ instead reduces it to the definition of biseparability. Thus, all GME states have $\mathcal{D}_\text{GME}>1$ but the precise value of the GME-dimension additionally reveals knowledge of their dimensionality. For fixed $(n,d)$, we write $\mathcal{G}_{d_\text{GME}}$ for the set of states obeying $\mathcal{D}_\text{GME}(\rho)\leq d_\text{GME}$. The GME-dimension can also be viewed as the smallest element of the so-called Schmidt number vector \cite{Huber_2013}. We note that while Schmidt number vectors cannot in general be compared, the GME-dimension forms an ordered hierarchy of nested convex sets, ranging from states that fail to be GME ($\mathcal{D}_\text{GME}=1$) up to GME states that are geuninely $d$-dimensionally entangled ($\mathcal{D}_\text{GME}=d$). This is illustrated in Figure 3.  
	
		\begin{figure}[t!]
			\centering
			\includegraphics[width=0.9\columnwidth]{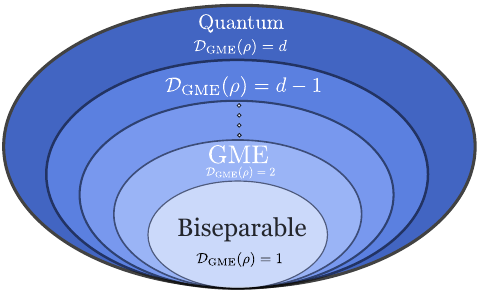}
			\caption{\textbf{Hierarchy of GME-dimension states.} States with $\mathcal{D}_{\text{GME}} = 1$ are biseparable. Any state with $\mathcal{D}_\text{GME}>1$ is GME.  States with  $\mathcal{D}_{\text{GME}} = d$ cannot be simulated by any lower-dimensional entanglement.}
			\label{FigGMEd}
		\end{figure}
		

	\section{Results}
	
	\subsection{Fidelity witnesses}
	Measuring the fidelity,  $F_\psi(\rho)=\bracket{\psi}{\rho}{\psi}$, between a pure target state $\psi$ and a mixed lab state $\rho$ is a standard way to detect GME in the vicinity of $\psi$. Therefore, we begin with identifying how the fidelity also can be used to detect the GME-dimension. 
	\begin{result}[Fidelity witness]\label{Result1}
		\label{res_fidelity-bound}
		For any $n$-partite pure target state $\psi$ with local dimension $d$, its fidelity with any state $\rho$ with GME-dimension no larger than $d_\text{GME}$ is bounded as
		\begin{equation}
			\label{eq: fidelity-bound}
			\max_{\rho \in \mathcal{G}_{d_\text{GME}}} F_\psi(\rho) \leq \max_{\{S,\bar{S}\}} \sum_{i=1}^{d_\text{GME}} \lambda_i(\psi_{S}),
		\end{equation}
		where the maximisation is over all bipartitions of the subsystems and $\lambda(\psi_{S})$ is the spectrum of the reduced state  $\psi_{S}=\tr_{\bar{S}}\left(\psi\right)$, ordered non-increasingly.
	\end{result}
	
	To prove this, a straightforward argument builds directly on extending previously known properties of the bipartite fidelity to the multipartite case. For $n=2$, for which there is only one bipartition, it is known that \eqref{eq: fidelity-bound} holds \cite{Fickler_2014}. Due to the linearity of the fidelity function and the convexity of the set $\mathcal{G}_{d_\text{GME}}$, $F_\psi(\rho)$ is optimised for a pure state $\rho=\ketbra{\phi}\in \mathcal{G}_{d_\text{GME}}$. Thus, we may separately consider each bipartition and select the maximal value; that is Eq.~\eqref{eq: fidelity-bound}. An alternative proof method is possible without relying on directly extending the bipartite case.  This is based on using the strong duality of linear programming and it is detailed Supplementary Material section I. 
	Notice that choosing $d_\text{GME}=1$ reduces Result~\ref{Result1} to the standard fidelity witness for GME \cite{Bourennane_2004}. Moreover, note that if we instead were to assign a separate Schmidt number to each bipartition, namely $d_{S|\bar{S}}$, the only change in Eq.~\eqref{res_fidelity-bound} will be that the demarcation of the sum becomes $d_{S|\bar{S}}$ instead of $d_\text{GME}$.

	It is important to consider Result~\ref{Result1} in practically relevant cases. We have evaluated it explicitly, for arbitrary choices of $(n,d)$, for three different seminal families of states, namely GHZ states,  cluster states and absolutely maximally entangled states (assuming they exist, see \cite{Huber_2018,Huber_table}). In all three cases, the fidelity witness for the GME-dimension is identical; it reads
	\begin{align}\label{fidexample}
		F_{\psi}(\rho) \leq \frac{d_\text{GME}}{d},
	\end{align}
	which is notably independent of $n$. Our proof for the former and latter state is straightforward, but less so for cluster states; see details in SM section II. Importantly, while \eqref{fidexample} provides a simple criterion, it also means that fidelity methods  cannot distinguish between these three classes of states. We note that Result~\ref{Result1} does not reduce to Eq.~\eqref{fidexample} for arbitrary choices of $\psi$.

	\subsection{Minimal fidelity witnesses}
	With multipartite high-dimensional states, it is often practically relevant to probe the system using as few measurements as  possible. Therefore, we now aim to detect the GME-dimension via fidelity estimation from only two complementary  basis measurements. This is the minimal setting for detecting the GME-dimension.

	First, we target states in the vicinity of the GHZ state,
	\begin{equation}
		\ket{\text{ghz}_{n,d}}=\frac{1}{\sqrt{d}}\sum_{i=0}^{d-1} \ket{i}^{\otimes n},
	\end{equation}	
	for arbitrary $(n,d)$. This state is natural to focus on since it has been the goal of most  multi-particle entanglement experiments in the literature.  Let $\{\ket{j}\}_{j=0}^{d-1}$ be the computational basis and $\{\ket{e_j}\}_{j=0}^{d-1}$  the Fourier basis,  $\ket{e_j}=\frac{1}{\sqrt{d}}\sum_{k=0}^{d-1}\omega^{jk}\ket{k}$, where $\omega=e^{\frac{2\pi i}{d}}$. Consider that we (i) measure all subsystems in the computational basis and compute the total probability of all $n$ local outcomes being identical, and (ii) measure all subsystems in the Fourier basis and compute the total probability that the sum over all local outcomes is divisible by $d$. The Hermitian operator describing the sum of these two events takes the form
	\begin{equation}
		\label{GHZ_witness_operator}
		O^\text{ghz}_{n,d}=\sum_{j=0}^{d-1} \ketbra{j}{j}^{\otimes n} + \sum_{\substack{j_1,\ldots,j_n=0}}^{d-1} \bigotimes_{l=1}^n \ketbra{e_{j_l}}{e_{j_l}}\delta_{j_1 \oplus \ldots \oplus j_n,0},
	\end{equation}
	where $\oplus$ denotes addition modulo $d$. The entanglement witness is the expectation value, $\mathcal{W}^\text{ghz}_{n,d}(\rho)=\tr\left(\rho	O^\text{ghz}_{n,d}\right)$. It is immediate that the GHZ state has perfect correlations for the event (i) and a direct calculation shows the same also for  event (ii). Hence, $\mathcal{W}^\text{ghz}_{n,d}(\text{ghz}_{n,d})=2$. Our next result shows how this witness detects the GME-dimension.
	
	\begin{result}[Minimal GHZ state witness]\label{Result2}
		For any $n$-partite state $\rho$ of local dimension $d$ with GME-dimension no larger than $d_\text{GME}$,
		\begin{equation}\label{minghz}
			\mathcal{W}^\text{ghz}_{n,d}(\rho)\leq 1+\frac{d_\text{GME}}{d},
		\end{equation}
		Moreover, any observed value of $\mathcal{W}^\text{ghz}_{n,d}$ implies a GHZ-fidelity bound $F_\text{ghz}(\rho)\geq \mathcal{W}^\text{ghz}_{n,d}(\rho) -1$.
	\end{result}
	The proof is based on the observation that the witness operator \eqref{GHZ_witness_operator} can be decomposed as a sum of a projector and the GHZ state. See SM section III for details. 
	
	Naturally, since we use only two bases, this criterion is weaker than the exact fidelity criterion, but it is practically advantageous. Importantly, it can perform well for the two most relevant noise models, namely depolarisation and dephasing.  Take first depolarising (white) noise, corresponding to  $\rho_v^\text{ghz}=v\ketbra{\text{ghz}_{n,d}}+\frac{1-v}{d^{n}}\openone$, where $v\in[0,1]$ is the visibility. The critical visibility for violating inequality~\eqref{minghz} is 
	\begin{equation}\label{Minimal_fidelity_crit_vis}
		v_\text{crit}=\frac{1-d^{n-2}(d_\text{GME}+d-1)}{1-d^{n-2}(2d-1)}.
	\end{equation}
	For instance, take a system of four qutrits; the threshold for detecting $d_\text{GME}=3$ becomes $v_\text{crit}= 79.5\%$, which is in the regime relevant for state-of-the-art experiments. Next, consider instead dephasing noise, corresponding to $\tau_v^\text{ghz}=v\ketbra{\text{ghz}_{n,d}}+\frac{1-v}{d}\sum_{i=0}^{d-1}\ketbra{i}^{\otimes n}$. The critical visibility from Result~\ref{Result2} becomes 
	\begin{equation}
		v_\text{crit}=\frac{d_\text{GME}-1}{d-1},
	\end{equation} 
	independently of $n$. We observe that this  equals what is obtained from  Eq.~\eqref{fidexample} by using complete fidelity measurements. In fact, we prove in SM section IV  that $v_\text{crit}$ is the exact threshold for the GME-dimension of $\tau_v^\text{ghz}$, i.e.~our minimal witness is actually necessary and sufficient.


	Another important class of states are cluster states. Interest in them draws mainly from that they are a universal resource for one-way quantum computing \cite{Raussendorf2001,Raussendorf2003}.  High-dimensional cluster states are generated in lattices of $n$ qudits with Ising-type interaction; in a linear lattice they take the form 
	\begin{equation}
		\ket{C_{n,d}} = \frac{1}{\sqrt{d^n}}\bigotimes_{a=1}^{n}\left(\sum_{k=0}^{d-1}\ket{k}_{a}Z^{k}_{a+1}\right),
	\end{equation}
	with $Z = \sum_{k=1}^{d} \omega^k\ketbra{k}{k}$ and the convention $Z_{n+1} = \openone$ \cite{PhysRevA.68.062303}. Using only two global product bases, we construct a GME-dimension witness targeting $\ket{C_{n,d}}$. 
	
	Consider that we (i) measure the odd subsystems in the Fourier basis and the even ones in the computational basis and (ii) measure the odd subsystems in the computational basis and the even ones in the Fourier basis. The relevant outcome combinations are different from those used before for GHZ. Specifically, they correspond to the Hermitian operator,
	\begin{equation}
		\label{Witness_operator-cluster}
		\begin{aligned}
			O^\text{cluster}_{n,d} &= \bigotimes_{l \text{ odd}}^{n}\, \sum_{q_{l},p_{l+1} = 0}^{d-1} \ketbra{e_{q_{l}}p_{l+1}}{e_{q_{l}}p_{l+1}}\delta_{p_{l+1}\oplus p_{l-1} \ominus q_{l},0}\\
			& + \bigotimes_{l \text{ odd}}^{n} \,\sum_{p_{l},q_{l+1} = 0}^{d-1} \ketbra{p_{l}e_{q_{l+1}}}{p_{l}e_{q_{l+1}}}\delta_{p_{l} \oplus p_{l+2} \ominus q_{l+1},0}.
		\end{aligned}	
	\end{equation}
	The total probability associated to observing these outcomes in events (i) and (ii) corresponds to the entanglement witness, i.e.~ $\mathcal{W}^\text{cluster}_{n,d}(\rho)=\tr\left(\rho	O^\text{cluster}_{n,d}\right)$. The witness is constructed so that the cluster state exhibits perfect correlations for both (i) and (ii), and hence $\mathcal{W}^\text{cluster}_{n,d}(C_{n,d}) = 2$. The next result shows how this witness detects the GME-dimension.

	\begin{result}[Minimal cluster state witness]\label{Result3}
		For any $n$-partite state $\rho$ of local dimension $d$ with GME-dimension no larger than $d_\text{GME}$, 
		\begin{equation}\label{mincluster}
			\mathcal{W}^\text{cluster}_{n,d}(\rho)\leq 1 + \dfrac{d_\text{GME}}{d}
		\end{equation}
		Moreover, any observed value of $\mathcal{W}^\text{cluster}_{n,d}$ implies a cluster state fidelity bound $F_\text{cluster}(\rho)\geq \mathcal{W}^\text{cluster}_{n,d}(\rho) - 1$.
	\end{result}
	The proof of this result is given in SM section VI and largely parallels the ideas used to derive Result~\ref{Result2}.
	
	In analogy with the analysis of the GHZ witness, we apply Result~\ref{Result3} to cluster states with depolarising and dephasing noise, namely $\rho_{v}^{\text{cluster}} = v\ketbra{C_{n,d}} + \frac{1-v}{d^n} \openone$ and $\tau_{v}^{\text{cluster}} = v\ketbra{C_{n,d}} + \frac{1-v}{d}\sum_{i = 0}^{d-1}\ketbra{i}^{\otimes n}$, respectively. The two critical noise thresholds turn out to be identical, namely
	\begin{equation}
		v_{\text{crit}} = \frac{b - d^{\frac{n}{2}+a} - d^{\frac{n}{2} + a - 1}d_{\text{GME}}}{b - 2 d^{\frac{n}{2}+a}},
	\end{equation}
	where $b=1 + d^{2a}$ and $ a = (1 + (-1)^{n+1})/4$. Continuing the example of four qutrits from earlier, with a maximal GME-dimension, this becomes $v_{\text{crit}} = 81.3 \%$, which is  reasonable for practical purposes.

	Lastly, we note that if instead of the GME-dimension, we assign a separate Schmidt number, $\{d_{S|\bar{S}}\}$, to each bipartition and ask whether the state is compatible with this hypothesis, we obtain necessary criteria directly by modifying Result~\ref{Result2} and Result~\ref{Result3}. The reason is that both of these ultimately rely on fidelity bounds, as previously mentioned in the context of Result~\ref{Result1}. For example, Result~\ref{Result2} will remain unchanged since only the largest element of $\{d_{S|\bar{S}}\}$ will be relevant.

	\subsection{Convex programming method}
	We now go beyond fidelity-based criteria and detect the GME-dimension via efficiently computable convex programming relaxations  \cite{tavakoli2023semidefinite}. To this end, let $\Lambda$ be an $r$-positive trace-preserving map. Such maps are positive when applied to one share of every bipartite state with Schmidt number at most $r$ but non-positive for some states  with larger Schmidt number \cite{Terhal_2000}. For any $r$-positive map, we define the semidefinite program
	\begin{equation}\label{SDP}
		\begin{aligned}
			\max_{v,\tilde{\sigma}} & \quad v\\
			\text{s.t.}&\quad v\rho+\frac{1-v}{d^{n}}\openone =  \sum_{ S|\bar{S}} \tilde{\sigma}_{S|\bar{S}},\\
			& \quad \tilde{\sigma}_{S|\bar{S}} \geq 0 \quad \forall (S|\bar{S}),\\
			& \quad (\Lambda_S\otimes \openone_{\bar{S}})[\tilde{\sigma}_{S|\bar{S}}] \geq 0 \quad \forall (S|\bar{S}),
		\end{aligned}
	\end{equation}
	where $\tilde{\sigma}_{S|\bar{S}}=q_{S|\bar{S}}\sigma_{S|\bar{S}}$ are unnormalised states. Here, we have chosen to introduce depolarising noise on the state $\rho$. This serves as one of several possible quantifiers of the separation of $\rho$ w.r.t.~the selected relaxation of $\mathcal{G}_r$. Thus, obtaining any value $v<1$ implies that $\rho\notin \mathcal{G}_r$ and hence  $\mathcal{D}_\text{GME}(\rho)>r$. For example, selecting $\Lambda$ as the partial transpose map, which is $1$-positive, reduces Eq.~\eqref{SDP} to the approach first outlined in Ref.~\cite{Jungnitsch2011} for  GME. This partial transpose map can be adapted for our higher-dimensional analysis by introducing auxiliary Hilbert spaces of dimension $r$ for each qudit \cite{Hulpke2004,Weilenmann2020}. However, this is a  costly approach for the GME-dimension because the extra dimensions accumulate exponentially in $n$ when $r>1$, causing a large computational overhead. Therefore, we instead propose to use the generalised reduction map, $\Lambda(X) = \tr X\openone - \alpha X$, which is known to be $r$-positive when $\alpha=\frac{1}{r}$ \cite{Tomiyama1985,Terhal_2000}. Note that applying the map on the system $S$, as in Eq.~\eqref{SDP}, is not necessarily equivalent to applying it on system $\bar{S}$. They correspond to different relaxations, and one can even apply it to both subsystems separately to obtain a more accurate relaxation of $\mathcal{G}_r$.
	
		\begin{table}[t!]
		\begin{tabular}{|c|c|c|c|c|}
			\hline
			$\,$ (n,d) $\,$ & $\mathcal{D}_{\text{GME}}(\rho)$ & $\,$  $v_{\text{ghz}}$ (LP)  $\,$ & $\,$  $v_{\text{cluster}}$ (LP) $\,$ & $v$ (fidelity)  \\
			\hline
			(3,3) & 1 & $0.2500$ & $ * $ & $0.3077$\\
			\hline
			(3,3) & 2 & $0.5909$ & $ * $ & $0.6538$\\
			\hline
			(4,3) & 1 & $0.2203$ & $0.2174$ & $0.3250$\\
			\hline
			(4,3) & 2 & $0.6029$ & $0.5129$ & $0.6625$\\
			\hline
			(4,4) & 3 & $0.6503$ & $0.534$ & $0.7490$\\
			\hline
		\end{tabular}
		\caption{\textbf{Results from convex programming method. }Critical visibility, $v$, for the GME-dimension computed by re-formulating the SDP in Eq.~\eqref{SDP} as a linear program (LP). Case study presented for target GHZ states and cluster states under white noise. Note that these states are local-unitary equivalent for $n = 3$ and hence the results are identical (marked by $``*"$ in the table). The results are compared with the visibility obtained from the fidelity bound \eqref{fidexample}.}
		\label{table:visibility-fidelity}
	\end{table}
	
	The program \eqref{SDP} can many times be considerably reduced. If $\rho$ is a pure state and we use the generalised reduction map, we can express \eqref{SDP} as just a linear program (LP) by representing it in a basis in which $\rho$ is diagonal (if $\rho$ is a graph state, similar reductions are possible also for  GME tests based on the partial transpose map  \cite{Jungnitsch_2011}); see SM section VIII. This speeds up computations. Furthermore, depending on the choice of $\rho$, additional symmetries may be available, which can be used to further reduce the program. For instance, for the GHZ state, one can also exploit that the state is invariant under permutations of its particle labels. This reduces the exponentially many diagonal matrix variables to just $n-1$. 

	We have evaluated these programs explicitly. On a standard computer, we could evaluate the LP up to  e.g.~$(n,d)= (5,4)$ for the GHZ state and $(n,d) = (4,4)$ for the cluster state. Our implementation is available at \cite{github-code}. In Table~\ref{table:visibility-fidelity} we display some of the resulting bounds on the critical visibility; more extensive results are given in SM section VIII.  The main observation is that the visibilities are considerably smaller than those obtained from the fidelity witness \eqref{fidexample}. This showcases the relevance of this method. 
	
	Moreover, with small modifications, the program \eqref{SDP} can also be adapted to the scenario with individual Schmidt numbers for each bipartition. To this end, it sufficies, for each bipartition $(S|\bar{S})$, to appropriately choose the value of $\alpha$ in the definition of $\Lambda$ as $\alpha=\frac{1}{d_{S|\bar{S}}}$. Further details are given in SM section IX together with case studies for GHZ states.
	
	Furthermore, the method can also be used to give resource-efficient criteria that are more noise-robust than the minimal fidelity estimation method discussed earlier. To this end, we no longer  simulate the whole state $\rho$ but only its statistics when a small number of measurements are performed on it. In SM section X, we show how to adapt the program \eqref{SDP} for this purpose. To showcase its usefulness, consider that we measure just two or three global product measurements, where the local bases are  mutually unbiased bases (MUBs) \cite{DURT_2010}. We denote by $E_i$ the set of constraints on the statistics for each choice of global product measurements. The results are displayed in Table~\ref{table:visibility-fidelity-prob} based on the GHZ state and a maximal GME-dimension. We see that using two bases (the computational basis and the Fourier basis, $E_C$ + $E_F$) is more noise-robust than the minimal fidelity witnesses, even though the measurements are identical. Remarkably, the noise tolerance considerably improves by just imposing one additional set of constraints, $E_M$, that involve a specific choice of local MUBs (see SM section X). By adding further appropriately selected global product MUBs, one can further improve the visibility.

	\begin{table}[t!]
		\resizebox{\columnwidth}{!}{%
			\begin{tabular}{|c|c|c|c|c|c|}
				\hline
				$\,$ (n,d) $\,$ & $\mathcal{D}_{\text{GME}}(\rho)$ &  $v$ (fidelity) & $v$ (min fid) & $v$ ($E_C$ + $E_F$)&  $v$ ($E_{CF}$ + $E_M$)\\
				\hline
				(3,3) & 2 & $0.6538$ & $0.7857$ & $0.7500$ & $0.6667$\\
				\hline
				(3,4) & 3 & $0.7460$ & $0.8518$ & $0.8222$ & $0.7576$\\
				\hline
				(4,3) & 2 & $0.6625$ & $0.7954$ & $0.7750$ & $0.7097$\\
				\hline
			\end{tabular}%
		}
		\caption{\textbf{Results from SDP with constraints on the statistics. }Comparison between values of critical visibility, $v$, for different GME-dimensions obtained from: fidelity criterion \eqref{fidexample}, minimal fidelity witness \eqref{Minimal_fidelity_crit_vis}, convex programming relaxations based on simulating statistics from two, ($E_C$ + $E_F$), and three global product measurements, ($E_C$ + $E_F$ + $E_M$). Case study presented for target GHZ states under white noise.}
		\label{table:visibility-fidelity-prob}
	\end{table}

	\section{Discussion}
	We have employed the GME-dimension as a quantity for benchmarking genuinely high-dimensional forms of genuine multi-particle entanglement. This naturally extends the established entanglement concepts of GME and Schmidt number into the multi-particle high-dimensional regime, see Figure 2. In order to detect the GME-dimension of initially unknown states, we have put forward three classes of criteria. 
	
	Firstly, we use the fidelity with a target state. This has the advantage of being applicable for any dimension and particle number, and to arbitrary target states. As we have shown for several of the most relevant families of quantum states, the fidelity method offers promising, yet not optimal, noise-robustness. The fidelity is also in itself a natural quantifier of the quality of a state preparation. In addition, measuring the fidelity is also far more sparse in terms of the number of necessary measurements than is quantum state tomography. Nevertheless, it is likely to be time-expensive, for instance on optical platforms where the multi-photon coincidence rate decreases with the particle  number and visibility \cite{Saggio2019}. 
	
	This motivated our second class of criteria, namely those using the smallest possible number of measurements (two) to estimate the fidelity. We have constructed such criteria for arbitrary dimension and particle number, for both the seminal GHZ states and the cluster states. In spite of the resource-minimal approach, our criteria are signficantly, and sometimes even optimally, noise tolerant. They perform particularly well for high-quality sources. They are much less time-expensive to estimate and can readily be applied to experiments on any physical platform.

	Thirdly, we develop convex programming criteria that can approximate the GME-dimension of arbitrary states. We show that this method is effectively computable for systems of a few particles with low dimension, and that it can considerably outperform fidelity criteria in terms of noise-robustness. This method is therefore practically relevant, because  most experiments \cite{Malik2016,Erhard2018,Imany2019,Reimer2019,Cervera2022,Bao2023,Xing2023} presently concern choices of $(n,d>2)$  that we are able to treat explicitly, even without enhanced computing resources.  This method can also systematically produce resource-efficient detection criteria, using any measurements considered convenient by the experimenter. Using the duality theory of semidefinite programming, one can also extract explicit witnesses for the GME-dimension from this method. 
	
	It is an interesting open problem  how  much noise various states tolerate before their GME-dimension is reduced. It is well-known that high-dimensional GME can persist under diverging noise rates \cite{Huber_2010,Gao2011,Ananth2015} but it is an open problem whether anything similar is possible for larger, or even maximal, GME-dimension. Connected to this is also the question of which classes of states that are most strongly entangled in terms of the GME-dimension. For instance, Table~\ref{table:visibility-fidelity} already suggests that the  cluster state is more strongly high-dimensionally GME than the GHZ state. From this point of view, there may be even more interesting   high-dimensional multi-particle states to study. Lastly, we  remark that there may also be other interesting ways to approach the characterisation of genuinely high-dimensional genuine multi-particle entanglement. Exploring the alternative paths is of evident interest.

	\begin{acknowledgments}
		We thank Otfried G\"uhne and Marcus Huber for useful comments and Emmanuel Zambrini Cruzeiro for computation support.
		
		\textbf{Funding:} This work is supported by the Wenner-Gren Foundation and by the Knut and Alice Wallenberg Foundation through the Wallenberg Center for Quantum Technology (WACQT).
		
		\textbf{Author contributions:} A.T. had the idea and proposed the basic concept. A.T. and G.C. developed the theory. All authors participated in the writing of the manuscript.
		
		\textbf{Competing interests:} The authors declare that they have no competing interests.
		
		\textbf{Data and materials availability:} All data are available in the main text or the supplementary materials. Our implementation of the linear programs is available at \url{https://doi.org/10.5281/zenodo.13123607}.

	\end{acknowledgments}

	\newpage
	\onecolumngrid
	\appendix
	\renewcommand{\thesection}{\Roman{section}}
	\renewcommand{\appendixname}{SM}
	
	
	\section{Proof of the fidelity bound for GME-D}
	\label{Appendix_A: Fidelity_bound}
	
	We show a linear programming based proof of \textit{Result} 1. For clarity, we organise it in different steps:
	\begin{itemize}
		\item \textit{Convexity argument}
		
		The starting point is to note that, because of the \textit{convexity} of the set of quantum states [81], the following inequality holds:
		\begin{equation}
			\max_{\rho: \,\mathcal{D}_\text{GME}(\rho) = d} F(\rho,\psi) \leq \max_{\phi: \, \mathcal{D}_\text{GME}(\phi) = d} F(\phi,\psi),
		\end{equation}
		where $\phi$ is a generic $n$-partite pure state with $\mathcal{D}_\text{GME}(\phi) = d$.
		\item \textit{Monotonicity of fidelity}
		
		Since $\mathcal{D}_\text{GME}(\psi) = D$ and $\mathcal{D}_\text{GME}(\phi) = d$, for what follows from \textit{Definition} 1 of the GME-Dimension, it is possible to find a bipartition $\alpha_{2}^{*} = (S|\bar{S})$ of the set $\lbrace 1, \dots, n \rbrace$ for which:
		\begin{equation}
			\psi = \sum_{k=1}^{D} \lambda_k \ket{\lambda_k}_{S}\ket{\lambda'_k}_{\bar{S}}\prescript{}{S}{\bra{\lambda_k}}\prescript{}{\bar{S}}{\bra{\lambda'_k}}, \quad \phi = \sum_{j=1}^{d} \mu_{j} \ket{\mu_{j}}_{S}\ket{\mu'_{j}}_{\bar{S}}\prescript{}{S}{\bra{\mu_{j}}}\prescript{}{\bar{S}}{\bra{\mu'_{j}}},
		\end{equation}
		where $\displaystyle \sum_{k=1}^{D} \lambda_k = \sum_{j=1}^{d} \mu_j = 1$ and the states $\lbrace \ket{\lambda_k} \rbrace$, $\lbrace \ket{\lambda'_k} \rbrace$,  $\lbrace \ket{\mu_j} \rbrace$,  $\lbrace \ket{\mu'_j} \rbrace$ form four orthogonal sets. Therefore, the reduction over one of the subsystems can be written as:
		\begin{equation}
			\label{App_A: States_decomposition}
			\psi^{S} = \sum_{k=1}^{D} \lambda_k \ket{\lambda_k}\bra{\lambda_k}, \qquad \phi^{S} = \sum_{j=1}^{d} \mu_{j} \ket{\mu_{j}}\bra{\mu_{j}},
		\end{equation}
		where for simplicity we have removed the subsystem index from the braket notation.\\
		Then, from the \textit{monotonicity of fidelity} [82, 83], we can derive that:
		\begin{equation}
			\label{App_A: Monotonicity_inequality}
			F(\phi,\psi) \leq F\left(\phi^{S},\psi^{S}\right).
		\end{equation}
		
		\item \textit{Symmetry and trace inequality}
		
		Now we can maximize the RHS of the inequality \eqref{App_A: Monotonicity_inequality} by starting from the definition of fidelity and its symmetry:
		\begin{equation}
			F\left(\phi^{S},\psi^{S}\right) = \left(\Tr\left[\sqrt{\sqrt{\phi^{S}}\psi^{S}\sqrt{\phi^{S}}}\right]\right)^2 = F\left(\psi^{S},\phi^{S}\right) = \left(\Tr\left[\sqrt{\sqrt{\psi^{S}}\phi^{S}\sqrt{\psi^{S}}}\right]\right)^2.
		\end{equation}
		Let's insert the decompositions showed in \eqref{App_A: States_decomposition}:
		\begin{equation}
			F\left(\phi^{S},\psi^{S}\right) = \left(\Tr\left[\sqrt{\sqrt{\sum_{k=1}^{D} \lambda_k \ket{\lambda_k}\bra{\lambda_k}}\sum_{j=1}^{d} \mu_{j} \ket{\mu_{j}}\bra{\mu_{j}}\sqrt{\sum_{k'=1}^{D} \lambda_{k'} \ket{\lambda_{k'}}\bra{\lambda_{k'}}}}\right]\right)^2.
		\end{equation}
		Since the states $\left\lbrace\ket{\lambda_k} \right\rbrace$ and $\left\lbrace \ket{\mu_{j}}\right\rbrace$ form two orthogonal sets, we can write:
		\begin{equation}
			\sqrt{\sum_{k=1}^{D} \lambda_k \ket{\lambda_k}\bra{\lambda_k}} = \sum_{k=1}^{D} \sqrt{\lambda_k}\ket{\lambda_k}\bra{\lambda_k}, \qquad \sqrt{\sum_{j=1}^{d} \mu_{j} \ket{\mu_{j}}\bra{\mu_{j}}} = \sum_{j=1}^{d} \sqrt{\mu_{j}} \ket{\mu_{j}}\bra{\mu_{j}},
		\end{equation}
		from which we find:
		\begin{equation}
			\begin{aligned}
				\label{App_A: Fidelity_decomposition}
				F(\phi^{S},\psi^{S})&=\left(\Tr\left[\sqrt{\sum_{k=1}^{D} \sqrt{\lambda_k}\ket{\lambda_k}\bra{\lambda_k}\sum_{j=1}^{d} \mu_{j} \ket{\mu_{j}}\bra{\mu_{j}}\sum_{k'=1}^{D} \sqrt{\lambda_{k'}}\ket{\lambda_{k'}}\bra{\lambda_{k'}}}\right]\right)^2 \\
				&= \left(\Tr\left[\sqrt{\sum_{j=1}^{d} \mu_{j}\sum_{k,k'=1}^{D} \sqrt{\lambda_{k} \lambda_{k'}}\ket{\lambda_k}\braket{\lambda_k}{\mu_{j}}\braket{\mu_{j}}{\lambda_{k'}}\bra{\lambda_{k'}}}\right]\right)^2.
			\end{aligned}
		\end{equation}
		Now we can define the following non-normalized states:
		\begin{equation}
			\label{App_A: v_state_definition}
			\left\lbrace \ket{\Tilde{v}_j}: \quad \ket{\Tilde{v}_j} = \sum_{k=1}^{D} \sqrt{\lambda_{k}}\braket{\lambda_k}{\mu_{j}}\ket{\lambda_k} , \quad j = 1,\dots,d\right\rbrace,
		\end{equation}
		and use them in \eqref{App_A: Fidelity_decomposition} to get:
		\begin{equation}
			\label{App_A: Trace_inequality_Fidelity}
			F\left(\phi^{S},\psi^{S}\right) = \left(\Tr\left[\sqrt{\sum_{j=1}^{d} \mu_{j}\ket{\Tilde{v}_j}\bra{\Tilde{v}_j}}\right]\right)^2 \leq \left(\Tr\left[\sum_{j=1}^{d}\sqrt{ \mu_{j}\ket{\Tilde{v}_j}\bra{\Tilde{v}_j}}\right]\right)^2,
		\end{equation}
		where in the last inequality we have used that for a set of \textit{positive semidefinite} operators $A_i \geq 0$, the following holds:
		\begin{equation}
			\Tr\left[\sqrt{\sum_{i} A_{i}}\right]\leq \Tr\left[\sum_{i}\sqrt{A_{i}}\right].
		\end{equation}
		Since the trace will affect just the operators, we can write:
		\begin{equation}
			\left(\Tr\left[\sum_{j=1}^{d}\sqrt{ \mu_{j}\ket{\Tilde{v}_j}\bra{\Tilde{v}_j}}\right]\right)^2 = \left(\sum_{j=1}^{d}\sqrt{ \mu_{j}}\Tr\sqrt{\ket{\Tilde{v}_j}\bra{\Tilde{v}_j}}\right)^2.
		\end{equation}
		Then, by normalizing the states defined in \eqref{App_A: v_state_definition}, we can realize that:
		\begin{equation}
			\ket{v_{j}}\coloneqq \dfrac{\ket{\Tilde{v}_{j}}}{\sqrt{\braket{\Tilde{v}_{j}}{\Tilde{v}_{j}}}} \qquad \implies \qquad \sqrt{\ket{\Tilde{v}_{j}}\bra{\Tilde{v}_{j}}} = \sqrt{\braket{\Tilde{v}_{j}}{\Tilde{v}_{j}}}\ket{v_j}\bra{v_j},
		\end{equation}
		from which we can derive that:
		\begin{equation}
			\left(\sum_{j=1}^{d}\sqrt{ \mu_{j}}\Tr\sqrt{\ket{\Tilde{v}_j}\bra{\Tilde{v}_j}}\right)^2 = \left(\sum_{j=1}^{d}\sqrt{ \mu_{j}\braket{\Tilde{v}_{j}}{\Tilde{v}_{j}}}\Tr\left[\ket{v_j}\bra{v_j}\right]\right)^2=\left(\sum_{j=1}^{d}\sqrt{ \mu_{j}\braket{\Tilde{v}_{j}}{\Tilde{v}_{j}}}\right)^2.
		\end{equation}
		Again, by using the definition \eqref{App_A: v_state_definition}, we can find:
		\begin{equation}
			\braket{\Tilde{v}_j}{\Tilde{v}_j} = \sum_{k=1}^{D} \lambda_{k}|\braket{\lambda_k}{ \mu_{j}}|^{2},
		\end{equation}
		and therefore we can maximize \eqref{App_A: Trace_inequality_Fidelity} as:
		\begin{equation}
			\label{App_A: Fidelity_for_maxim}
			F(\phi^S,\psi^S) \leq \left(\sum_{j=1}^{d}\sqrt{ \mu_{j}}\sqrt{\sum_{k=1}^{D} \lambda_{k}|\braket{\lambda_k}{ \mu_{j}}|^{2}}\right)^2.
		\end{equation}
		
		\item \textit{Lagrange multipliers method} $-$ $\mu_j$ variables
		
		Now we can define $P_{k}^{j} \coloneqq |\braket{\lambda_k}{\mu_{j}}|^2$ and focus on maximizing the RHS of \eqref{App_A: Fidelity_for_maxim}. Since the power function is monotone, we can actually just look at its argument. Therefore, the maximization problem can be rewritten as:
		\begin{equation}
			\label{App_A: Maximization_problem}
			\max_{\mu_{j},P_{k}^{j}} f(\mu_{j},P_{k}^{j}) = \max_{\mu_{j},P_{k}^{j}} \left[\sum_{j=1}^{d}\sqrt{ \mu_{j}}\sqrt{\sum_{k=1}^{D} \lambda_{k}P_{k}^{j}}\right] \qquad s.t. \qquad \begin{cases}
				\displaystyle\sum_{j = 1}^{d} \mu_{j} = 1, \\
				P_{k}^{j} \geq 0, \\
				\displaystyle\sum_{k=1}^{D} P_{k}^{j} = 1, \\
				\displaystyle\sum_{j=1}^{d} P_{k}^{j} \leq 1.
			\end{cases}
		\end{equation}
		From the constraints we can notice that $\mu_{j}$ and $P_{k}^{j}$ are independent variables. Therefore, we can start maximizing the function just with respect to $\mu_{j}$. We can use the method of \textit{Lagrange multipliers} [84] by defining the following constraint:
		\begin{equation}
			g(\mu_{j}) \coloneqq \sum_{j = 1}^{d} \mu_{j} = 1,
		\end{equation}
		and by finding the solution of the following set of equations:
		\begin{equation}
			\label{App_A: Lagrangian_condition}
			\nabla f(\mu_{j},P_{k}^{j}) = \gamma \nabla g(\mu_{j}), \qquad \forall j = 1,\dots,d.
		\end{equation}
		For the generic $\mu_l$ we get:
		\begin{equation}
			\begin{cases}
				\displaystyle\dfrac{\partial}{\partial \mu_{l}}f(\mu_{j},P_{k}^{j}) = \dfrac{\sqrt{\sum_{k=1}^{D}\lambda_{k}P_{k}^{j}}}{2\sqrt{\mu_{l}}},\\
				\displaystyle\dfrac{\partial}{\partial \mu_{l}}g(\mu_{j}) = \gamma.
			\end{cases}
		\end{equation}
		Since the RHS of \eqref{App_A: Lagrangian_condition} results to be the same for any $j$, we can conclude that the values $\lbrace\mu_{j}^{max}\rbrace$ that solve the problem \eqref{App_A: Maximization_problem} satisfy:
		\begin{equation}
			\sqrt{\dfrac{\sum_{k=1}^{D}\lambda_{k}P_{k}^{1}}{2 \mu_{1}^{max}}} = ... = \sqrt{\dfrac{\sum_{k=1}^{D}\lambda_{k}P_{k}^{d}}{2 \mu_{d}^{max}}}.
		\end{equation}
		Therefore, for two generic $i,j$ we find:
		\begin{equation}
			\dfrac{\mu_{i}^{max}}{\displaystyle\sum_{k=1}^{D}\lambda_{k}P_{k}^{i}} = \dfrac{\mu_{j}^{max}}{\displaystyle\sum_{k=1}^{D}\lambda_{k}P_{k}^{j}} \quad \implies \quad  \dfrac{\mu_{i}^{max}}{\mu_{j}^{max}} = \dfrac{\displaystyle\sum_{k=1}^{D}\lambda_{k}P_{k}^{i}}{\displaystyle\sum_{k=1}^{D}\lambda_{k}P_{k}^{j}}.
		\end{equation}
		We can sum both sides of the equation for $i=1,...,d$ and use the constraint on the $\lbrace \mu_{i} \rbrace$ variables in \eqref{App_A: Maximization_problem} to get:
		\begin{equation}
			\sum_{i=1}^{d} \dfrac{\mu_{i}^{max}}{\mu_{j}^{max}} = \sum_{i=1}^{d}\dfrac{\displaystyle\sum_{k=1}^{D}\lambda_{k}P_{k}^{i}}{\displaystyle\sum_{k=1}^{D}\lambda_{k}P_{k}^{j}} \quad \implies \quad \dfrac{1}{\mu_{j}^{max}} = \sum_{i=1}^{d}\dfrac{\displaystyle\sum_{k=1}^{D}\lambda_{k}P_{k}^{i}}{\displaystyle\sum_{k=1}^{D}\lambda_{k}P_{k}^{j}},
		\end{equation}
		from which we find:
		\begin{equation}
			\label{App_A: Maximum_mu}
			\mu_{j}^{max} = \dfrac{\displaystyle\sum_{k=1}^{D}\lambda_{k}P_{k}^{j}}{\displaystyle\sum_{i=1}^{d}\sum_{k=1}^{D}\lambda_{k}P_{k}^{i}}.
		\end{equation}
		
		\item \textit{Duality theorem} $-$ $P_{k}^{j}$ variables
		
		We can substitute the maximum value $\mu_j^{max}$ into \eqref{App_A: Maximization_problem} and focus now on the maximization with respect to the other variables $P_{k}^{j}$. The new problem becomes:
		\begin{equation}
			\label{App_A: Maximization_problem_2}
			\max_{P_{k}^{j}} \sqrt{\sum_{j=1}^{d}\sum_{k=1}^{D} \lambda_{k}P_{k}^{j}} = \sqrt{\max_{P_{k}^{j}}\sum_{j=1}^{d}\sum_{k=1}^{D} \lambda_{k}P_{k}^{j}} \qquad s.t. \qquad \begin{cases}
				P_{k}^{j} \geq 0, \\
				\displaystyle\sum_{k=1}^{D} P_{k}^{j} = 1, \\
				\displaystyle\sum_{j=1}^{d} P_{k}^{j} \leq 1.
			\end{cases}
		\end{equation}
		where, since the square root is a monotonic map, we can just focus on the maximization of its argument. In order to solve the problem we can write that:
		\begin{equation}
			\label{App_A: Rewriting_function}
			\sum_{j=1}^{d}\sum_{k=1}^{D} \lambda_{k}P_{k}^{j} = \sum_{k=1}^{D-1}\sum_{j=1}^{d}\lambda_{k}P_{k}^{j} + \lambda_{D}\sum_{j=1}^{d}\left(1-\sum_{l=1}^{D-1} P_{l}^{j}\right) = \sum_{k=1}^{D-1}\sum_{j=1}^{d}(\lambda_{k}-\lambda_{D})P_{k}^{j} + d \lambda_{D},
		\end{equation}
		where we have used the second constraint in \eqref{App_A: Maximization_problem_2}. Therefore, since $d \lambda_{D}$ is constant, our maximization problem becomes:
		\begin{equation}
			\label{App_A: Maximization_problem_3}
			\max_{P_{k}^{j}}\, g(P_{k}^{j}) = \sum_{k=1}^{D-1}\sum_{j=1}^{d}(\lambda_{k}-\lambda_{D})P_{k}^{j}\qquad s.t. \qquad \begin{cases}
				P_{k}^{j} \geq 0, \\
				\displaystyle\sum_{k=1}^{D-1} P_{k}^{j} \leq 1, \\
				\displaystyle\sum_{j=1}^{d} P_{k}^{j} \leq 1.
			\end{cases}
		\end{equation}
		This is now a \textit{linear program} (LP) that can be recasted in the following matrix form:
		\begin{equation}
			\label{App_A: Primal_LP}
			\max_{\mathbf{x}}\, g=\mathbf{c}^{T}\mathbf{x}\qquad
			s.t. \quad \begin{cases}
				\mathbf{Ax} \leq \mathbf{b},\\
				\mathbf{x} \geq 0,
			\end{cases}
		\end{equation}
		where we have defined:
		\begin{equation}
			\label{App_A: Primal_vectors}
			\mathbf{x} = 
			\begin{tikzpicture}[baseline={-0.5ex},mymatrixenv]
				\matrix [mymatrix,inner sep=4pt] (m)  
				{
					P_{1}^{1}\\
					\vdots \\
					P_{1}^{d}\\
					P_{2}^{1}\\
					\vdots \\
					P_{2}^{d}\\
					\vdots \\
					P_{D-1}^{1} \\
					\vdots \\
					P_{D-1}^{d}\\
				};
				
				\mymatrixbraceleft{1}{3}{$d$}
				\mymatrixbraceleft{4}{6}{$d$}
				\mymatrixbraceleft{8}{10}{$d$}
			\end{tikzpicture}, \qquad
			\mathbf{c} = 
			\begin{tikzpicture}[baseline={-0.5ex},mymatrixenv]
				\matrix [mymatrix,inner sep=4pt] (m)  
				{
					\lambda_{1}-\lambda_{D}\\
					\vdots \\
					\lambda_{1}-\lambda_{D}\\
					\lambda_{2}-\lambda_{D}\\
					\vdots \\
					\lambda_{2}-\lambda_{D}\\
					\vdots \\
					\lambda_{D-1}-\lambda_{D} \\
					\vdots \\
					\lambda_{D-1}-\lambda_{D} \\
				};
				
				\mymatrixbraceleft{1}{3}{$d$}
				\mymatrixbraceleft{4}{6}{$d$}
				\mymatrixbraceleft{8}{10}{$d$}
			\end{tikzpicture}, \qquad
			\mathbf{b} = 
			\begin{tikzpicture}[baseline={-0.5ex},mymatrixenv]
				\matrix [mymatrix,inner sep=4pt] (m)  
				{
					1 \\
					1 \\
					\vdots \\
					1\\
					1\\
					1 \\
					\vdots \\
					1\\
				};
				
				\mymatrixbraceleft{1}{4}{$D-1$}
				\mymatrixbraceleft{5}{8}{$d$}
			\end{tikzpicture},
		\end{equation}
		\begin{equation}
			\label{App_A: Primal_matrix}
			\mathbf{A} = 
			\begin{tikzpicture}[baseline={-0.5ex},mymatrixenv]
				\matrix [mymatrix,inner sep=4pt] (m)  
				{
					\, 1 1 \dots 1 & 0 0 \dots 0 & \dots & 0 0 \dots 0 \, \\
					\, 0 0 \dots 0 & 1 1 \dots 1 & \dots & 0 0 \dots 0 \, \\
					\vdots & \vdots & & \vdots \\
					\, 0 0 \dots 0 & 0 0 \dots 0 & \dots & 1 1 \dots 1 \, \\
					\, 1 0 \dots 0 & 1 0 \dots 0 & \dots & 1 0 \dots 0 \, \\
					\, 0 1 \dots 0 & 0 1 \dots 0 & \dots & 0 1 \dots 0 \, \\
					\vdots & \vdots & & \vdots \\
					\, 0 0 \dots 1 & 0 0 \dots 1 & \dots & 0 0 \dots 1 \, \\
				};
				
				\mymatrixbracetop{1}{1}{$d$}
				\mymatrixbracetop{2}{2}{$d$}
				\mymatrixbracetop{4}{4}{$d$}
				\mymatrixbraceleft{1}{4}{$D-1$}
				\mymatrixbraceleft{5}{8}{$d$}
			\end{tikzpicture}
		\end{equation}
		The recasting is useful, because it allows us to use the \textit{strong duality theorem} [85, 86] to solve the problem.\\
		Indeed, we can derive the following \textit{dual} LP:
		\begin{equation}
			\label{App_A: Dual_LP}
			\min_{\mathbf{y}}\, g^{*} = \mathbf{b}^{T}\mathbf{y}\qquad
			s.t. \quad \begin{cases}
				\mathbf{A}^{T}\mathbf{y} \geq \mathbf{c},\\
				\mathbf{y} \geq 0,
			\end{cases}
		\end{equation}
		where we have defined:
		\begin{equation}
			\label{App_A: Dual_vectors}
			\mathbf{A}^{T} = 
			\begin{tikzpicture}[baseline={-0.5ex},mymatrixenv]
				\matrix [mymatrix,inner sep=4pt] (m)  
				{
					\, 1 & 0 & 0\dots 0 & 1 & 0 \dots 0 \, \\
					\vdots & \vdots &\vdots & \vdots & \vdots \, \\
					\, 1 & 0 & 0\dots 0 & 0 & 0 \dots 1 \, \\
					\, 0 & 1 & 0\dots 0 & 1 & 0 \dots 0 \, \\
					\vdots & \vdots &\vdots & \vdots & \vdots \, \\
					\, 0 & 1 & 0 \dots 0 & 0 & 0 \dots 1 \, \\
					\vdots & \vdots &\vdots & \vdots & \vdots \, \\
					\, 0 & 0 & 0\dots 1 & 1 & 0 \dots 0 \, \\
					\vdots & \vdots &\vdots & \vdots & \vdots \, \\
					\, 0 & 0 & 0\dots 1 & 0 & 0 \dots 1 \, \\
				};
				
				\mymatrixbracetop{1}{3}{$D-1$}
				\mymatrixbracetop{4}{5}{$d$}
				\mymatrixbraceleft{1}{3}{$d$}
				\mymatrixbraceleft{4}{6}{$d$}
				\mymatrixbraceleft{8}{10}{$d$}
			\end{tikzpicture}, \qquad
			\mathbf{y} = 
			\begin{tikzpicture}[baseline={-0.5ex},mymatrixenv]
				\matrix [mymatrix,inner sep=4pt] (m)  
				{
					y_1 \\
					y_2 \\
					\vdots \\
					y_{D-1}\\
					y_{D}\\
					\vdots \\
					y_{D-1+d}\\
				};
				
				\mymatrixbraceleft{1}{4}{$D-1$}
				\mymatrixbraceleft{5}{7}{$d$}
			\end{tikzpicture}.
		\end{equation}
		Therefore, explicitly the \textit{dual} LP in \eqref{App_A: Dual_LP} becomes:
		\begin{equation}
			\label{App_A: Dual_LP_explicit}
			\min_{\mathbf{y}}\, g^{*} = \sum_{k=1}^{D-1+d} y_{k} \qquad s.t \quad \begin{cases}
				\begin{rcases}
					y_1 + y_D \geq \lambda_{1}-\lambda_{D}, \\
					y_1 + y_{D+1} \geq \lambda_{1}-\lambda_{D}, \\
					\qquad \vdots \\
					y_1 + y_{D-1+d} \geq \lambda_{1}-\lambda_{D},
				\end{rcases} d \\
				\qquad \vdots \\
				\begin{rcases}
					y_{D-1} + y_D \geq \lambda_{D-1}-\lambda_{D}, \\
					y_{D-1} + y_{D+1} \geq \lambda_{D-1}-\lambda_{D}, \\
					\qquad \vdots \\
					y_{D-1} + y_{D-1+d} \geq \lambda_{D-1}-\lambda_{D}.
				\end{rcases} d
			\end{cases}
		\end{equation}
		In order to minimize the function $g^{*}$ we can expand the summation:
		\begin{equation}
			g^{*} = \sum_{k=1}^{D-1+d} y_{k} = y_{1} + y_{2} + \dots + y_{d} + \dots + y_{D-1} + y_{D} + \dots + y_{D-1+d},
		\end{equation}
		and group the terms in the following way:
		\begin{align}
			\label{App_A: Minimization_g*}
			\begin{split}
				g^{*} &= (y_{1} + y_{D}) + (y_{2} + y_{D+1}) + \dots + (y_{d} + y_{D-1+d}) + \sum_{k=d+1}^{D-1} y_{k} \\
				&\geq (\lambda_{1}-\lambda_{D}) + (\lambda_{2}-\lambda_{D}) + \dots + (\lambda_{d}-\lambda_{D}) + \sum_{k=d+1}^{D-1} y_{k},
			\end{split}
		\end{align}
		where in the last inequality we have used the constraints in \eqref{App_A: Dual_LP_explicit}.\\
		Without loss of generality we can decompose the state $\psi$ in \eqref{App_A: States_decomposition} such that its spectrum:
		\begin{equation}
			\label{App_A: Decomposition_choice}
			\lambda(\psi) = \lbrace \lambda_{k}: \, \lambda_{1} \geq \lambda_{2} \geq ... \geq \lambda_{D} \rbrace.
		\end{equation}
		With such a decomposition we will have that:
		\begin{equation}
			\lambda_{1}-\lambda_D \geq \lambda_2-\lambda_D \geq ... \geq \lambda_{D-1}-\lambda_D,
		\end{equation}
		and therefore we can minimize $g^{*}$ by choosing $y_{d+1} = \dots = y_{D-1} = 0$, since this will always satisfy the constraints in \eqref{App_A: Dual_LP_explicit}. For example, let's choose $y_{d+1} = 0$. Then from the constraints of problem \eqref{App_A: Dual_LP_explicit} we will have:
		\begin{equation}
			y_{d+1} + y_{D} \geq \lambda_{d+1}-\lambda_{D} \quad \implies \quad y_{D} \geq \lambda_{d+1}-\lambda_{D}.
		\end{equation}
		This imposes a lower bound on the value of $y_D$, that shouldn't be in contrast with all the other constraints. However, since the spectrum of the state satisfies \eqref{App_A: Decomposition_choice}, the following constraints are still fulfilled:
		\begin{equation}
			y_{i} + y_{D} \geq \lambda_{i} - \lambda_{D} \qquad \forall i = \lbrace 1,...,d,...,D-1 \rbrace.
		\end{equation}
		The same will happen with the values $y_{D+1},\dots,y_{D-1+d}$, once we impose the other constraints: also in those cases we will find lower bounds that are not in contrast with each other.\\
		In this way we can cancel the last term in \eqref{App_A: Minimization_g*} and derive the following lower bound:
		\begin{equation}
			\label{App_A: Minimization_Dual}
			g^{*} \geq \sum_{i=1}^{d} \lambda_{i} - d\lambda_{D}.
		\end{equation}
		Now we can go back to the \textit{primal} LP in \eqref{App_A: Primal_LP} and observe that there exists a specific choice of values for the variables $\lbrace P_{k}^{j}\rbrace$ such that:
		\begin{equation}
			\label{App_A: Optimal_solution_primal}
			g(P_{k}^{j}) = \sum_{i=1}^{d} \lambda_{i} - d\lambda_{D}, \qquad \text{for the choice: } \quad \begin{cases}
				\displaystyle P_{k}^{j} = \dfrac{1}{d}, \quad \forall k = 1,...,d,\\
				\displaystyle P_{k}^{j} = 0,\quad  otherwise.
			\end{cases} 
		\end{equation}
		Therefore, there exists a feasible solution \eqref{App_A: Optimal_solution_primal} of the \textit{primal} LP and a feasible solution \eqref{App_A: Minimization_Dual} of the \textit{dual} LP for which the \textit{primal} and the \textit{dual} coincide. Since from the strong duality theorem we know that this is possible only in presence of an \textit{optimal} solution for both problems, we can realize that the one found in \eqref{App_A: Optimal_solution_primal} is precisely the solution of the maximization problem in \eqref{App_A: Maximization_problem_3}.\\
		Now, by inserting this solution into \eqref{App_A: Rewriting_function}, we can solve the original problem \eqref{App_A: Maximization_problem_2}:
		\begin{equation}
			\max_{P_{k}^{j}} \sqrt{\sum_{j=1}^{d}\sum_{k=1}^{D} \lambda_{k}P_{k}^{j}} = \sqrt{\sum_{i=1}^{d}\lambda_{i}}.
		\end{equation}
		By remembering that the LHS of the latter is the result of the maximization over $\lbrace \mu_{j} \rbrace$ of \eqref{App_A: Maximization_problem}, we can finally maximize the fidelity in \eqref{App_A: Fidelity_for_maxim} and find that:
		\begin{equation}
			F(\phi^{S},\psi^{S}) \leq \sum_{i=1}^{d} \lambda_{i},
		\end{equation}
		that, combined with \eqref{App_A: Monotonicity_inequality} gives the \textit{Result 1}.
		
	\end{itemize}
	
	\newpage
	
	\section{Fidelity witness for GHZ, cluster and AME states}
	\label{Appendix_Fidelity-evaluation}
	In this section we evaluate the bound on the fidelity (Result 1) for any state $\rho$ with GME-dimension no larger than $d_{\text{GME}}$, with three different seminal families of $n$-partite pure target states: GHZ states, AME states and linear cluster states. In all the cases we need to find the bipartition $(S|\bar{S})$ for which the sum of the first $d_{\text{GME}}$ eigenvalues, ordered non-increasingly, of the reduced state is maximum.
	\begin{itemize}
		\item \textit{GHZ state}
		
		Since the $n$-partite GHZ state $\ket{\text{ghz}_{n,d}} = \frac{1}{\sqrt{d}} \sum_{k = 0}^{d-1} \ket{k}^{\otimes n}$ is already expressed in its Schmidt decomposition, for any bipartition $(S|\bar{S})$ the spectrum of the reductions has always the same non-zero values $\lbrace \lambda_{i} = 1/d, \quad i = 1,\dots,d \rbrace$. Therefore, the fidelity bound in this case becomes
		\begin{equation}
			\label{Appendix_Fid-evaluation: GHZ-bound}
			F_{\text{ghz}_{n,d}}(\rho) \leq \max_{\lbrace S|\bar{S} \rbrace}\sum_{i = 1}^{d_{\text{GME}}} \lambda_i = \frac{d_{\text{GME}}}{d}.
		\end{equation}
		
		\item \textit{AME states}
		
		Also in this case the fidelity bound can be found by looking at the definition of the state. A $n$-partite state $\ket{\psi}$ is called \textit{absolutely maximally entangled} (AME) if for any bipartition $(S|\bar{S})$ its reduction is maximally mixed, i.e. if we choose $S$ to be always the smallest subset of the bipartition, we have that the spectrum of the reductions is $\lbrace \lambda_{i}^{S} = 1/d^{|S|}, \quad i = 1,\dots, d^{|S|} \rbrace$ for any $(S|\bar{S})$. Therefore, the bipartitions that maximize the sum of the first $d_{\text{GME}}$ eigenvalues of the reductions are the ones such that $|S| = 1$. Then the fidelity bound reads:
		\begin{equation}
			F_{\text{AME}}(\rho) \leq \max_{\lbrace S|\bar{S} \rbrace} \sum_{i = 1}^{d_{\text{GME}}} \lambda_i^{S} = \frac{d_{\text{GME}}}{d}.
		\end{equation}
		
		\item \textit{Linear cluster state}
		
		In this case the derivation of the fidelity bound is less straightforward than before, since it is not possible to obtain the spectrum of all possible reductions simply by looking at the generic expression of the linear cluster state, 
		\begin{align}\nonumber
			\label{Appendix_Wcluster: Cluster_state}
			\ket{C_{n,d}} = \frac{1}{d^{n/2}}\bigotimes_{a=1}^{n}\left(\sum_{k=0}^{d-1}\ket{k}_{a}Z^{k}_{a+1}\right) &= \frac{1}{d^{n/2}} \sum_{k_{1} = 0}^{d-1}\ket{k_1} \otimes \sum_{k_2 = 0}^{d-1} Z^{k_1} \ket{k_2} \otimes \cdots \otimes \sum_{k_n = 0}^{d-1} Z^{k_{n-1}}\ket{k_n}\\
			&=\frac{1}{d^{n/2}} \sum_{k_{1},\dots,k_{n} = 0}^{d-1}\omega^{k_{1}k_{2} + \dots + k_{n-1}k_n} \ket{k_1 \dots k_n}.
		\end{align}
		Therefore, we can proceed in a different way. Fix the number of parties $n$. Let us group all possible bipartitions $(S|\bar{S})$ whose smallest subset $S$ has the same cardinality $|S|$ in collections $\mathcal{B}_{|S|}$ (e.g. for $n = 4$ we have two collections: $\mathcal{B}_{1} = \lbrace (1|234),(2|134),(3|124),(4|123) \rbrace$ and $\mathcal{B}_{2} = \lbrace (12|34), (13|24), (14|23) \rbrace$).
		We can realize that for all the bipartitions living in the same collection, the reductions can only have a finite number of possible structures.\\
		As an example, let us fix $n = 6$ and consider three possible bipartitions from the collection $\mathcal{B}_{3}$: $(123|456), (156|	234)$ and $(145|236)$. The reductions over the $S$ subset are:
		\begin{equation}
			\begin{aligned}
				\tr_{123}(C_{6,d}) = \frac{1}{d^3} \sum_{k_4 = 0}^{d-1}\sum_{k_5,k'_5 = 0}^{d-1}\sum_{k_6,k'_6 = 0}^{d-1}\omega^{k_4(k_5 - k'_5)}\, \omega^{k_5k_6 - k'_5k'_6} \ketbra{k_4 k_5 k_6}{k_4 k'_5 k'_6},
			\end{aligned}
		\end{equation}
		\begin{equation}
			\tr_{156}(C_{6,d}) = \frac{1}{d^3} \sum_{k_2 = 0}^{d-1}\sum_{k_3,k'_3 = 0}^{d-1} \sum_{k_4 = 0}^{d-1} \omega^{(k_2+k_4)(k_3 - k'_3)} \ketbra{k_2 k_3 k_4}{k_2 k'_3 k_4},
		\end{equation}
		\begin{equation}
			\tr_{145}(C_{6,d}) = \frac{1}{d^3}\sum_{k_2,k_3,k_6 = 0}^{d-1}\ketbra{k_2 k_3 k_6}.
		\end{equation}
		These reductions have three different structures that can be distinguished by the number of $k'$ terms in the tensor product of the basis vectors. By computing the other possible reductions in the collection $\mathcal{B}_3$, we can realize that all the structures fall into one of these three cases. Indeed, in general, fixed the number of parties in our reduction, the whole set of possible matrices can be constructed by changing the position and the number of the $k'$ terms.\\
		Therefore, given a generic collection $\mathcal{B}_{|S|}$, where we label the elements in every subset $S$ as $\lbrace S_1,\dots,S_p\rbrace$, the reductions over the subsets $\bar{S}$ can only have the following shapes:
		\begin{equation}
			\label{Appendix_Fidelity-evaluation: Reduction_cases}
			\tr_{\bar{S}}(C_{n,d}) = \frac{1}{d^{|S|}}
			\begin{cases}
				\displaystyle
				\sum_{S_1,\dots,S_p = 0}^{d-1} \ketbra{S_1 \dots S_p},\\
				\displaystyle
				\sum_{S \setminus \lbrace S_i \rbrace} \sum_{S_i,S'_i = 0}^{d-1} \omega^{(S_{i-1} + S_{i+1})(S_i - S'_i)} \ketbra{S_1 \dots S_i \dots S_p}{S_1 \dots S'_i \dots S_p},\\
				\quad \quad \quad \vdots\\
				\displaystyle
				\sum_{S_1 = 0}^{d-1} \sum_{S_2,S'_2 = 0}^{d-1} \dots \sum_{S_p,S'_p = 0}^{d-1} \omega^{S_1(S_2 - S'_2)}\cdots \omega^{S_{p-1}S_p - S'_{p-1}S'_p}\ketbra{S_1 S_2 \dots S_p}{S_1 S'_2 \dots S'_p}.
			\end{cases}
		\end{equation}
		To study the spectrum of all the reductions in the collection $\mathcal{B}_{|S|}$, then we need to diagonalize each of the possible matrices. 
		By defining with $A'_{m} = \lbrace S'_{i_1},\dots,S'_{i_m} \rbrace$ the generic set of the $S'$ terms and with $A_{m} = \lbrace S_{i_1},\dots,S_{i_m} \rbrace$ the set of the corresponding elements in the original subset, we can group all the cases in \eqref{Appendix_Fidelity-evaluation: Reduction_cases} into a single expression:
		\begin{equation}
			\label{Appendix_Fidelity-evaluation: generic_reduction}
			\rho_{S}^{m} = \frac{1}{d^{|S|}}\sum_{S}\sum_{S'_{i_1}, \dots, S'_{i_m} = 0}^{d-1} \omega^{(S_{i_1-1} + S_{i_1+1})(S_{i_1} - S'_{i_1})}\, \dots \, \omega^{(S_{i_m-1} + S_{i_m+1})(S_{i_m} - S'_{i_m})}\ketbra{S}{(S \setminus A) \cup A'},
		\end{equation}
		where $0 \leq m \leq |S|-1$. Indeed, because of the shape of the linear cluster state in \eqref{Appendix_Wcluster: Cluster_state}, for a reduction of $|S|$ parties it is not possible to have a number of $S'$ terms larger than $|S|-1$. Then it is possible to show that the following states
		\begin{equation}
			\ket{g_{q_1q_2\dots q_{|S|}}} = \frac{1}{\sqrt{d^{|S|}}}\sum_{l_1,\dots,l_{|S|} = 0}^{d-1}\omega^{l_1 l_2 + \dots + l_{|S|-1}l_{|S|}} \, \omega^{q_1 l_1 + \dots + q_{|S|-2}l_{|S|-2}}\ket{l_1\dots l_{|S|-2}\,(l_{|S|-1} - q_{|S|-1})\,(l_{|S|} - q_{|S|})}
		\end{equation}
		with $q_1,q_2,\dots,q_{|S|} \in \lbrace 0,\dots,d-1 \rbrace$ diagonalize the matrix \eqref{Appendix_Fidelity-evaluation: generic_reduction} such that
		\begin{equation}
			\rho_{S}^{m}\ket{g_{q_1 q_2\dots q_{|S|}}} = \frac{1}{d^{|S|-m}}\ket{g_{q_1 q_2 \dots q_{|S|}}}.
		\end{equation}
		With this result we can now calculate the fidelity bound. Indeed, we have shown that for any bipartition $(S|\bar{S})$ the spectrum of the reduction is always uniform with values within $\lbrace 1/d,\dots,1/d^{|S|} \rbrace$. Therefore, the sum of the first $d_{\text{GME}}$ eigenvalues is maximum for the bipartitions $(S|\bar{S})$ in which $\lbrace \lambda_i^S = 1/d, \quad i = 1,\dots,d\rbrace$ and the fidelity bound reads:
		\begin{equation}
			\label{Appendix_Fidelity-evaluation: Cluster-bound}
			F_{\text{cluster}}(\rho) \leq \max_{\lbrace S|\bar{S} \rbrace} \sum_{i = 1}^{d_{\text{GME}}}\lambda_{i} = \frac{d_{\text{GME}}}{d}.
		\end{equation}

	\end{itemize}
	
	\newpage
	
	\section{Proof of Result 2}
	\label{Appendix_WGHZ}
	Let us consider the following operator:
	\begin{equation}
		\label{Appendix_WGHZ: Operator_definition}
		O^\text{ghz}_{n,d}=\sum_{j=0}^{d-1} \ketbra{j}{j}^{\otimes n} + \sum_{\substack{j_1,\ldots,j_n=0}}^{d-1} \bigotimes_{l=1}^n \ketbra{e_{j_l}}{e_{j_l}}\delta_{j_1 \oplus \ldots \oplus j_n,0},
	\end{equation}
	where $\oplus$ denotes addition modulo $d$, $\{\ket{j}\}_{j=0}^{d-1}$ is the computational basis and  $\{\ket{e_j}\}_{j=0}^{d-1}$ is the Fourier basis obtained from
	\begin{equation}
		\label{Appendix_WGHZ: Fourier-basis}
		\ket{e_j}=\frac{1}{\sqrt{d}}\sum_{k=0}^{d-1}\omega^{jk}\ket{k}, \qquad \text{with: } \omega=e^{\frac{2\pi i}{d}}.
	\end{equation}
	The proof of our result for the entanglement witness $\mathcal{W}^\text{ghz}_{n,d}(\rho)=\tr\left(\rho	O^\text{ghz}_{n,d}\right)$ can be provided in three steps:
	\begin{itemize}
		\item \textit{Show that $\ket{\text{ghz}_{n,d}}$ has perfect correlations for $O^\text{ghz}_{n,d}$ and therefore $\mathcal{W}^\text{ghz}_{n,d}(\ket{\text{ghz}_{n,d}}) = 2$.} \\
		\newline
		For simplicity, let us separately consider the measurement operators in $O^\text{ghz}_{n,d}$:
		\begin{equation}
			A = \sum_{j=0}^{d-1} \ketbra{j}{j}^{\otimes n}, \qquad B = \sum_{\substack{j_1,\ldots,j_n=0}}^{d-1} \bigotimes_{l=1}^n \ketbra{e_{j_l}}{e_{j_l}}\delta_{j_1 \oplus \ldots \oplus j_n,0}.
		\end{equation}
		and show that $\ket{\text{ghz}_{n,d}}$ is an eigenstate of both $A$ and $B$.\\
		For $A$ we can use the orthonormality of the computational basis $\braket{l}{x}=\delta_{lx}$ to get that $\ket{\text{ghz}_{n,d}}$ is an eigenstate.\\
		For $B$ the situation is slightly different:
		\begin{equation}
			B\ket{\text{ghz}_{n,d}} = \left(\sum_{\substack{j_1,\ldots,j_n=0}}^{d-1} \bigotimes_{l=1}^n \ketbra{e_{j_l}}{e_{j_l}}\delta_{j_1 \oplus \ldots \oplus j_n,0}\right)\dfrac{1}{\sqrt{d}}\sum_{x=0}^{d-1}\ket{x}^{\otimes n}. 
		\end{equation}
		By using the definition \eqref{Appendix_WGHZ: Fourier-basis} of the Fourier basis, we get:
		\begin{equation}
			\begin{aligned}
				\label{Appendix_WGHZ: B_eigenstate}
				B\ket{\text{ghz}_{n,d}}&=\dfrac{1}{\sqrt{d}}\sum_{j_1 \oplus \dots \oplus j_n = 0}\,\sum_{x=0}^{d-1} \left(\dfrac{1}{\sqrt{d^n}}\,\sum_{k_1,\dots,k_{n}=0}^{d-1}\omega^{-k_{1}j_{1}}\,\dots\,\omega^{-k_{n}j_{n}} \, \underbrace{\braket{k_{1}\dots k_{n}}{x\dots x}}_{\delta_{k_{i}x}}\right)\ket{e_{j_{1}}\dots e_{j_{n}}}\\
				&=\dfrac{1}{\sqrt{d^{n+1}}}\sum_{j_1 \oplus \dots \oplus j_n = 0}\,\sum_{x=0}^{d-1}\,\omega^{-\sum_{\gamma=1}^{n}j_{\gamma}x}\ket{e_{j_{1}}\dots e_{j_{n}}}=\dfrac{1}{\sqrt{d^{n-1}}}\sum_{j_1 \oplus \dots \oplus j_n = 0}\ket{e_{j_{1}}\dots e_{j_{n}}}.
			\end{aligned}
		\end{equation}
		By performing a change of basis, we can show that this state is actually the original GHZ state. Indeed, if we consider that
		\begin{equation}
			\ket{x}=\sum_{j=0}^{d-1}\ket{e_{j}}\braket{e_{j}}{x}=\sum_{j=0}^{d-1}\ket{e_{j}}\, \dfrac{1}{\sqrt{d}}\sum_{k=0}^{d-1}\,\omega^{-kj}\,\underbrace{\braket{k}{x}}_{\delta_{kx}}=\dfrac{1}{\sqrt{d}}\sum_{j=0}^{d-1}\,\omega^{-jx}\,\ket{e_{j}},
		\end{equation}
		we can rewrite
		\begin{equation}
			\label{Appendix_WGHZ: Change_basis_GHZ}
			\ket{\text{ghz}_{n,d}}=\dfrac{1}{\sqrt{d}}\,\sum_{x=0}^{d-1}\ket{x}^{\otimes n}=\dfrac{1}{\sqrt{d}}\, \dfrac{1}{\sqrt{d^n}}\sum_{j_{1},\dots,j_{n}=0}^{d-1}\left(\sum_{x=0}^{d-1}\omega^{-\sum_{\gamma}j_{\gamma}x}\right)\ket{e_{j_{1}}\dots e_{j_{n}}} = \dfrac{1}{\sqrt{d^{n-1}}}\sum_{j_1 \oplus \dots \oplus j_n = 0}\ket{e_{j_{1}}...e_{j_{n}}},
		\end{equation}
		where in the last equality we have used that the sum over $x$ is a geometric series:
		\begin{equation}
			\label{Appendix_WGHZ: Geometric_series}
			\sum_{x=0}^{d-1}\,\omega^{-Jx}=\dfrac{1-\omega^{-Jd}}{1-\omega^{-J}}=\begin{cases}
				0 \quad \leftrightarrow \, J\neq 0,\\
				d \quad \leftrightarrow \, J=0\,(\text{mod } d),
			\end{cases}
		\end{equation}
		with $J\coloneqq \sum_{\gamma=1}^{n}j_{\gamma}$. We can realize that the final state in \eqref{Appendix_WGHZ: Change_basis_GHZ} is precisely the one obtained in \eqref{Appendix_WGHZ: B_eigenstate} and	therefore we can conclude that:
		\begin{equation}
			\label{Appendix_WGHZ: Eigenstate_GHZ_W}
			\begin{cases}
				A \ket{\text{ghz}_{n,d}}=\ket{\text{ghz}_{n,d}},\\
				B \ket{\text{ghz}_{n,d}}=\ket{\text{ghz}_{n,d}}
			\end{cases}
			\implies O^\text{ghz}_{n,d} \ket{\text{ghz}_{n,d}}=2\,\ket{\text{ghz}_{n,d}}.
		\end{equation}
		\item \textit{Show that the spectral decomposition of $O^\text{ghz}_{n,d}$ is $\lambda ( O^\text{ghz}_{n,d}) = \lbrace \lambda_{\text{ghz}_{n,d}}, \lambda_2, \dots, \lambda_{d^n} \rbrace$ such that $\lambda_{i}\le 1 \quad \forall i \neq \text{ghz}_{n,d}$}\\
		\newline
		Since $O^\text{ghz}_{n,d} \ge 0$, it admits a spectral decomposition:
		\begin{equation}
			O^\text{ghz}_{n,d}=\sum_{i=1}^{d^n}\lambda_{i}\ket{\lambda_{i}}\bra{\lambda_{i}}.
		\end{equation}
		From the previous step we know that one of the possible eigenstates is the $\ket{\text{ghz}_{n,d}}$ state for which $\lambda_{\text{ghz}_{n,d}} = 2$. Therefore, we can write that
		\begin{equation}
			\label{Appendix_WGHZ: Operator_spectral_decomposition}
			O^\text{ghz}_{n,d}=2\,\ketbra{\text{ghz}_{n,d}}{\text{ghz}_{n,d}}+\sum_{i=2}^{d^n}\lambda_{i}\ketbra{\lambda_{i}}{\lambda_{i}}.
		\end{equation}
		For simplicity, let us define $\text{ghz}_{n,d} = \ketbra{\text{ghz}_{n,d}}{\text{ghz}_{n,d}}$ and let us decompose the previous expression as
		\begin{equation}
			\label{Appendix_WGHZ: Witness_add_rem_GHZ}
			O^\text{ghz}_{n,d}=\text{ghz}_{n,d} + \left(\text{ghz}_{n,d}+\sum_{i=2}^{d^n}\lambda_{i}\ketbra{\lambda_{i}}{\lambda_{i}}\right),
		\end{equation}
		so that we can define the following shifted operator:
		\begin{equation}
			\Tilde{O}^\text{ghz}_{n,d} =	O^\text{ghz}_{n,d} - \text{ghz}_{n,d}.
		\end{equation}
		We now prove that this shifted operator is a projector, and consequently that all its eigenvalues are either zero or one.\\
		To do that we calculate its square:
		\begin{equation}
			\label{Appendix_WGHZ: Tilde_W_square}
			(\Tilde{O}^\text{ghz}_{n,d})^2=(O^\text{ghz}_{n,d})^2 + \text{ghz}_{n,d} - \lbrace O^\text{ghz}_{n,d},\text{ghz}_{n,d} \rbrace,
		\end{equation}
		where the anticommutator can be computed by using \eqref{Appendix_WGHZ: Eigenstate_GHZ_W}:
		\begin{equation}
			\lbrace O^\text{ghz}_{n,d},\text{ghz}_{n,d} \rbrace = O^\text{ghz}_{n,d} \ketbra{\text{ghz}_{n,d}}{\text{ghz}_{n,d}} + \ketbra{\text{ghz}_{n,d}}{\text{ghz}_{n,d}} O^\text{ghz}_{n,d} = 4\, \text{ghz}_{n,d}.
		\end{equation}
		The $(O^\text{ghz}_{n,d})^2$ term instead can be calculated by using the definition \eqref{Appendix_WGHZ: Operator_definition}:
		\begin{equation}
			\begin{aligned}
				(O^\text{ghz}_{n,d})^2=&\left[\sum_{j=0}^{d-1} \ketbra{j}{j}^{\otimes n} + \sum_{\substack{j_1,\ldots,j_n=0}}^{d-1} \bigotimes_{l=1}^n \ketbra{e_{j_l}}{e_{j_l}}\delta_{j_1 \oplus \ldots \oplus j_n,0}\right]  \\
				& \cdot\left[\sum_{j'=0}^{d-1} \ketbra{j'}{j'}^{\otimes n} + \sum_{\substack{j'_1,\ldots,j'_n=0}}^{d-1} \bigotimes_{l=1}^n \ketbra{e_{j'_l}}{e_{j'_l}}\delta_{j'_1 \oplus \ldots \oplus j'_n,0}\right],
			\end{aligned}
		\end{equation}
		in which we can simplify the non-mixing terms by using the orthonormality of both computational and Fourier basis:
		\begin{equation}
			\begin{aligned}
				&(O^\text{ghz}_{n,d})^2=\underbrace{\sum_{j=0}^{d-1} \ketbra{j}{j}^{\otimes n} + \sum_{\substack{j_1,\ldots,j_n=0}}^{d-1} \bigotimes_{l=1}^n \ketbra{e_{j_l}}{e_{j_l}}\delta_{j_1 \oplus \ldots \oplus j_n,0}}_{O^\text{ghz}_{n,d}}\\
				&+\sum_{\substack{j=0\\j'_{1}\oplus \dots \oplus j'_{n}=0}}^{d-1} \ket{j \dots j}\underbrace{\braket{j \dots j}{e_{j'_{1}}\dots e_{j'_{n}}}}_{\frac{1}{\sqrt{d^n}}\omega^{J'j}}\bra{e_{j'_{1}}\dots e_{j'_{n}}} + \sum_{\substack{j'=0\\j_{1}\oplus \dots \oplus j_{n}=0}}^{d-1}\ket{j'\dots j'}\underbrace{\braket{j'\dots j'}{e_{j_{1}}\dots e_{j_{n}}}}_{\frac{1}{\sqrt{d^n}}\omega^{-Jj'}}\bra{e_{j_{1}}\dots e_{j_{n}}},
			\end{aligned}
		\end{equation}
		where we are using $J\coloneqq \sum_{\gamma=1}^{n}j_{\gamma}$ and $J'\coloneqq \sum_{\gamma=1}^{n}j'_{\gamma}$. We end up with
		\begin{equation}
			(O^\text{ghz}_{n,d})^2=O^\text{ghz}_{n,d} + \sum_{\substack{j=0\\j'_{1}\oplus \dots \oplus j'_{n}=0}}^{d-1} \dfrac{1}{\sqrt{d}}\,\ket{j \dots j}\bra{e_{j'_{1}}\dots e_{j'_{n}}}\,\dfrac{1}{\sqrt{d^{n-1}}}+ \sum_{\substack{j'=0\\j_{1}\oplus \dots \oplus j_{n}=0}}^{d-1} \dfrac{1}{\sqrt{d^{n-1}}}\,\ket{e_{j_{1}}\dots e_{j_{n}}}\bra{j'\dots j'}\,\dfrac{1}{\sqrt{d}}.
		\end{equation}
		By looking at \eqref{Appendix_WGHZ: Change_basis_GHZ} we can realize that the previous expression can be simply written as
		\begin{equation}
			(O^\text{ghz}_{n,d})^2 = O^\text{ghz}_{n,d} + 2 \,\text{ghz}_{n,d}.
		\end{equation}
		Now we can insert this result into \eqref{Appendix_WGHZ: Tilde_W_square} and get
		\begin{equation}
			(\Tilde{O}^\text{ghz}_{n,d})^2 = (O^\text{ghz}_{n,d})^2 + \text{ghz}_{n,d} - 4\,\text{ghz}_{n,d} = O^\text{ghz}_{n,d}-\text{ghz}_{n,d} = \Tilde{O}^\text{ghz}_{n,d},
		\end{equation}
		which proves that $\Tilde{O}^\text{ghz}_{n,d}$ is a projector.
		
		\item \textit{Derivation of the upper bound for $\mathcal{W}^\text{ghz}_{n,d}(\rho)=\tr\left(\rho	O^\text{ghz}_{n,d}\right)$}\\
		\newline
		In order to maximize the witness, we can now use the spectral decomposition \eqref{Appendix_WGHZ: Witness_add_rem_GHZ} of the operator $O^\text{ghz}_{n,d}$:
		\begin{equation}
			\label{Appendix_WGHZ: Witness_bound_derivation}
			\mathcal{W}^\text{ghz}_{n,d}(\rho) = \tr\left(\rho	O^\text{ghz}_{n,d}\right) = F_{\text{ghz}_{n,d}}(\rho) + \tr\left(\rho \left(\text{ghz}_{n,d} + \sum_{i=2}^{d^n}\lambda_{i}\ketbra{\lambda_{i}}{\lambda_{i}}\right)\right) \leq F_{\text{ghz}_{n,d}}(\rho) + 1,
		\end{equation}
		where in the last inequality we have used that
		\begin{equation}
			\lambda(O^\text{ghz}_{n,d}) = \lbrace 2, \lambda_{2},\dots ,\lambda_{d^n}\, :\, \lambda_{i} \le 1\rbrace \quad \implies \quad \text{ghz}_{n,d} + \sum_{i=2}^{d^n}\lambda_{i}\ketbra{\lambda_{i}}{\lambda_{i}} \leq \openone.
		\end{equation}
		We can realize that $F_{\text{ghz}_{n,d}}(\rho)$ is the fidelity between a $n$-partite pure GHZ state and a state $\rho$ with $\mathcal{D}_{\textit{GME}}(\rho) \leq d_{\textit{GME}}$, that can be bounded using Result 1 as shown in \eqref{Appendix_Fid-evaluation: GHZ-bound}.
		By inserting this result into \eqref{Appendix_WGHZ: Witness_bound_derivation} we get:
		\begin{equation}
			\mathcal{W}^\text{ghz}_{n,d}(\rho) \leq 1 + \dfrac{d_{\text{GME}}}{d},
		\end{equation}
		that is the result we wanted to prove. Moreover, by looking at the last inequality in \eqref{Appendix_WGHZ: Witness_bound_derivation} we can realize that
		\begin{equation}
			F_{\text{ghz}_{n,d}}(\rho)\geq \mathcal{W}^\text{ghz}_{n,d}(\rho) - 1.
		\end{equation}
		
	\end{itemize}
	
	\newpage
	\section{Optimality of minimal witness for dephased GHZ state}
	\label{Appendix_Fidel-Tigh}
	The proof of this result parallels the approach used in [87] to prove the tightness of the fidelity bound in the case of a dephased bipartite maximally entangled state. We consider  a dephased high-dimensional $n$-partite GHZ state,
	\begin{equation}
		\label{Appendix_Fidel-Tigh: Dephased_GHZ}
		\rho = v\ketbra{\text{ghz}_{n,d}} + \frac{1-v}{d}\sum_{i = 0}^{d-1}\ketbra{i}^{\otimes n} = \frac{1}{d} \sum_{i = 0}^{d-1} \ketbra{i}^{\otimes n} + \frac{v}{d}\sum_{i\neq j = 0}^{d-1} \ketbra{i}{j}^{\otimes n},
	\end{equation}
	where in the second equality we wrote the state by expanding GHZ in the computational basis for later convenience.\\
	From Result 1 we know that, if the GME-dimension of $\rho$ is no larger than $d_{\text{GME}}$, then its fidelity with the GHZ state is bounded by
	\begin{equation}
		F_{\text{ghz}}(\rho) \leq \max_{\lbrace S|\bar{S} \rbrace}\sum_{i = 1}^{d_{\text{GME}}} \lambda_{i} = \frac{d_{\text{GME}}}{d},
	\end{equation}
	where we used that the spectrum of the reductions of $\text{ghz}_{n,d}$ over all possible bipartitions is always $\lbrace \lambda_i = 1/d,\quad i = 1,\dots,d\rbrace$.\\
	Now we can prove that a dephased GHZ state fulfilling this bound necessarily has a GME-dimension no larger than $d_{\text{GME}}$.\\
	We can start by considering that for any $d_{\text{GME}}$ it is always possible to find a critical visibility $v^{*}$ and the associated state $\rho^{*}$ such that $F_{\text{ghz}}(\rho^{*}) = d_{\text{GME}}/d$, since
	\begin{equation}
		F_{\text{ghz}}(\rho^{*}) = v^{*} + \frac{1-v^{*}}{d} = \frac{d_{\text{GME}}}{d} \quad \implies \quad v^{*} = \frac{d_{\text{GME}}-1}{d-1}.
	\end{equation}
	Then, to prove our result, we only need to show that this state $\rho^{*}$ has a GME-dimension no larger than $d_{\text{GME}}$.\\
	We can consider the family of all pure states with GME-dimension equal to $d_{\text{GME}}$:
	\begin{equation}
		\ket{\Phi_{\alpha}} = \frac{1}{\sqrt{|\alpha|}} \sum_{i \in \alpha} \ket{i}^{\otimes n},
	\end{equation}
	where $\alpha \subset \lbrace 0,1,\dots,d-1 \rbrace$ with cardinality $|\alpha| = d_{\text{GME}}$. They are in total ${ d \choose d_{\text{GME}} } $, so that we can take their incoherent mixture
	\begin{equation}
		\rho_{d_\text{GME}} = \frac{1}{ {d \choose d_{\text{GME}} }} \sum_{\alpha \, \text{s.t.} \, |\alpha| = d_{\text{GME}}} \ketbra{\Phi_{\alpha}} = \frac{1}{ {d \choose d_{\text{GME}} }} \sum_{\alpha \, \text{s.t.} \, |\alpha| = d_{\text{GME}}} \frac{1}{|\alpha|} \left[ \sum_{i \in \alpha} \ketbra{i}^{\otimes n} + \sum_{i \neq j \in \alpha} \ketbra{i}{j}^{\otimes n}\right].
	\end{equation}
	We can realize that each diagonal term on the RHS will appear in ${ d-1 \choose d_{\text{GME}}-1 }$ possible $\alpha$ subsets while each off-diagonal term will appear in $ {d-2 \choose d_{\text{GME}}-2} $ possible $\alpha$ subsets. Therefore, we will have:
	\begin{equation}
		\begin{aligned}
			\rho_{d_\text{GME}} &= \frac{1}{ {d \choose d_{\text{GME}} }} \frac{1}{|\alpha|} \left[ { d-1 \choose d_{\text{GME}-1} } \sum_{i = 0}^{d-1} \ketbra{i}^{\otimes n} + {d-2 \choose d_{\text{GME}}-2} \sum_{i \neq j = 0}^{d-1} \ketbra{i}{j}^{\otimes n} \right] \\
			& = \frac{1}{d} \sum_{i = 0}^{d-1} \ketbra{i}^{\otimes n} + \left(\frac{d_{\text{GME}}-1}{d-1} \right)\frac{1}{d} \sum_{i \neq j = 0}^{d-1} \ketbra{i}{j}^{\otimes n}
		\end{aligned}
	\end{equation}
	that, considering the expression shown in \eqref{Appendix_Fidel-Tigh: Dephased_GHZ} of a generic dephased GHZ state, turns out to be precisely the state $\rho^{*}$. We can now remember that this state was constructed by taking a convex combination of pure states with GME-dimension equal to $d_{\text{GME}}$ and therefore cannot have GME-dimension larger that $d_{\text{GME}}$. Then we conclude that any dephased GHZ state $\rho^{*}$ whose $F_{\text{ghz}}(\rho^{*}) \leq d_{\text{GME}}/d$ has GME-dimension no larger than $d_{\text{GME}}$.

	\newpage
	
	\section{Three-partite witness for odd-dimensional GHZ states}\label{App:10basisghz}
	Define the computational basis $\{\ket{l}\}_{l=0}^{d-1}$ and another $d$ bases as $ \ket{e_{l}^{(j)}} = \frac{1}{\sqrt{d}}\sum_{m = 0}^{d-1}\omega^{m(l+jm)}\ket{m}$ with basis index $j = 0, \dots, d-1$ and element index $l=0,\ldots,d-1$. In odd prime dimensions, this happens to consitute a complete set of mutually unbiased bases.  We consider a three-partite state in any odd local dimension $d$. We will use a total of ten global product measurements to witness the GME-dimension via fidelity estimation. The first global product measurement corresponds to the computational basis. The other nine correspond to the following tuples indexing the basis index $j$ for each of the three parties; 
	\begin{equation}
		L=\{(0,0,0),(1,1,d-2),(d-1,d-1,2),(0,1,d-1),(1,0,d-1),(0,d-1,1),(1,d-1,0),(d-1,0,1),(d-1,1,0)\}.
	\end{equation}
	Based on these measurements, we define the entanglement witness operator
	\begin{equation}
		\mathcal{W}_d= 3\sum_{l=0}^{d-1} \ketbra{l}{l}^{\otimes 3}+\sum_{(j_1,j_2,j_3)\in L} \sum_{l_1\oplus l_2\oplus l_3=0}^{d-1} \bigotimes_{k=1}^3 \ketbra{e_{l_k}^{(j_k)}}{e_{l_k}^{(j_k)}}.
	\end{equation}
	The corresponding entanglement witness quantity is therefore $W_d=\tr(\mathcal{W}_d \rho)$. In the case of a GHZ state, one has perfect correlations in each of the ten measurements and hence obtains $W_d=12$.  For an arbitrary three-partite state $\rho$ of equal odd local dimension $d$, it holds that 
	\begin{equation}\label{stepres}
		W_d\leq 3+\frac{9\mathcal{D}_\text{GME}(\rho)}{d}.
	\end{equation}
	Moreover, any observed witness value implies a GHZ-fidelity bound of 
	\begin{equation}
		F_\text{ghz}(\rho)\geq\frac{W_d-3}{9}.
	\end{equation}
	We have shown this explicitly for odd $d\leq 17$ and conjecture it to hold for odd $d$ in general. The proof is based on explicitly computing the spectrum of $\mathcal{W}_d$. All eigenvalues are non-negative by construction. Moreover, one eigenvector is the GHZ state with eigenvalue 12. For $d=3,5,\ldots,17$ we observe that all other eigenvalues are no larger than 3. Hence, we can write
	
	\begin{equation}
		\mathcal{W}_d \preceq 9 \ketbra{\text{ghz}_{n,d}}+3\openone. 
	\end{equation}
	Taking the expectation value of both sides w.r.t.~the state $\rho$ and using that $F_\text{ghz}(\rho)\leq \frac{d_\text{GME}}{d}$ gives Eq.~\eqref{stepres}.

	When applied to the noisy GHZ state $\rho_v^\text{ghz} = v \ketbra{\text{ghz}_{n,d}}+ \frac{1-v}{d^3} \openone$, the critical visibility for detecting the GME-dimension is
	\begin{equation}
		v_\text{crit}^\text{10 bases}=\frac{d^2-1+3d(d_\text{GME}-1)}{4d^2-3d-1}.
	\end{equation}
	This can be compared to 
	\begin{equation}
		v_\text{crit}^\text{exact}=\frac{d_\text{GME}d^2-1}{d^3-1},
	\end{equation}
	for exact fidelity measurements on a three-partite state. Note that for $d=3$, both formulas are identical. For $d>3$ the trade-off in resource-efficiency and detection power can be quantified via the impact parameter $\Delta=\frac{1-v_\text{crit}^\text{10 bases}}{1-v_\text{crit}^\text{exact}}$. For the most interesting case when $d_\text{GME}=d-1$, this becomes
	\begin{equation}
		\Delta=\frac{3(1+d+d^2)}{d(1+4d)}.
	\end{equation}
	As expected, this is monotonically decreasing, since for larger $d$ the number of extra measurements needed for the exact fidelity grows. However, even in the limit of $d\rightarrow \infty$ we  have $\Delta=\frac{3}{4}$, showing that even for very large $d$, using just ten bases is still a decent  bound on the true fidelity.
	
	\newpage
	\section{Proof of Result 3}
	\label{Appendix_Wcluster}
	Let us consider the following operator:
	\begin{equation}
		\label{Appendix_Wcluster: Operator_definition}
		O^\text{cluster}_{n,d} = \bigotimes_{l \text{ odd}}^{n} \, \sum_{q_{l},p_{l+1} = 0}^{d-1} \ketbra{e_{q_{l}}p_{l+1}}{e_{q_{l}}p_{l+1}}\delta_{p_{l+1}\oplus p_{l-1} \ominus q_{l},0} + \bigotimes_{l \text{ odd}}^{n} \, \sum_{p_{l},q_{l+1} = 0}^{d-1} \ketbra{p_{l}e_{q_{l+1}}}{p_{l}e_{q_{l+1}}}\delta_{p_{l} \oplus p_{l+2} \ominus q_{l+1},0},
	\end{equation}
	where $\oplus$ denotes addition modulo $d$, $\{\ket{j}\}_{j=0}^{d-1}$ is the computational basis and  $\{\ket{e_j}\}_{j=0}^{d-1}$ is the Fourier basis defined in \eqref{Appendix_WGHZ: Fourier-basis}.\\
	By recalling the definition of the cluster state $\ket{C_{n,d}}$ given in \eqref{Appendix_Wcluster: Cluster_state}, we prove Result 3 in three steps as we did for Result 2 in section III:
	\begin{itemize}
		\item \textit{Show that $\ket{C_{n,d}}$ has perfect correlations for $O^\text{cluster}_{n,d}$ and therefore $\mathcal{W}^\text{cluster}_{n,d}(\ket{C_{n,d}}) = 2$.}\\
		\newline
		In principle we can use  the same procedure presented in the first step of section III, i.e.~to directly show that $\ket{C_{n,d}}$ is an eigenstate of the operator $O^\text{cluster}_{n,d}$. However, we here take the other, more informative, route, where we calculate the probability distribution for the two sets of global product bases used to construct the witness \eqref{Appendix_Wcluster: Operator_definition}.\\
		To do that, we separately consider the case where $n$ is an even number and the case where $n$ is odd.
		\begin{itemize}
			\item $n$-even\\
			We can start by calculating the probability distribution for the measurements in the first term of \eqref{Appendix_Wcluster: Operator_definition}. In this case the $n/2$ subsystems $\lbrace 2,4,\dots,n \rbrace$ are measured in the computational basis and the remaining $n/2$ subsystems $\lbrace 1,3,\dots,n-1 \rbrace$ are measured in the Fourier basis. Therefore, we get:
			\begin{equation}
				\small
				\begin{aligned}
					&P(e_{q_1}, p_{2}, e_{q_{3}}, \dots , p_{n}|C_{n,d}) = |\braket{e_{q_1} p_{2} e_{q_{3}} \dots p_{n}}{C_{n,d}}|^2\\
					&= \dfrac{1}{d^n} \left|\sum_{k_{1},\dots,k_{n} = 0}^{d-1} \omega^{k_{1}k_{2} + \dots + k_{n-1}k_n} \dfrac{1}{\sqrt{d^{n/2}}}\sum_{l_1,l_3,\dots,l_{n-1} = 0}^{d-1}\omega^{-(l_{1}q_{1} + l_{3}q_{3} + \dots + l_{n-1}q_{n-1})}\braket{l_{1}p_{2}l_{3}\dots p_{n}}{k_1 k_2 k_3 \dots k_n}\right|^2\\
					&= \dfrac{1}{d^n} \dfrac{1}{d^{n/2}} \left|\sum_{k_{1},k_{3}\dots,k_{n-1} = 0}^{d-1} \omega^{k_{1}p_{2} + p_{2}k_{3} + k_{3}p_{4} + \dots + k_{n-1}p_{n}} \omega^{-(k_{1}q_{1} + k_{3}q_{3} + \dots + k_{n-1}q_{n-1})}\right|^2\\
					&= \dfrac{1}{d^n} \dfrac{1}{d^{n/2}}\left| \sum_{k_1 = 0}^{d-1} \omega^{k_{1}(p_2 - q_1)} \sum_{k_{3} = 0}^{d-1} \omega^{k_3(p_{2} + p_{4} - q_{3})}\cdots \sum_{k_{n-1} = 0}^{d-1} \omega^{k_{n-1}(p_{n-2} + p_{n} - q_{n-1})}\right|^2,
				\end{aligned}
			\end{equation}
			where in the last equality we can evaluate each of the sums with \eqref{Appendix_WGHZ: Geometric_series} to end up with
			\begin{equation}
				\label{Appendix_WCluster: Statistics_operator1_even}
				P(e_{q_1}, p_{2}, e_{q_{3}}, \dots , p_{n}|C_{n,d}) = \dfrac{1}{d^{n/2}} \, \delta_{p_2 \ominus q_1,0}\, \delta_{p_2 \oplus p_4 \ominus q_{3},0} \dots \delta_{p_{n-2} \oplus p_{n} \ominus q_{n-1},0}.
			\end{equation}
			Note that this probability distribution is uniform of $d^{n/2}$ possible outcomes and zero over the remaining $d^{n/2}$ outcomes. The former events coincide exactly with those used in the first term of the witness \eqref{Appendix_Wcluster: Operator_definition}. Hence,
			\begin{equation}
				\label{Appendix_Wcluster: Cluster_eigenstate1}
				\left(\bigotimes_{l \text{ odd}}^{n}\, \sum_{q_{l},p_{l+1} = 0}^{d-1} \ketbra{e_{q_{l}}p_{l+1}}{e_{q_{l}}p_{l+1}}\delta_{p_{l+1}\oplus p_{l-1} \ominus q_{l},0}\right) \ket{C_{n,d}} = \ket{C_{n,d}}.
			\end{equation}
			The same approach applies to the second term in \eqref{Appendix_Wcluster: Operator_definition}. In this case we consider the $n/2$ subsystems $\lbrace 2,4,\dots,n \rbrace$ to be measured in the Fourier basis and the remaining $n/2$ subsystems $\lbrace 1,3,\dots,n-1 \rbrace$ to be measured in the computational basis; then the probability distribution is
			\begin{equation}
				\small
				\begin{aligned}
					\label{Appendix_WCluster: Statistics_operator2_even}
					&P(p_{1}, e_{q_{2}}, p_{3} \dots , e_{q_{n}}|C_{n,d}) = |\braket{p_{1} e_{q_{2}} p_{3} \dots e_{q_{n}}}{C_{n,d}}|^2\\
					&= \dfrac{1}{d^n} \left|\sum_{k_{1},\dots,k_{n} = 0}^{d-1} \omega^{k_{1}k_{2} + \dots + k_{n-1}k_n} \dfrac{1}{\sqrt{d^{n/2}}}\sum_{l_2,l_4,\dots,l_{n} = 0}^{d-1}\omega^{-(l_{2}q_{2} + l_{4}q_{4} + \dots + l_{n}q_{n})}\braket{p_{1}l_{2}p_{3}\dots l_{n}}{k_1 k_2 k_3 \dots k_n}\right|^2\\
					&= \dfrac{1}{d^{n/2}}\, \delta_{p_1 \oplus p_3 \ominus q_{2},0} \, \delta_{p_3 \oplus p_5 \ominus q_{4},0} \dots \, \delta_{p_{n-1} \ominus q_n,0}.
				\end{aligned}
			\end{equation}
			We observe an analogous structure, which corresponds to the second term in \eqref{Appendix_Wcluster: Operator_definition}, thus implying
			\begin{equation}
				\label{Appendix_Wcluster: Cluster_eigenstate2}
				\left(\bigotimes_{l \text{ odd}}^{n} \,\sum_{p_{l},q_{l+1} = 0}^{d-1} \ketbra{p_{l}e_{q_{l+1}}}{p_{l}e_{q_{l+1}}}\delta_{p_{l} \oplus p_{l+2} \ominus q_{l+1},0}\right) \ket{C_{n,d}} = \ket{C_{n,d}}.
			\end{equation}
			
			\item $n$-odd\\
			For the first term in \eqref{Appendix_Wcluster: Operator_definition} we calculate the probability distribution for the $\frac{n+1}{2}$ subsystems $\lbrace 1,3,\dots,n \rbrace$ measured in the Fourier basis and the remaining $\frac{n-1}{2}$ subsystems $\lbrace 2,4,\dots,n-1 \rbrace$ measured in the computational basis. We get:
			\begin{equation}
				\label{Appendix_WCluster: Statistics_operator1_odd}
				\small
				\begin{aligned}
					&P(e_{q_1}, p_{2}, e_{q_{3}}, \dots , p_{n-1},e_{q_n}|C_{n,d}) = |\braket{e_{q_1} p_{2} e_{q_{3}} \dots p_{n-1} e_{q_n}}{C_{n,d}}|^2\\
					&= \dfrac{1}{d^n} \left|\sum_{k_{1},\dots,k_{n} = 0}^{d-1} \omega^{k_{1}k_{2} + \dots + k_{n-1}k_n} \dfrac{1}{\sqrt{d^{(n+1)/2}}}\sum_{l_1,l_3,\dots,l_{n} = 0}^{d-1}\omega^{-(l_{1}q_{1} + l_{3}q_{3} + \dots + l_{n}q_{n})}\braket{l_{1}p_{2}l_{3}\dots l_{n}}{k_1 k_2 k_3 \dots k_n}\right|^2,\\
					&= \dfrac{1}{d^{\frac{n-1}{2}}} \, \delta_{p_{2} \ominus q_{1},0}\, \delta_{p_{2} \oplus p_{4} \ominus q_{3},0} \dots \delta_{p_{n-1}\ominus q_{n},0}.
				\end{aligned}
			\end{equation}
			As for even $n$,  this probability distribution is non-zero for all outcome tuples appearing in the first term in \eqref{Appendix_Wcluster: Operator_definition}: the only difference with the $n$-even case is in the number of non-zero terms in the distribution. Therefore, also for $n$-odd we recover the result in \eqref{Appendix_Wcluster: Cluster_eigenstate1}.
			
			For the second term in \eqref{Appendix_Wcluster: Operator_definition}, we calculate the probability distribution related to measurements of the $\frac{n+1}{2}$ subsystems $\lbrace 1,3,\dots,n \rbrace$ in the computational basis and measurements of the remaining $\frac{n-1}{2}$ subsystems $\lbrace 2,4,\dots,n-1 \rbrace$ in the Fourier basis:
			\begin{equation}
				\label{Appendix_WCluster: Statistics_operator2_odd}
				\small
				\begin{aligned}
					&P(p_{1}, e_{q_{2}}, p_{3} \dots , e_{q_{n-1}},p_{n}|C_{n,d}) = |\braket{p_{1} e_{q_{2}} p_{3} \dots e_{q_{n-1}}p_{n}}{C_{n,d}}|^2\\
					&= \dfrac{1}{d^n} \left|\sum_{k_{1},\dots,k_{n} = 0}^{d-1} \omega^{k_{1}k_{2} + \dots + k_{n-1}k_n} \dfrac{1}{\sqrt{d^{(n-1)/2}}}\sum_{l_2,l_4,\dots,l_{n-1} = 0}^{d-1}\omega^{-(l_{2}q_{2} + l_{4}q_{4} + \dots + l_{n-1}q_{n-1})}\braket{p_{1}l_{2}p_{3}\dots p_{n}}{k_1 k_2 k_3 \dots k_n}\right|^2,\\
					&= \dfrac{1}{d^{\frac{n+1}{2}}}\, \delta_{p_1 \oplus p_3 \ominus q_{2},0} \, \delta_{p_3 \oplus p_5 \ominus q_{4},0} \dots \, \delta_{p_{n-2} \oplus p_{n} \ominus q_{n-1},0}.
				\end{aligned}
			\end{equation}
			This allows us to recover \eqref{Appendix_Wcluster: Cluster_eigenstate2} also for the $n$-odd case.
		\end{itemize}
		
		\item \textit{Show that the spectral decomposition of $O^\text{cluster}_{n,d}$ is $\lambda ( O^\text{cluster}_{n,d}) = \lbrace \lambda_{C_{n,d}}, \lambda_2, \dots, \lambda_{d^n} \rbrace$ such that $\lambda_{i}\le 1 \quad \forall i \neq C_{n,d}$}\\
		\newline
		We can use the same procedure presented in the second step of section III for the GHZ witness operator. Therefore, after realizing that
		\begin{equation}
			O^\text{cluster}_{n,d} = 2 \ketbra{C_{n,d}}{C_{n,d}} + \sum_{i=2}^{d^n}\lambda_i \ketbra{\lambda_i}{\lambda_i},
		\end{equation}
		we can prove that the shifted operator $\tilde{O}^\text{cluster}_{n,d} = O^\text{cluster}_{n,d} - C_{n,d}$ is a projector. Also in this case we can calculate its square,
		\begin{equation}
			\label{Appendix_WCluster: Shifted_operator_squared}
			(\tilde{O}^\text{cluster}_{n,d})^2 = (O^\text{cluster}_{n,d})^2 + C_{n,d} - \lbrace O^\text{cluster}_{n,d}, C_{n,d} \rbrace = (O^\text{cluster}_{n,d})^2 - 3\, C_{n,d},
		\end{equation}
		and focus on the $(O^\text{cluster}_{n,d})^2$ term. From the definition \eqref{Appendix_Wcluster: Operator_definition} we get
		\begin{equation}
			\begin{aligned}
				(O^\text{cluster}_{n,d})^2 = &\left[\bigotimes_{l \text{ odd}}^{n}\,\sum_{q_{l},p_{l+1} = 0}^{d-1} \ketbra{e_{q_{l}}p_{l+1}}{e_{q_{l}}p_{l+1}}\delta_{p_{l+1}\oplus p_{l-1} \ominus q_{l},0} + \bigotimes_{l \text{ odd}}^{n}\,\sum_{p_{l},q_{l+1} = 0}^{d-1} \ketbra{p_{l}e_{q_{l+1}}}{p_{l}e_{q_{l+1}}}\delta_{p_{l} \oplus p_{l+2} \ominus q_{l+1},0}\right]\\
				& \cdot \left[\bigotimes_{l \text{ odd}}^{n}\,\sum_{q'_{l},p'_{l+1} = 0}^{d-1} \ketbra{e_{q'_{l}}p'_{l+1}}{e_{q'_{l}}p'_{l+1}}\delta_{p'_{l+1}\oplus p'_{l-1} \ominus q'_{l},0} + \bigotimes_{l \text{ odd}}^{n}\,\sum_{p'_{l},q'_{l+1} = 0}^{d-1} \ketbra{p'_{l}e_{q'_{l+1}}}{p'_{l}e_{q'_{l+1}}}\delta_{p'_{l} \oplus p'_{l+2} \ominus q'_{l+1},0}\right],
			\end{aligned}
		\end{equation}
		where the non-mixing terms produce again $O^\text{cluster}_{n,d}$ because of the orthonormality of both computational and Fourier basis.\\
		We can separately look at the mixing terms. From the first one we get
		\begin{equation}
			\label{Appendix_WCluster: First_mixing}
			\begin{aligned}
				&\bigotimes_{l \text{ odd}}^{n} \, \, \sum_{\substack{q_{l},p_{l+1} = 0\\ p'_{l},q'_{l+1} = 0}}^{d-1} \braket{p'_{l}e_{q'_{l+1}}}{e_{q_{l}}p_{l+1}} \ketbra{e_{q_{l}}p_{l+1}}{p'_{l}e_{q'_{l+1}}}\delta_{p_{l+1}\oplus p_{l-1} \ominus q_{l},0}\, \delta_{p'_{l} \oplus p'_{l+2} \ominus q'_{l+1},0},\\
				& = \dfrac{1}{\sqrt{d^n}}\,\bigotimes_{l \text{ odd}}^{n} \, \, \sum_{\substack{q_{l},p_{l+1} = 0\\ p'_{l},q'_{l+1} = 0}}^{d-1} 
				\omega^{q_{l}p'_{l} - p_{l+1}q'_{l+1}} \ketbra{e_{q_{l}}p_{l+1}}{p'_{l}e_{q'_{l+1}}}\delta_{p_{l+1}\oplus p_{l-1} \ominus q_{l},0}\, \delta_{p'_{l} \oplus p'_{l+2} \ominus q'_{l+1},0}.
			\end{aligned}
		\end{equation}
		We can use the conditions imposed by the delta functions and rewrite the exponential term as
		\begin{equation}
			\omega^{q_{l}p'_{l}-p_{l+1}q'_{l+1}} = \omega^{(p_{l+1}+p_{l-1})p'_{l} - p_{l+1}q'_{l+1} \pm p_{l+1}p'_{l+2}} =  \omega^{p_{l+1}(p'_{l}+p'_{l+2}-q'_{l+1})} \,\omega^{p_{l-1}p'_{l}- p_{l+1}p'_{l+2}},
		\end{equation}
		where the first exponential simplifies when we insert it into \eqref{Appendix_WCluster: First_mixing} because of the condition imposed by the second delta, ending up with
		\begin{equation}
			\label{Appendix_WCluster: First_mixing2}
			\dfrac{1}{\sqrt{d^n}}\,\bigotimes_{l \text{ odd}}^{n} \, \, \sum_{\substack{q_{l},p_{l+1} = 0\\ p'_{l},q'_{l+1} = 0}}^{d-1} 
			\omega^{p_{l-1}p'_{l}- p_{l+1}p'_{l+2}} \ketbra{e_{q_{l}}p_{l+1}}{p'_{l}e_{q'_{l+1}}}\delta_{p_{l+1}\oplus p_{l-1} \ominus q_{l},0}\, \delta_{p'_{l} \oplus p'_{l+2} \ominus q'_{l+1},0}.
		\end{equation}
		In addition, we can realize that, once we take the tensor product over the odd $l$, all the exponents simplify:
		\begin{equation}
			\begin{aligned}
				n\text{-odd}:&\quad p_{l-1}p'_{l} - p_{l+1}p'_{l+2} = \underbrace{0-p_{2}p'_{3}}_{l = 1} + \underbrace{p_{2}p'_{3} + p_{4}p'_{5}}_{l = 3} + \dots + \underbrace{p_{n-3}p'_{n-2} - p_{n-1}p'_{n}}_{l = n-2} + \underbrace{p_{n-1}p'_{n} - 0}_{l = n} = 0,\\
				n\text{-even}:&\quad p_{l-1}p'_{l} - p_{l+1}p'_{l+2} = \underbrace{0-p_{2}p'_{3}}_{l = 1} + \underbrace{p_{2}p'_{3} + p_{4}p'_{5}}_{l = 3} + \dots + \underbrace{p_{n-4}p'_{n-3} - p_{n-2}p'_{n-1}}_{l = n-3} + \underbrace{p_{n-2}p'_{n-1} - 0}_{l = n-1} = 0.\\
			\end{aligned}
		\end{equation}
		Therefore, the whole exponential term in \eqref{Appendix_WCluster: First_mixing2} simplifies and we end up with
		\begin{equation}
			\dfrac{1}{\sqrt{d^{n}}} \bigotimes_{l \text{ odd}}^{n} \, \, \left(\sum_{q_{l},p_{l+1} = 0}^{d-1} \delta_{p_{l+1}\oplus p_{l-1} \ominus q_{l},0}\, \ket{e_{q_{l}}p_{l+1}}\right) \left(\sum_{p'_{l},q'_{l+1} = 0}^{d-1} \delta_{p'_{l} \oplus p'_{l+2} \ominus q'_{l+1},0} \bra{p'_{l}e_{q'_{l+1}}}\right).
		\end{equation}
		By looking at the statistics derived in \eqref{Appendix_WCluster: Statistics_operator1_even} and \eqref{Appendix_WCluster: Statistics_operator2_even} for $n$ even and in \eqref{Appendix_WCluster: Statistics_operator1_odd} and \eqref{Appendix_WCluster: Statistics_operator2_odd} for $n$ odd, we can realize that the last result is precisely the $\ketbra{C_{n,d}}{C_{n,d}}$ projector.\\
		The same procedure can be used for the second mixing term:
		\begin{equation}
			\bigotimes_{l \text{ odd}}^{n}\,
			\sum_{\substack{p_{l},q_{l+1} = 0 \\ q'_{l},p'_{l+1} = 0}}^{d-1} \braket{p_{l}e_{q_{l+1}}}{e_{q'_{l}}p'_{l+1}} \ketbra{p_{l}e_{q_{l+1}}}{e_{q'_{l}}p'_{l+1}} \delta_{p_{l} \oplus p_{l+2} \ominus q_{l+1},0}\, \delta_{p'_{l+1}\oplus p'_{l-1} \ominus q'_{l},0},
		\end{equation}
		since we can realize that it is precisely the first mixing term with $(p,q) \rightarrow (p',q')$.\\
		Therefore, by inserting these results into \eqref{Appendix_WCluster: Shifted_operator_squared}, we get
		\begin{equation}
			(\tilde{O}^\text{cluster}_{n,d})^2 = (O^\text{cluster}_{n,d})^2 - 3\, C_{n,d} = O^\text{cluster}_{n,d} - C_{n,d} = \tilde{O}^\text{cluster}_{n,d},
		\end{equation}
		which proves that $\tilde{O}^\text{cluster}_{n,d}$ is a projector.
		
		\item \textit{Derivation of the upper bound for $\mathcal{W}^\text{cluster}_{n,d}(\rho)=\tr\left(\rho	O^\text{cluster}_{n,d}\right)$}\\
		\newline
		The procedure for this step is identical to the one presented in Appendix \ref{Appendix_WGHZ}, since also in this case it is possible to use that
		\begin{equation}
			\lambda(O^{cluster}_{n,d}) = \lbrace 2,\lambda_2,\dots,\lambda_{d^n}: \, \lambda_i \leq 1 \rbrace \quad \implies \quad C_{n,d} + \sum_{i=2}^{d^n} \lambda_i \ketbra{\lambda_i}{\lambda_i} \leq \openone
		\end{equation}
		to bound the witness as
		\begin{equation}
			\label{Appendix_WCluster: Witness_bound_derivation}
			\mathcal{W}^\text{cluster}_{n,d}(\rho) = \tr\left(\rho	O^\text{cluster}_{n,d}\right) = F_{\text{cluster}}(\rho) + \tr\left(\rho \left(C_{n,d} + \sum_{i=2}^{d^n}\lambda_{i}\ketbra{\lambda_{i}}{\lambda_{i}}\right)\right) \leq F_{\text{cluster}}(\rho) + 1.
		\end{equation}
		We can realize that $F_{\text{cluster}}(\rho)$ is the fidelity between a $n$-partite pure linear cluster state and a state $\rho$ with $\mathcal{D}_{\text{GME}}(\rho) \leq d_{\text{GME}}$ that can be bounded by using Result 1 as shown in \eqref{Appendix_Fidelity-evaluation: Cluster-bound}.
			By inserting this result into \eqref{Appendix_WCluster: Witness_bound_derivation} we get:
			\begin{equation}
				\mathcal{W}^\text{cluster}_{n,d}(\rho) \leq 1 + \dfrac{d_{\text{GME}}}{d}.
			\end{equation}
			Moreover, by looking at the last inequality in \eqref{Appendix_WCluster: Witness_bound_derivation} we can realize that:
			\begin{equation}
				F_{\text{cluster}}(\rho)\geq \mathcal{W}^\text{cluster}_{n,d}(\rho) - 1.
			\end{equation}

		\end{itemize}

		\newpage
		\section{Reduction of the SDP to LP}
		\label{Appendix_SDP-LP}

		Define the $r$-positive generalised reduction map $\Lambda(X) = \tr X \openone - \frac{1}{r}X$ and the unnormalised states $\tilde{\sigma}_{S|\bar{S}} = q_{S|\bar{S}}\sigma_{S|\bar{S}}$. Then it is possible to show that, if $\rho$ is a pure state, the following semidefinite program,
		\begin{equation}\label{Appendix_SDP-LP: SDP}
			\begin{aligned}
				\max_{v,\tilde{\sigma}} & \quad v\\
				\text{s.t.}&\quad v\rho+\frac{1-v}{d^{n}}\openone =  \sum_{ S|\bar{S}} \tilde{\sigma}_{S|\bar{S}},\\
				& \quad \tilde{\sigma}_{S|\bar{S}} \geq 0 \quad \forall (S|\bar{S}),\\
				& \quad (\Lambda_S\otimes \openone_{\bar{S}})[\tilde{\sigma}_{S|\bar{S}}] \geq 0 \quad \forall (S|\bar{S}),
			\end{aligned}
		\end{equation}
		can be reduced to a linear program by representing it in a basis $\lbrace \ket{a_1 \dots a_n}^{D} \rbrace$, with $\lbrace a_1, \dots, a_n \rbrace \in \lbrace 0,\dots,d-1 \rbrace$, that we define as state-diagonal since the first element $\ket{0\dots 0}^{D}\prescript{D}{}{\bra{0\dots 0}}$ is equal to the state $\rho$. Let $U$ be the unitary that maps $\rho$ expressed in the computational basis into the same state $\rho$ expressed in this new basis, i.e. $U \rho U^{\dagger} = \ket{0\dots 0}^{D}\prescript{D}{}{\bra{0\dots 0}}$. Then \eqref{Appendix_SDP-LP: SDP} is equivalent to the following program,
		\begin{equation}\label{Appendix_SDP-LP: LP}
			\begin{aligned}
				\max_{v,\tilde{\sigma}} & \quad v\\
				\text{s.t.}&\quad v \, U\rho U^{\dagger}+\frac{1-v}{d^{n}}\openone =  \sum_{ S|\bar{S}} U\tilde{\sigma}_{S|\bar{S}} U^{\dagger},\\
				& \quad U\tilde{\sigma}_{S|\bar{S}} U^{\dagger} \geq 0 \quad \forall (S|\bar{S}),\\
				& \quad (\Lambda_S\otimes \openone_{\bar{S}})[U \tilde{\sigma}_{S|\bar{S}} U^{\dagger}] \geq 0 \quad \forall (S|\bar{S}).
			\end{aligned}
		\end{equation}
		Here, $U \rho U^{\dagger}$ is diagonal and w.l.g.~therefore optimally chooses $U \tilde{\sigma}_{S|\bar{S}} U^{\dagger}$ diagonal as well. Since all the matrices are diagonal, this is in fact a linear program.
		
		To prove this result it is enough to show that all the constraints in \eqref{Appendix_SDP-LP: SDP} imply the constraints in \eqref{Appendix_SDP-LP: LP}. Since for the first constraint this can be shown by simply using the unitarity of $U$, we will directly focus on the other two:
		\begin{itemize}
			\item $\tilde{\sigma}_{S|\bar{S}} \geq 0 \quad \implies \quad U \tilde{\sigma}_{S|\bar{S}} U^{\dagger} \geq 0$
			
			We can remember that the terms $U \tilde{\sigma}_{S|\bar{S}} U^{\dagger}$ are diagonal in the new basis, therefore:
			\begin{equation}
				U \tilde{\sigma}_{S|\bar{S}} U^{\dagger} = \sum_{a_1,\dots, a_n = 0}^{d-1} \prescript{D}{}{\braket{a_1\dots a_n}{\tilde{\sigma}_{S|\bar{S}}|a_1 \dots a_n}}^{D} \quad \ketbra{a_1 \dots a_n} \geq 0,
			\end{equation}
			since, being $\tilde{\sigma}_{S|\bar{S}} \geq 0$ positive semidefinite, for any basis element we have $\prescript{D}{}{\braket{a_1 \dots a_n}{\tilde{\sigma}_{S|\bar{S}}|a_1 \dots a_n}}^{D} \geq 0$.
			
			\item $(\Lambda_S \otimes \openone_{\bar{S}})[\tilde{\sigma}_{S|\bar{S}}] \geq 0 \quad \implies \quad (\Lambda_S \otimes \openone_{\bar{S}})[U \tilde{\sigma}_{S|\bar{S}} U^{\dagger}] \geq 0$
			
			First, we can use the definition of our $r$-positive generalised reduction map and write
			\begin{equation}
				\label{Appendix_SDP-LP: gen-red-map}
				(\Lambda_S \otimes \openone_{\bar{S}})[\tilde{\sigma}_{S|\bar{S}}] = \openone_{S} \otimes \tr_{S}(\tilde{\sigma}_{S|\bar{S}}) - \frac{1}{r} \tilde{\sigma}_{S|\bar{S}} \geq 0.
			\end{equation}
			Then, we for any bipartition $(S|\bar{S})$ the states in the new basis can be expressed as $\ket{a_1\dots a_n}^D = \ket{a_1\dots a_{S}}_{S}^D \otimes \ket{a_{S+1}\dots a_n}_{\bar{S}}^D$ and therefore we can explicitly calculate
			\begin{equation}
				\begin{aligned}
					&\openone_{S} \otimes \tr_{S}(U \tilde{\sigma}_{S|\bar{S}} U^{\dagger}) = \openone_{S} \otimes \sum_{a_1, \dots, a_n = 0}^{d-1} \prescript{D}{}{\braket{a_1 \dots a_n}{\tilde{\sigma}_{S|\bar{S}}|a_1 \dots a_n}}^{D} \,\tr_{S}(\ket{a_1 \dots a_n}^{D} \prescript{D}{}{\bra{a_1 \dots a_n}})\\
					&= \sum_{b_1, \dots, b_S = 0}^{d-1} \ket{b_1\dots b_S}_{S}^{D}\prescript{D}{S}{\bra{b_1\dots b_{S}}} \otimes \sum_{a_1,\dots,a_n = 0}^{d-1}
					\prescript{D}{}{\braket{a_1\dots a_n}{\tilde{\sigma}_{S|\bar{S}}|a_1\dots a_n}}^{D} \quad \ket{a_{S+1}\dots a_n}_{\bar{S}}^{D}\prescript{D}{\bar{S}}{\bra{a_{S+1}\dots a_n}},
				\end{aligned}
			\end{equation}
			in which we can rename $\ket{a_1 \dots a_S}_{S}^{D} \leftrightarrow \ket{b_1 \dots b_S}_{S}^{D}$ to get
			\begin{equation}
				\begin{aligned}
					&= \sum_{a_1, \dots, a_n = 0}^{d-1} \quad \sum_{b_1, \dots,b_{S} = 0}^{d-1} \prescript{D}{S}{\bra{b_1 \dots b_{S}}}\prescript{D}{\bar{S}}{\bra{a_{S+1}\dots a_{n}}} \tilde{\sigma}_{S|\bar{S}} \ket{b_{1}\dots b_{S}}_{S}^{D} \ket{a_{S+1}\dots a_{n}}_{\bar{S}}^{D} \quad \ket{a_1 \dots a_n}^{D}\prescript{D}{}{\bra{a_1 \dots a_n}}\\
					& = \sum_{a_1,\dots,a_n = 0}^{d-1} \prescript{D}{}{\braket{a_1\dots a_n}{\openone_{S}\otimes \tr_{S}(\tilde{\sigma}_{S|\bar{S}})|a_1\dots a_n}}^{D} \, \ket{a_1 \dots a_n}^{D}\prescript{D}{}{\bra{a_1 \dots a_n}},
				\end{aligned}
			\end{equation}
			where in the last equality we used the definition of the partial trace. In conclusion,
			\begin{equation}
				\begin{aligned}
					&(\Lambda_S \otimes \openone_{\bar{S}})[U \tilde{\sigma}_{S|\bar{S}} U^{\dagger}] = \openone_{S} \otimes \tr_{S}(U \tilde{\sigma}_{S|\bar{S}} U^{\dagger}) - \frac{1}{r} U \tilde{\sigma}_{S|\bar{S}} U^{\dagger}\\
					& = \sum_{a_1,\dots,a_n = 0}^{d-1} \quad \prescript{D}{}{\bra{a_1\dots a_n}} \left(\openone_{S} \otimes \tr_{S}(\tilde{\sigma}_{S|\bar{S}}) - \frac{1}{r} \tilde{\sigma}_{S|\bar{S}}\right) \ket{a_1\dots a_n}^{D} \quad \ket{a_1 \dots a_n}^{D}\prescript{D}{}{\bra{a_1 \dots a_n}},
				\end{aligned}
			\end{equation}
			which implies that $(\Lambda_S \otimes \openone_{\bar{S}})[U \tilde{\sigma}_{S|\bar{S}} U^{\dagger}] \geq 0$ if $(\Lambda_S \otimes \openone_{\bar{S}})[\tilde{\sigma}_{S|\bar{S}}] \geq 0$.
			
		\end{itemize}
		
		\newpage
		\section{Tables of critical visibilities from linear programming}
		\label{Appendix_LP-results}

		In the  table below, we present the bounds on the critical visibility obtained from solving the linear program presented in the main text and Appendix~\ref{Appendix_SDP-LP}. All computations were made on a standard desktop computer (CPU: AMD Ryzen 9 5900X 12-Core Processor 3.70 GHz; RAM: 2 $\times$ 16 GB Samsung DDR4-3200 (1600 MHz)).
		Our case studies are based on the $n$-partite $d$-dimensional GHZ states $\ket{\text{ghz}_{n,d}}$ and linear cluster states $\ket{C_{n,d}}$.  
		
		For GHZ states and cluster states respectively, we define a state-diagonal basis, indicated as $\lbrace \ket{a_1\dots a_n}^{D} \rbrace$ with $\lbrace a_1, \dots, a_n \rbrace \in \lbrace 0,\dots, d-1 \rbrace$, in which $\ket{\text{ghz}_{n,d}}$ and $\ket{C_{n,d}}$ are associated respectively to the first element, $\ket{0\dots 0}^{D}$. By indicating with $\lbrace \ket{\psi_{a_1 \dots a_n}}\rbrace$ the vectors of the state-diagonal basis expressed in the computational basis, we can define the unitary, $U$, that maps a state from the computational basis to its state-diagonal basis:
		\begin{equation}
			\begin{cases}
				\text{GHZ-case: } \quad \ket{\psi_{a_{1}\dots a_{n}}} = Z^{a_1} \otimes X^{a_2} \otimes \dots \otimes X^{a_n}\ket{\text{ghz}_{n,d}}\\
				\text{Cluster-case: } \quad \ket{\psi_{a_{1}\dots a_{n}}} = Z^{a_1} \otimes \dots \otimes Z^{a_{n-2}} \otimes X^{a_{n-1}} \otimes X^{a_n} \ket{C_{n,d}}
			\end{cases}
			\quad \implies \quad U = \sum_{a_1,\dots, a_{n} = 0}^{d-1}\ketbra{a_{1} \dots a_{n}}{\psi_{a_{1}\dots a_{n}}}.
		\end{equation}
		Moreover, to get more accurate results, we applied the generalised reduction map on both sets of each bipartition, i.e. we evaluated
		\begin{equation}
			\begin{aligned}
				\max_{v,\tilde{\sigma}} & \quad v\\
				\text{s.t.}&\quad v \, U\rho U^{\dagger}+\frac{1-v}{d^{n}}\openone =  \sum_{ S|\bar{S}} U\tilde{\sigma}_{S|\bar{S}} U^{\dagger},\\
				& \quad U\tilde{\sigma}_{S|\bar{S}} U^{\dagger} \geq 0 \quad \forall (S|\bar{S}),\\
				& \quad (\Lambda_S\otimes \openone_{\bar{S}})[U \tilde{\sigma}_{S|\bar{S}} U^{\dagger}] \geq 0 \quad \forall (S|\bar{S}),\\
				& \quad (\openone_{S} \otimes \Lambda_{\bar{S}})[U \tilde{\sigma}_{S|\bar{S}} U^{\dagger}] \geq 0 \quad \forall (S|\bar{S}).
			\end{aligned}
		\end{equation}
		From the table of results, it is seen that in all considered cases, the cluster state is more noise tolerant than the GHZ state.
		\newpage
		\begin{table}[H]
			\label{tab: LP_results}
			\centering
			\caption{\textbf{LP results for GHZ state and cluster state.}}
			\begin{tabular}{|c|c|c|c|c|}
				\hline
				$\,$ (n,d) $\,$ & $\mathcal{D}_{\text{GME}}(\rho)$ & $\,$  $v_{\text{ghz}}$ (LP) & $\,$ $v_{\text{cluster}}$ (LP)  $\,$ & $v$ (fidelity)  \\
				\hline
				(3,2) & 1 & $0.4286$ & $*$ & $0.4286$\\
				\hline
				(3,3) & 1 & $0.2500$ & $*$ & $0.3077$\\
				\hline
				(3,3) & 2 & $0.5909$ & $*$  & $0.6538$\\
				\hline
				(3,4) & 1 & $0.1579$ & $*$  & $0.2381$\\
				\hline
				(3,4) & 2 & $0.3725$ & $*$  & $0.4921$\\
				\hline
				(3,4) & 3 & $0.6444$ & $*$  & $0.7460$\\
				\hline
				(3,5) & 1 & $0.1071$ & $*$  & $0.1935$\\
				\hline
				(3,5) & 2 & $0.2500$ & $*$  & $0.3951$\\
				\hline
				(3,5) & 3 & $0.4318$ & $*$  & $0.5967$\\
				\hline
				(3,5) & 4 & $0.6711$ & $*$  & $0.7984$\\
				\hline
				(4,2) & 1 & $0.4667$ & $0.4146$  & $0.4667$\\
				\hline
				(4,3) & 1 & $0.2703$ & $0.2174$  & $0.3250$\\
				\hline
				(4,3) & 2 & $0.6029$ & $0.5129$  & $0.6625$\\
				\hline
				(4,4) & 1 & $0.1688$ & $0.128$  & $0.2471$\\
				\hline
				(4,4) & 2 & $0.3816$ & $0.294$  & $0.4980$\\
				\hline
				(4,4) & 3 & $0.6503$ & $0.534$  & $0.7490$\\
				\hline
				(4,5) & 1 & $0.1135$ & $-$  & $0.1987$\\
				\hline
				(4,5) & 2 & $0.2560$ & $-$  & $0.3990$\\
				\hline
				(4,5) & 3 & $0.4369$ & $-$  & $0.5994$\\
				\hline
				(4,5) & 4 & $0.6745$ & $-$  & $0.7997$\\
				\hline
				(5,2) & 1 & $0.4839$ & $-$  & $0.4839$\\
				\hline
				(5,3) & 1 & $0.2358$ & $-$  & $0.3306$\\
				\hline
				(5,3) & 2 & $0.5549$ & $-$  & $0.6653$\\
				\hline
				(5,4) & 1 & $0.1203$ & $-$  & $0.2493$\\
				\hline
				(5,4) & 2 & $0.2918$ & $-$  & $0.4995$\\
				\hline
				(5,4) & 3 & $0.5532$ & $-$  & $0.7498$\\
				\hline
			\end{tabular}
			\caption*{Critical visibility, $v$, from convex programming relaxation for the GME-dimension. Case study presented for target GHZ states and cluster states under white noise. These states are local-unitary equivalent for $n = 3$, while in all the other cases, the cluster state is more noise tolerant than GHZ. The $``-"$ indicates the cases in which our desktop was not able to evaluate the linear program.}
		\end{table}

		\newpage
		\section{Convex programming method for detection of Schmidt number across all bipartitions}
		The convex programming method presented in the main text to detect the GME-dimension can also be used to study finer structures, in which one looks at the Schmidt number across all bipartitions. In this case, the semidefinite program is very similar to the one in \eqref{Appendix_SDP-LP: SDP}, except that the $r$-positive map constraint will be different for each bipartition.
		
		Consider, for example, that we want to detect the entanglement structure $(d_1,d_2,d_3)$ in a three-particle system, where $d_1,d_2,d_3$ are the Schmidt numbers across the bipartitions $A|BC$, $B|AC$ and $C|AB$, respectively. Then, we can use the following semidefinite program
		\begin{equation}\label{SDP_SV3}
			\begin{aligned}
				\max_{v,\tilde{\sigma}} & \quad v\\
				\text{s.t.}&\quad v\rho+\frac{1-v}{d^{3}}\openone =  \tilde{\sigma}_{A|BC} + \tilde{\sigma}_{B|AC} + \tilde{\sigma}_{C|AB},\\
				& \quad \tilde{\sigma}_{A|BC} \geq 0, \quad \tilde{\sigma}_{B|AC} \geq 0, \quad \tilde{\sigma}_{C|AB} \geq 0,\\
				& \quad (\Lambda_A^{(d_1)}\otimes \openone_{BC})[\tilde{\sigma}_{A|BC}] \geq 0, \\
				& \quad (\Lambda_B^{(d_2)}\otimes \openone_{AC})[\tilde{\sigma}_{B|AC}] \geq 0, \\
				& \quad (\Lambda_C^{(d_3)}\otimes \openone_{AB})[\tilde{\sigma}_{C|AB}] \geq 0, \\
			\end{aligned}
		\end{equation}
		where $\tilde{\sigma}_{A|BC}, \tilde{\sigma}_{B|AC},\tilde{\sigma}_{C|AB}$ are unnormalised states and the $r$-positive map is $\Lambda_S^{(r)}(X) = \tr X \openone - \frac{1}{r} X $. More generally, to detect an entanglement structure denoted by a set of Schmidt numbers $\lbrace d_{S|\bar{S}} \rbrace$, the semidefinite program will be
		\begin{equation}\label{SDP_SV}
			\begin{aligned}
				\max_{v,\tilde{\sigma}} & \quad v\\
				\text{s.t.}&\quad v\rho+\frac{1-v}{d^{n}}\openone =  \sum_{ S|\bar{S}} \tilde{\sigma}_{S|\bar{S}},\\
				& \quad \tilde{\sigma}_{S|\bar{S}} \geq 0 \quad \forall (S|\bar{S}),\\
				& \quad (\Lambda_S^{(d_{S|\bar{S}})}\otimes \openone_{\bar{S}})[\tilde{\sigma}_{S|\bar{S}}] \geq 0 \quad \forall (S|\bar{S}).
			\end{aligned}
		\end{equation}
		Thus, any value $v < 1$ denotes the critical visibility above which the multipartite state $v\rho+\frac{1-v}{d^{n}}\openone$ cannot have Schmidt numbers $\lbrace d_{S|\bar{S}} \rbrace$ across the bipartitions $\lbrace S|\bar{S} \rbrace$. We have evaluated this program in the case of a noisy three-particle GHZ state with local dimension $d = \lbrace 3,4 \rbrace$ (see Table \ref{table: SDP_SV}). Given the invariance of the GHZ states under permutations of the parties, we only need to consider the Schmidt number sets, $(d_1,d_2,d_3)$, that are not equivalent under permutation (e.g. $(1,2,2)$ and $(2,1,2)$ are equivalent and therefore give the same result).
		
		\begin{table}[h!]
			\begin{tabular}{|c|c|c|}
				\hline
				$\,$ (n,d) $\,$ & $\,$ $d_{S|\bar{S}}$ $\,$ & $\,$  $v_{\text{ghz}}$ (LP)\\
				\hline
				\rowcolor{LightCyan}
				(3,3) & $\,$ $(1,1,1)$$\,$ & $0.2500$\\
				\hline
				(3,3) & $(1,1,2)$ & $0.4375$\\
				\hline
				(3,3) & $(1,2,2)$ & $0.5263$\\
				\hline
				\rowcolor{LightCyan}
				(3,3) & $(2,2,2)$ & $0.5909$\\
				\hline
				\rowcolor{LightCyan}
				(3,4) & $(1,1,1)$ & $0.1579$\\
				\hline
				(3,4) & $(1,1,2)$ & $0.2558$\\
				\hline
				(3,4) & $(1,1,3)$ & $0.4483$\\
				\hline
				(3,4) & $(1,2,2)$ & $0.3191$\\
				\hline
				(3,4) & $(1,2,3)$ & $0.4839$\\
				\hline
				(3,4) & $(1,3,3)$ & $0.5676$\\
				\hline
				\rowcolor{LightCyan}
				(3,4) & $(2,2,2)$ & $0.3725$\\
				\hline
				(3,4) & $(2,2,3)$ & $0.5152$\\
				\hline
				(3,4) & $(2,3,3)$ & $0.5897$\\
				\hline
				(3,4) & $(3,3,1)$ & $0.5676$\\
				\hline
				\rowcolor{LightCyan}
				(3,4) & $(3,3,3)$ & $0.6444$\\
				\hline
			\end{tabular}
			\caption{\textbf{LP results for the detection of Schmidt number across all bipartitions. }Critical visibility, $v$, computed by reformulating \eqref{SDP_SV3} as a linear program for the detection of Schmidt number structures $(d_1,d_2,d_3)$. Case study presented for target three-particle GHZ states of local dimension $d = \lbrace 3,4 \rbrace$ under white noise. We considered only nonequivalent structures under permutation. The highlighted cases correspond to the results already obtained for the detection of the GME-dimension.}
			\label{table: SDP_SV}
		\end{table}

		\color{black}	
		\newpage
		\section{Convex programming method for simulation with few measurements}	
		In this appendix, we show how the convex programming method presented in the main text can be used to give resource-efficient criteria beyond the minimal fidelity estimation. This is particularly natural for experimental considerations, where both noise and coincidence rates can be a significant challenge.
		
		The semidefinite program to be implemented now is very similar to the one in \eqref{Appendix_SDP-LP: SDP}, except that we no longer want to simulate the whole state $\rho$, but only its statistics when a small number of measurements are performed on it.
		For this, we can choose any set of global product projections $E_j = \{\mathcal{M}_{\vec{k}}^{(j)}\}$ and bound the white noise tolerance by evaluating the SDP
		\begin{equation}\label{SDP_prob}
			\begin{aligned}
				\max_{v,\tilde{\sigma}} & \quad v\\
				\text{s.t.}& \, \trace\left(\left(v\rho+\frac{1-v}{d^{n}}\openone \right) \mathcal{M}_{\vec{k}}^{(j)}\right) =  \trace\left(\sum_{ S|\bar{S}} \tilde{\sigma}_{S|\bar{S}} \, \mathcal{M}_{\vec{k}}^{(j)}\right) \, \forall {\vec{k}},j,\\
				& \quad \tilde{\sigma}_{S|\bar{S}} \geq 0 \quad \forall (S|\bar{S}),\\
				& \quad (\Lambda_S\otimes \openone_{\bar{S}})[\tilde{\sigma}_{S|\bar{S}}] \geq 0 \quad \forall (S|\bar{S}).
			\end{aligned}
		\end{equation}
		To showcase the method in the main text we consider that $\lbrace M_{\vec{k}}^{(j)} \rbrace$ correspond to global product measurements in which all local measurements are mutually unibiased bases (MUBs). In particular we considered
		\begin{equation}
			\begin{aligned}
				E_C &= \lbrace \mathcal{M}_{\vec{k}}^{(C)} = \ketbra{k_1,\dots,k_n}, \quad \forall \, k_l = 0,\dots,d-1 \rbrace, \\
				E_F &= \left\lbrace \mathcal{M}_{\vec{k}}^{(F)} = \ketbra{e_{k_1}^{(0)}, \dots, e_{k_n}^{(0)}}, \quad \forall \, k_l = 0,\dots,d-1 \right\rbrace, \\
				E_M &= \left\lbrace \mathcal{M}_{\vec{k}}^{(M)} = \ketbra{e_{k_1}^{(j_1)},\dots,e_{k_n}^{(j_n)}}\delta_{j_1 \oplus \dots \oplus j_n,0}, \quad \forall \, k_l = 0,\dots,d-1  \right\rbrace,
			\end{aligned}
		\end{equation}
		where $\lbrace \ket{k} \rbrace_{k=0}^{d-1}$ denotes the computational basis and $\left\lbrace  \ket{e_l^{(j)}}\right\rbrace_{l = 0}^{d-1}$ the remaining $j = 0,\dots,d-1$ MUBs.\\
		Already by imposing constraints only for $E_C$ and $E_F$ (i.e. using only the computational and the Fourier basis), this method is more noise-robust than the minimal fidelity witnesses. The results significantly improve by adding constraints on the measurements in the set $E_M$.\\
		Moreover, from these convex programming methods it is also possible to extract explicit witnesses for the GME-dimension by using the duality theory of semidefinite programming.


\begin{thebibliography}{80}%
		\makeatletter
		\providecommand \@ifxundefined [1]{%
			\@ifx{#1\undefined}
		}%
		\providecommand \@ifnum [1]{%
			\ifnum #1\expandafter \@firstoftwo
			\else \expandafter \@secondoftwo
			\fi
		}%
		\providecommand \@ifx [1]{%
			\ifx #1\expandafter \@firstoftwo
			\else \expandafter \@secondoftwo
			\fi
		}%
		\providecommand \natexlab [1]{#1}%
		\providecommand \enquote  [1]{``#1''}%
		\providecommand \bibnamefont  [1]{#1}%
		\providecommand \bibfnamefont [1]{#1}%
		\providecommand \citenamefont [1]{#1}%
		\providecommand \href@noop [0]{\@secondoftwo}%
		\providecommand \href [0]{\begingroup \@sanitize@url \@href}%
		\providecommand \@href[1]{\@@startlink{#1}\@@href}%
		\providecommand \@@href[1]{\endgroup#1\@@endlink}%
		\providecommand \@sanitize@url [0]{\catcode `\\12\catcode `\$12\catcode
			`\&12\catcode `\#12\catcode `\^12\catcode `\_12\catcode `\%12\relax}%
		\providecommand \@@startlink[1]{}%
		\providecommand \@@endlink[0]{}%
		\providecommand \url  [0]{\begingroup\@sanitize@url \@url }%
		\providecommand \@url [1]{\endgroup\@href {#1}{\urlprefix }}%
		\providecommand \urlprefix  [0]{URL }%
		\providecommand \Eprint [0]{\href }%
		\providecommand \doibase [0]{https://doi.org/}%
		\providecommand \selectlanguage [0]{\@gobble}%
		\providecommand \bibinfo  [0]{\@secondoftwo}%
		\providecommand \bibfield  [0]{\@secondoftwo}%
		\providecommand \translation [1]{[#1]}%
		\providecommand \BibitemOpen [0]{}%
		\providecommand \bibitemStop [0]{}%
		\providecommand \bibitemNoStop [0]{.\EOS\space}%
		\providecommand \EOS [0]{\spacefactor3000\relax}%
		\providecommand \BibitemShut  [1]{\csname bibitem#1\endcsname}%
		\let\auto@bib@innerbib\@empty
		\bibitem [{\citenamefont {Horodecki}\ \emph {et~al.}(2009)\citenamefont
			{Horodecki}, \citenamefont {Horodecki}, \citenamefont {Horodecki},\ and\
			\citenamefont {Horodecki}}]{Horodecki2009}%
		\BibitemOpen
		\bibfield  {author} {\bibinfo {author} {\bibfnamefont {R.}~\bibnamefont
				{Horodecki}}, \bibinfo {author} {\bibfnamefont {P.}~\bibnamefont
				{Horodecki}}, \bibinfo {author} {\bibfnamefont {M.}~\bibnamefont
				{Horodecki}},\ and\ \bibinfo {author} {\bibfnamefont {K.}~\bibnamefont
				{Horodecki}},\ }\bibfield  {title} {\bibinfo {title} {Quantum entanglement},\
		}\href {https://doi.org/10.1103/RevModPhys.81.865} {\bibfield  {journal}
			{\bibinfo  {journal} {Rev. Mod. Phys.}\ }\textbf {\bibinfo {volume} {81}},\
			\bibinfo {pages} {865} (\bibinfo {year} {2009})}\BibitemShut {NoStop}%
		\bibitem [{\citenamefont {Gühne}\ and\ \citenamefont
			{Tóth}(2009)}]{Guhne2009}%
		\BibitemOpen
		\bibfield  {author} {\bibinfo {author} {\bibfnamefont {O.}~\bibnamefont
				{Gühne}}\ and\ \bibinfo {author} {\bibfnamefont {G.}~\bibnamefont {Tóth}},\
		}\bibfield  {title} {\bibinfo {title} {Entanglement detection},\ }\href
		{https://doi.org/https://doi.org/10.1016/j.physrep.2009.02.004} {\bibfield
			{journal} {\bibinfo  {journal} {Physics Reports}\ }\textbf {\bibinfo {volume}
				{474}},\ \bibinfo {pages} {1} (\bibinfo {year} {2009})}\BibitemShut {NoStop}%
		\bibitem [{\citenamefont {Xu}\ \emph {et~al.}(2020)\citenamefont {Xu},
			\citenamefont {Ma}, \citenamefont {Zhang}, \citenamefont {Lo},\ and\
			\citenamefont {Pan}}]{Xu2020}%
		\BibitemOpen
		\bibfield  {author} {\bibinfo {author} {\bibfnamefont {F.}~\bibnamefont
				{Xu}}, \bibinfo {author} {\bibfnamefont {X.}~\bibnamefont {Ma}}, \bibinfo
			{author} {\bibfnamefont {Q.}~\bibnamefont {Zhang}}, \bibinfo {author}
			{\bibfnamefont {H.-K.}\ \bibnamefont {Lo}},\ and\ \bibinfo {author}
			{\bibfnamefont {J.-W.}\ \bibnamefont {Pan}},\ }\bibfield  {title} {\bibinfo
			{title} {Secure quantum key distribution with realistic devices},\ }\href
		{https://doi.org/10.1103/RevModPhys.92.025002} {\bibfield  {journal}
			{\bibinfo  {journal} {Rev. Mod. Phys.}\ }\textbf {\bibinfo {volume} {92}},\
			\bibinfo {pages} {025002} (\bibinfo {year} {2020})}\BibitemShut {NoStop}%
		\bibitem [{\citenamefont {Pirandola}\ \emph {et~al.}(2020)\citenamefont
			{Pirandola}, \citenamefont {Andersen}, \citenamefont {Banchi}, \citenamefont
			{Berta}, \citenamefont {Bunandar}, \citenamefont {Colbeck}, \citenamefont
			{Englund}, \citenamefont {Gehring}, \citenamefont {Lupo}, \citenamefont
			{Ottaviani}, \citenamefont {Pereira}, \citenamefont {Razavi}, \citenamefont
			{Shaari}, \citenamefont {Tomamichel}, \citenamefont {Usenko}, \citenamefont
			{Vallone}, \citenamefont {Villoresi},\ and\ \citenamefont
			{Wallden}}]{Pirandola2020}%
		\BibitemOpen
		\bibfield  {author} {\bibinfo {author} {\bibfnamefont {S.}~\bibnamefont
				{Pirandola}}, \bibinfo {author} {\bibfnamefont {U.~L.}\ \bibnamefont
				{Andersen}}, \bibinfo {author} {\bibfnamefont {L.}~\bibnamefont {Banchi}},
			\bibinfo {author} {\bibfnamefont {M.}~\bibnamefont {Berta}}, \bibinfo
			{author} {\bibfnamefont {D.}~\bibnamefont {Bunandar}}, \bibinfo {author}
			{\bibfnamefont {R.}~\bibnamefont {Colbeck}}, \bibinfo {author} {\bibfnamefont
				{D.}~\bibnamefont {Englund}}, \bibinfo {author} {\bibfnamefont
				{T.}~\bibnamefont {Gehring}}, \bibinfo {author} {\bibfnamefont
				{C.}~\bibnamefont {Lupo}}, \bibinfo {author} {\bibfnamefont {C.}~\bibnamefont
				{Ottaviani}}, \bibinfo {author} {\bibfnamefont {J.~L.}\ \bibnamefont
				{Pereira}}, \bibinfo {author} {\bibfnamefont {M.}~\bibnamefont {Razavi}},
			\bibinfo {author} {\bibfnamefont {J.~S.}\ \bibnamefont {Shaari}}, \bibinfo
			{author} {\bibfnamefont {M.}~\bibnamefont {Tomamichel}}, \bibinfo {author}
			{\bibfnamefont {V.~C.}\ \bibnamefont {Usenko}}, \bibinfo {author}
			{\bibfnamefont {G.}~\bibnamefont {Vallone}}, \bibinfo {author} {\bibfnamefont
				{P.}~\bibnamefont {Villoresi}},\ and\ \bibinfo {author} {\bibfnamefont
				{P.}~\bibnamefont {Wallden}},\ }\bibfield  {title} {\bibinfo {title}
			{Advances in quantum cryptography},\ }\href
		{https://doi.org/10.1364/AOP.361502} {\bibfield  {journal} {\bibinfo
				{journal} {Adv. Opt. Photon.}\ }\textbf {\bibinfo {volume} {12}},\ \bibinfo
			{pages} {1012} (\bibinfo {year} {2020})}\BibitemShut {NoStop}%
		\bibitem [{\citenamefont {Portmann}\ and\ \citenamefont
			{Renner}(2022)}]{Portman2022}%
		\BibitemOpen
		\bibfield  {author} {\bibinfo {author} {\bibfnamefont {C.}~\bibnamefont
				{Portmann}}\ and\ \bibinfo {author} {\bibfnamefont {R.}~\bibnamefont
				{Renner}},\ }\bibfield  {title} {\bibinfo {title} {Security in quantum
				cryptography},\ }\href {https://doi.org/10.1103/RevModPhys.94.025008}
		{\bibfield  {journal} {\bibinfo  {journal} {Rev. Mod. Phys.}\ }\textbf
			{\bibinfo {volume} {94}},\ \bibinfo {pages} {025008} (\bibinfo {year}
			{2022})}\BibitemShut {NoStop}%
		\bibitem [{\citenamefont {Giovannetti}\ \emph {et~al.}(2011)\citenamefont
			{Giovannetti}, \citenamefont {Lloyd},\ and\ \citenamefont
			{Maccone}}]{Giovannetti2011}%
		\BibitemOpen
		\bibfield  {author} {\bibinfo {author} {\bibfnamefont {V.}~\bibnamefont
				{Giovannetti}}, \bibinfo {author} {\bibfnamefont {S.}~\bibnamefont {Lloyd}},\
			and\ \bibinfo {author} {\bibfnamefont {L.}~\bibnamefont {Maccone}},\
		}\bibfield  {title} {\bibinfo {title} {Advances in quantum metrology},\
		}\href {https://doi.org/10.1038/nphoton.2011.35} {\bibfield  {journal}
			{\bibinfo  {journal} {Nature Photonics}\ }\textbf {\bibinfo {volume} {5}},\
			\bibinfo {pages} {222} (\bibinfo {year} {2011})}\BibitemShut {NoStop}%
		\bibitem [{\citenamefont {Degen}\ \emph {et~al.}(2017)\citenamefont {Degen},
			\citenamefont {Reinhard},\ and\ \citenamefont {Cappellaro}}]{Degen2017}%
		\BibitemOpen
		\bibfield  {author} {\bibinfo {author} {\bibfnamefont {C.~L.}\ \bibnamefont
				{Degen}}, \bibinfo {author} {\bibfnamefont {F.}~\bibnamefont {Reinhard}},\
			and\ \bibinfo {author} {\bibfnamefont {P.}~\bibnamefont {Cappellaro}},\
		}\bibfield  {title} {\bibinfo {title} {Quantum sensing},\ }\href
		{https://doi.org/10.1103/RevModPhys.89.035002} {\bibfield  {journal}
			{\bibinfo  {journal} {Rev. Mod. Phys.}\ }\textbf {\bibinfo {volume} {89}},\
			\bibinfo {pages} {035002} (\bibinfo {year} {2017})}\BibitemShut {NoStop}%
		\bibitem [{\citenamefont {Pezz\`e}\ \emph {et~al.}(2018)\citenamefont
			{Pezz\`e}, \citenamefont {Smerzi}, \citenamefont {Oberthaler}, \citenamefont
			{Schmied},\ and\ \citenamefont {Treutlein}}]{Pezze2018}%
		\BibitemOpen
		\bibfield  {author} {\bibinfo {author} {\bibfnamefont {L.}~\bibnamefont
				{Pezz\`e}}, \bibinfo {author} {\bibfnamefont {A.}~\bibnamefont {Smerzi}},
			\bibinfo {author} {\bibfnamefont {M.~K.}\ \bibnamefont {Oberthaler}},
			\bibinfo {author} {\bibfnamefont {R.}~\bibnamefont {Schmied}},\ and\ \bibinfo
			{author} {\bibfnamefont {P.}~\bibnamefont {Treutlein}},\ }\bibfield  {title}
		{\bibinfo {title} {Quantum metrology with nonclassical states of atomic
				ensembles},\ }\href {https://doi.org/10.1103/RevModPhys.90.035005} {\bibfield
			{journal} {\bibinfo  {journal} {Rev. Mod. Phys.}\ }\textbf {\bibinfo
				{volume} {90}},\ \bibinfo {pages} {035005} (\bibinfo {year}
			{2018})}\BibitemShut {NoStop}%
		\bibitem [{\citenamefont {Arute}\ \emph {et~al.}(2019)\citenamefont {Arute},
			\citenamefont {Arya}, \citenamefont {Babbush}, \citenamefont {Bacon},
			\citenamefont {Bardin}, \citenamefont {Barends}, \citenamefont {Biswas},
			\citenamefont {Boixo}, \citenamefont {Brandao}, \citenamefont {Buell},
			\citenamefont {Burkett}, \citenamefont {Chen}, \citenamefont {Chen},
			\citenamefont {Chiaro}, \citenamefont {Collins}, \citenamefont {Courtney},
			\citenamefont {Dunsworth}, \citenamefont {Farhi}, \citenamefont {Foxen},
			\citenamefont {Fowler}, \citenamefont {Gidney}, \citenamefont {Giustina},
			\citenamefont {Graff}, \citenamefont {Guerin}, \citenamefont {Habegger},
			\citenamefont {Harrigan}, \citenamefont {Hartmann}, \citenamefont {Ho},
			\citenamefont {Hoffmann}, \citenamefont {Huang}, \citenamefont {Humble},
			\citenamefont {Isakov}, \citenamefont {Jeffrey}, \citenamefont {Jiang},
			\citenamefont {Kafri}, \citenamefont {Kechedzhi}, \citenamefont {Kelly},
			\citenamefont {Klimov}, \citenamefont {Knysh}, \citenamefont {Korotkov},
			\citenamefont {Kostritsa}, \citenamefont {Landhuis}, \citenamefont
			{Lindmark}, \citenamefont {Lucero}, \citenamefont {Lyakh}, \citenamefont
			{Mandr{\`a}}, \citenamefont {McClean}, \citenamefont {McEwen}, \citenamefont
			{Megrant}, \citenamefont {Mi}, \citenamefont {Michielsen}, \citenamefont
			{Mohseni}, \citenamefont {Mutus}, \citenamefont {Naaman}, \citenamefont
			{Neeley}, \citenamefont {Neill}, \citenamefont {Niu}, \citenamefont {Ostby},
			\citenamefont {Petukhov}, \citenamefont {Platt}, \citenamefont {Quintana},
			\citenamefont {Rieffel}, \citenamefont {Roushan}, \citenamefont {Rubin},
			\citenamefont {Sank}, \citenamefont {Satzinger}, \citenamefont {Smelyanskiy},
			\citenamefont {Sung}, \citenamefont {Trevithick}, \citenamefont
			{Vainsencher}, \citenamefont {Villalonga}, \citenamefont {White},
			\citenamefont {Yao}, \citenamefont {Yeh}, \citenamefont {Zalcman},
			\citenamefont {Neven},\ and\ \citenamefont {Martinis}}]{Arute2019}%
		\BibitemOpen
		\bibfield  {author} {\bibinfo {author} {\bibfnamefont {F.}~\bibnamefont
				{Arute}}, \bibinfo {author} {\bibfnamefont {K.}~\bibnamefont {Arya}},
			\bibinfo {author} {\bibfnamefont {R.}~\bibnamefont {Babbush}}, \bibinfo
			{author} {\bibfnamefont {D.}~\bibnamefont {Bacon}}, \bibinfo {author}
			{\bibfnamefont {J.~C.}\ \bibnamefont {Bardin}}, \bibinfo {author}
			{\bibfnamefont {R.}~\bibnamefont {Barends}}, \bibinfo {author} {\bibfnamefont
				{R.}~\bibnamefont {Biswas}}, \bibinfo {author} {\bibfnamefont
				{S.}~\bibnamefont {Boixo}}, \bibinfo {author} {\bibfnamefont {F.~G. S.~L.}\
				\bibnamefont {Brandao}}, \bibinfo {author} {\bibfnamefont {D.~A.}\
				\bibnamefont {Buell}}, \bibinfo {author} {\bibfnamefont {B.}~\bibnamefont
				{Burkett}}, \bibinfo {author} {\bibfnamefont {Y.}~\bibnamefont {Chen}},
			\bibinfo {author} {\bibfnamefont {Z.}~\bibnamefont {Chen}}, \bibinfo {author}
			{\bibfnamefont {B.}~\bibnamefont {Chiaro}}, \bibinfo {author} {\bibfnamefont
				{R.}~\bibnamefont {Collins}}, \bibinfo {author} {\bibfnamefont
				{W.}~\bibnamefont {Courtney}}, \bibinfo {author} {\bibfnamefont
				{A.}~\bibnamefont {Dunsworth}}, \bibinfo {author} {\bibfnamefont
				{E.}~\bibnamefont {Farhi}}, \bibinfo {author} {\bibfnamefont
				{B.}~\bibnamefont {Foxen}}, \bibinfo {author} {\bibfnamefont
				{A.}~\bibnamefont {Fowler}}, \bibinfo {author} {\bibfnamefont
				{C.}~\bibnamefont {Gidney}}, \bibinfo {author} {\bibfnamefont
				{M.}~\bibnamefont {Giustina}}, \bibinfo {author} {\bibfnamefont
				{R.}~\bibnamefont {Graff}}, \bibinfo {author} {\bibfnamefont
				{K.}~\bibnamefont {Guerin}}, \bibinfo {author} {\bibfnamefont
				{S.}~\bibnamefont {Habegger}}, \bibinfo {author} {\bibfnamefont {M.~P.}\
				\bibnamefont {Harrigan}}, \bibinfo {author} {\bibfnamefont {M.~J.}\
				\bibnamefont {Hartmann}}, \bibinfo {author} {\bibfnamefont {A.}~\bibnamefont
				{Ho}}, \bibinfo {author} {\bibfnamefont {M.}~\bibnamefont {Hoffmann}},
			\bibinfo {author} {\bibfnamefont {T.}~\bibnamefont {Huang}}, \bibinfo
			{author} {\bibfnamefont {T.~S.}\ \bibnamefont {Humble}}, \bibinfo {author}
			{\bibfnamefont {S.~V.}\ \bibnamefont {Isakov}}, \bibinfo {author}
			{\bibfnamefont {E.}~\bibnamefont {Jeffrey}}, \bibinfo {author} {\bibfnamefont
				{Z.}~\bibnamefont {Jiang}}, \bibinfo {author} {\bibfnamefont
				{D.}~\bibnamefont {Kafri}}, \bibinfo {author} {\bibfnamefont
				{K.}~\bibnamefont {Kechedzhi}}, \bibinfo {author} {\bibfnamefont
				{J.}~\bibnamefont {Kelly}}, \bibinfo {author} {\bibfnamefont {P.~V.}\
				\bibnamefont {Klimov}}, \bibinfo {author} {\bibfnamefont {S.}~\bibnamefont
				{Knysh}}, \bibinfo {author} {\bibfnamefont {A.}~\bibnamefont {Korotkov}},
			\bibinfo {author} {\bibfnamefont {F.}~\bibnamefont {Kostritsa}}, \bibinfo
			{author} {\bibfnamefont {D.}~\bibnamefont {Landhuis}}, \bibinfo {author}
			{\bibfnamefont {M.}~\bibnamefont {Lindmark}}, \bibinfo {author}
			{\bibfnamefont {E.}~\bibnamefont {Lucero}}, \bibinfo {author} {\bibfnamefont
				{D.}~\bibnamefont {Lyakh}}, \bibinfo {author} {\bibfnamefont
				{S.}~\bibnamefont {Mandr{\`a}}}, \bibinfo {author} {\bibfnamefont {J.~R.}\
				\bibnamefont {McClean}}, \bibinfo {author} {\bibfnamefont {M.}~\bibnamefont
				{McEwen}}, \bibinfo {author} {\bibfnamefont {A.}~\bibnamefont {Megrant}},
			\bibinfo {author} {\bibfnamefont {X.}~\bibnamefont {Mi}}, \bibinfo {author}
			{\bibfnamefont {K.}~\bibnamefont {Michielsen}}, \bibinfo {author}
			{\bibfnamefont {M.}~\bibnamefont {Mohseni}}, \bibinfo {author} {\bibfnamefont
				{J.}~\bibnamefont {Mutus}}, \bibinfo {author} {\bibfnamefont
				{O.}~\bibnamefont {Naaman}}, \bibinfo {author} {\bibfnamefont
				{M.}~\bibnamefont {Neeley}}, \bibinfo {author} {\bibfnamefont
				{C.}~\bibnamefont {Neill}}, \bibinfo {author} {\bibfnamefont {M.~Y.}\
				\bibnamefont {Niu}}, \bibinfo {author} {\bibfnamefont {E.}~\bibnamefont
				{Ostby}}, \bibinfo {author} {\bibfnamefont {A.}~\bibnamefont {Petukhov}},
			\bibinfo {author} {\bibfnamefont {J.~C.}\ \bibnamefont {Platt}}, \bibinfo
			{author} {\bibfnamefont {C.}~\bibnamefont {Quintana}}, \bibinfo {author}
			{\bibfnamefont {E.~G.}\ \bibnamefont {Rieffel}}, \bibinfo {author}
			{\bibfnamefont {P.}~\bibnamefont {Roushan}}, \bibinfo {author} {\bibfnamefont
				{N.~C.}\ \bibnamefont {Rubin}}, \bibinfo {author} {\bibfnamefont
				{D.}~\bibnamefont {Sank}}, \bibinfo {author} {\bibfnamefont {K.~J.}\
				\bibnamefont {Satzinger}}, \bibinfo {author} {\bibfnamefont {V.}~\bibnamefont
				{Smelyanskiy}}, \bibinfo {author} {\bibfnamefont {K.~J.}\ \bibnamefont
				{Sung}}, \bibinfo {author} {\bibfnamefont {M.~D.}\ \bibnamefont
				{Trevithick}}, \bibinfo {author} {\bibfnamefont {A.}~\bibnamefont
				{Vainsencher}}, \bibinfo {author} {\bibfnamefont {B.}~\bibnamefont
				{Villalonga}}, \bibinfo {author} {\bibfnamefont {T.}~\bibnamefont {White}},
			\bibinfo {author} {\bibfnamefont {Z.~J.}\ \bibnamefont {Yao}}, \bibinfo
			{author} {\bibfnamefont {P.}~\bibnamefont {Yeh}}, \bibinfo {author}
			{\bibfnamefont {A.}~\bibnamefont {Zalcman}}, \bibinfo {author} {\bibfnamefont
				{H.}~\bibnamefont {Neven}},\ and\ \bibinfo {author} {\bibfnamefont {J.~M.}\
				\bibnamefont {Martinis}},\ }\bibfield  {title} {\bibinfo {title} {Quantum
				supremacy using a programmable superconducting processor},\ }\href
		{https://doi.org/10.1038/s41586-019-1666-5} {\bibfield  {journal} {\bibinfo
				{journal} {Nature}\ }\textbf {\bibinfo {volume} {574}},\ \bibinfo {pages}
			{505} (\bibinfo {year} {2019})}\BibitemShut {NoStop}%
		\bibitem [{\citenamefont {Zhong}\ \emph {et~al.}(2020)\citenamefont {Zhong},
			\citenamefont {Wang}, \citenamefont {Deng}, \citenamefont {Chen},
			\citenamefont {Peng}, \citenamefont {Luo}, \citenamefont {Qin}, \citenamefont
			{Wu}, \citenamefont {Ding}, \citenamefont {Hu}, \citenamefont {Hu},
			\citenamefont {Yang}, \citenamefont {Zhang}, \citenamefont {Li},
			\citenamefont {Li}, \citenamefont {Jiang}, \citenamefont {Gan}, \citenamefont
			{Yang}, \citenamefont {You}, \citenamefont {Wang}, \citenamefont {Li},
			\citenamefont {Liu}, \citenamefont {Lu},\ and\ \citenamefont
			{Pan}}]{Zhong2020}%
		\BibitemOpen
		\bibfield  {author} {\bibinfo {author} {\bibfnamefont {H.-S.}\ \bibnamefont
				{Zhong}}, \bibinfo {author} {\bibfnamefont {H.}~\bibnamefont {Wang}},
			\bibinfo {author} {\bibfnamefont {Y.-H.}\ \bibnamefont {Deng}}, \bibinfo
			{author} {\bibfnamefont {M.-C.}\ \bibnamefont {Chen}}, \bibinfo {author}
			{\bibfnamefont {L.-C.}\ \bibnamefont {Peng}}, \bibinfo {author}
			{\bibfnamefont {Y.-H.}\ \bibnamefont {Luo}}, \bibinfo {author} {\bibfnamefont
				{J.}~\bibnamefont {Qin}}, \bibinfo {author} {\bibfnamefont {D.}~\bibnamefont
				{Wu}}, \bibinfo {author} {\bibfnamefont {X.}~\bibnamefont {Ding}}, \bibinfo
			{author} {\bibfnamefont {Y.}~\bibnamefont {Hu}}, \bibinfo {author}
			{\bibfnamefont {P.}~\bibnamefont {Hu}}, \bibinfo {author} {\bibfnamefont
				{X.-Y.}\ \bibnamefont {Yang}}, \bibinfo {author} {\bibfnamefont {W.-J.}\
				\bibnamefont {Zhang}}, \bibinfo {author} {\bibfnamefont {H.}~\bibnamefont
				{Li}}, \bibinfo {author} {\bibfnamefont {Y.}~\bibnamefont {Li}}, \bibinfo
			{author} {\bibfnamefont {X.}~\bibnamefont {Jiang}}, \bibinfo {author}
			{\bibfnamefont {L.}~\bibnamefont {Gan}}, \bibinfo {author} {\bibfnamefont
				{G.}~\bibnamefont {Yang}}, \bibinfo {author} {\bibfnamefont {L.}~\bibnamefont
				{You}}, \bibinfo {author} {\bibfnamefont {Z.}~\bibnamefont {Wang}}, \bibinfo
			{author} {\bibfnamefont {L.}~\bibnamefont {Li}}, \bibinfo {author}
			{\bibfnamefont {N.-L.}\ \bibnamefont {Liu}}, \bibinfo {author} {\bibfnamefont
				{C.-Y.}\ \bibnamefont {Lu}},\ and\ \bibinfo {author} {\bibfnamefont {J.-W.}\
				\bibnamefont {Pan}},\ }\bibfield  {title} {\bibinfo {title} {Quantum
				computational advantage using photons},\ }\href
		{https://doi.org/10.1126/science.abe8770} {\bibfield  {journal} {\bibinfo
				{journal} {Science}\ }\textbf {\bibinfo {volume} {370}},\ \bibinfo {pages}
			{1460} (\bibinfo {year} {2020})},\ \Eprint
		{https://arxiv.org/abs/https://www.science.org/doi/pdf/10.1126/science.abe8770}
		{https://www.science.org/doi/pdf/10.1126/science.abe8770} \BibitemShut
		{NoStop}%
		\bibitem [{\citenamefont {Pan}\ \emph {et~al.}(2012)\citenamefont {Pan},
			\citenamefont {Chen}, \citenamefont {Lu}, \citenamefont {Weinfurter},
			\citenamefont {Zeilinger},\ and\ \citenamefont {\ifmmode~\dot{Z}\else
				\.{Z}\fi{}ukowski}}]{Pan2012}%
		\BibitemOpen
		\bibfield  {author} {\bibinfo {author} {\bibfnamefont {J.-W.}\ \bibnamefont
				{Pan}}, \bibinfo {author} {\bibfnamefont {Z.-B.}\ \bibnamefont {Chen}},
			\bibinfo {author} {\bibfnamefont {C.-Y.}\ \bibnamefont {Lu}}, \bibinfo
			{author} {\bibfnamefont {H.}~\bibnamefont {Weinfurter}}, \bibinfo {author}
			{\bibfnamefont {A.}~\bibnamefont {Zeilinger}},\ and\ \bibinfo {author}
			{\bibfnamefont {M.}~\bibnamefont {\ifmmode~\dot{Z}\else \.{Z}\fi{}ukowski}},\
		}\bibfield  {title} {\bibinfo {title} {Multiphoton entanglement and
				interferometry},\ }\href {https://doi.org/10.1103/RevModPhys.84.777}
		{\bibfield  {journal} {\bibinfo  {journal} {Rev. Mod. Phys.}\ }\textbf
			{\bibinfo {volume} {84}},\ \bibinfo {pages} {777} (\bibinfo {year}
			{2012})}\BibitemShut {NoStop}%
		\bibitem [{\citenamefont {Brunner}\ \emph {et~al.}(2014)\citenamefont
			{Brunner}, \citenamefont {Cavalcanti}, \citenamefont {Pironio}, \citenamefont
			{Scarani},\ and\ \citenamefont {Wehner}}]{Brunner2014}%
		\BibitemOpen
		\bibfield  {author} {\bibinfo {author} {\bibfnamefont {N.}~\bibnamefont
				{Brunner}}, \bibinfo {author} {\bibfnamefont {D.}~\bibnamefont {Cavalcanti}},
			\bibinfo {author} {\bibfnamefont {S.}~\bibnamefont {Pironio}}, \bibinfo
			{author} {\bibfnamefont {V.}~\bibnamefont {Scarani}},\ and\ \bibinfo {author}
			{\bibfnamefont {S.}~\bibnamefont {Wehner}},\ }\bibfield  {title} {\bibinfo
			{title} {Bell nonlocality},\ }\href
		{https://doi.org/10.1103/RevModPhys.86.419} {\bibfield  {journal} {\bibinfo
				{journal} {Rev. Mod. Phys.}\ }\textbf {\bibinfo {volume} {86}},\ \bibinfo
			{pages} {419} (\bibinfo {year} {2014})}\BibitemShut {NoStop}%
		\bibitem [{\citenamefont {Seevinck}\ and\ \citenamefont
			{Uffink}(2001)}]{Seevinck2001}%
		\BibitemOpen
		\bibfield  {author} {\bibinfo {author} {\bibfnamefont {M.}~\bibnamefont
				{Seevinck}}\ and\ \bibinfo {author} {\bibfnamefont {J.}~\bibnamefont
				{Uffink}},\ }\bibfield  {title} {\bibinfo {title} {Sufficient conditions for
				three-particle entanglement and their tests in recent experiments},\ }\href
		{https://doi.org/10.1103/PhysRevA.65.012107} {\bibfield  {journal} {\bibinfo
				{journal} {Phys. Rev. A}\ }\textbf {\bibinfo {volume} {65}},\ \bibinfo
			{pages} {012107} (\bibinfo {year} {2001})}\BibitemShut {NoStop}%
		\bibitem [{\citenamefont {Bourennane}\ \emph {et~al.}(2004)\citenamefont
			{Bourennane}, \citenamefont {Eibl}, \citenamefont {Kurtsiefer}, \citenamefont
			{Gaertner}, \citenamefont {Weinfurter}, \citenamefont {Gühne}, \citenamefont
			{Hyllus}, \citenamefont {Bruß}, \citenamefont {Lewenstein},\ and\
			\citenamefont {Sanpera}}]{Bourennane_2004}%
		\BibitemOpen
		\bibfield  {author} {\bibinfo {author} {\bibfnamefont {M.}~\bibnamefont
				{Bourennane}}, \bibinfo {author} {\bibfnamefont {M.}~\bibnamefont {Eibl}},
			\bibinfo {author} {\bibfnamefont {C.}~\bibnamefont {Kurtsiefer}}, \bibinfo
			{author} {\bibfnamefont {S.}~\bibnamefont {Gaertner}}, \bibinfo {author}
			{\bibfnamefont {H.}~\bibnamefont {Weinfurter}}, \bibinfo {author}
			{\bibfnamefont {O.}~\bibnamefont {Gühne}}, \bibinfo {author} {\bibfnamefont
				{P.}~\bibnamefont {Hyllus}}, \bibinfo {author} {\bibfnamefont
				{D.}~\bibnamefont {Bruß}}, \bibinfo {author} {\bibfnamefont
				{M.}~\bibnamefont {Lewenstein}},\ and\ \bibinfo {author} {\bibfnamefont
				{A.}~\bibnamefont {Sanpera}},\ }\bibfield  {title} {\bibinfo {title}
			{Experimental detection of multipartite entanglement using witness
				operators},\ }\bibfield  {journal} {\bibinfo  {journal} {Physical Review
				Letters}\ }\textbf {\bibinfo {volume} {92}},\ \href
		{https://doi.org/10.1103/physrevlett.92.087902}
		{10.1103/physrevlett.92.087902} (\bibinfo {year} {2004})\BibitemShut
		{NoStop}%
		\bibitem [{\citenamefont {H{\"a}ffner}\ \emph {et~al.}(2005)\citenamefont
			{H{\"a}ffner}, \citenamefont {H{\"a}nsel}, \citenamefont {Roos},
			\citenamefont {Benhelm}, \citenamefont {Chek-al kar}, \citenamefont
			{Chwalla}, \citenamefont {K{\"o}rber}, \citenamefont {Rapol}, \citenamefont
			{Riebe}, \citenamefont {Schmidt}, \citenamefont {Becher}, \citenamefont
			{G{\"u}hne}, \citenamefont {D{\"u}r},\ and\ \citenamefont
			{Blatt}}]{Haffner2005}%
		\BibitemOpen
		\bibfield  {author} {\bibinfo {author} {\bibfnamefont {H.}~\bibnamefont
				{H{\"a}ffner}}, \bibinfo {author} {\bibfnamefont {W.}~\bibnamefont
				{H{\"a}nsel}}, \bibinfo {author} {\bibfnamefont {C.~F.}\ \bibnamefont
				{Roos}}, \bibinfo {author} {\bibfnamefont {J.}~\bibnamefont {Benhelm}},
			\bibinfo {author} {\bibfnamefont {D.}~\bibnamefont {Chek-al kar}}, \bibinfo
			{author} {\bibfnamefont {M.}~\bibnamefont {Chwalla}}, \bibinfo {author}
			{\bibfnamefont {T.}~\bibnamefont {K{\"o}rber}}, \bibinfo {author}
			{\bibfnamefont {U.~D.}\ \bibnamefont {Rapol}}, \bibinfo {author}
			{\bibfnamefont {M.}~\bibnamefont {Riebe}}, \bibinfo {author} {\bibfnamefont
				{P.~O.}\ \bibnamefont {Schmidt}}, \bibinfo {author} {\bibfnamefont
				{C.}~\bibnamefont {Becher}}, \bibinfo {author} {\bibfnamefont
				{O.}~\bibnamefont {G{\"u}hne}}, \bibinfo {author} {\bibfnamefont
				{W.}~\bibnamefont {D{\"u}r}},\ and\ \bibinfo {author} {\bibfnamefont
				{R.}~\bibnamefont {Blatt}},\ }\bibfield  {title} {\bibinfo {title} {Scalable
				multiparticle entanglement of trapped ions},\ }\href
		{https://doi.org/10.1038/nature04279} {\bibfield  {journal} {\bibinfo
				{journal} {Nature}\ }\textbf {\bibinfo {volume} {438}},\ \bibinfo {pages}
			{643} (\bibinfo {year} {2005})}\BibitemShut {NoStop}%
		\bibitem [{\citenamefont {Lu}\ \emph {et~al.}(2007)\citenamefont {Lu},
			\citenamefont {Zhou}, \citenamefont {G{\"u}hne}, \citenamefont {Gao},
			\citenamefont {Zhang}, \citenamefont {Yuan}, \citenamefont {Goebel},
			\citenamefont {Yang},\ and\ \citenamefont {Pan}}]{Lu2007}%
		\BibitemOpen
		\bibfield  {author} {\bibinfo {author} {\bibfnamefont {C.-Y.}\ \bibnamefont
				{Lu}}, \bibinfo {author} {\bibfnamefont {X.-Q.}\ \bibnamefont {Zhou}},
			\bibinfo {author} {\bibfnamefont {O.}~\bibnamefont {G{\"u}hne}}, \bibinfo
			{author} {\bibfnamefont {W.-B.}\ \bibnamefont {Gao}}, \bibinfo {author}
			{\bibfnamefont {J.}~\bibnamefont {Zhang}}, \bibinfo {author} {\bibfnamefont
				{Z.-S.}\ \bibnamefont {Yuan}}, \bibinfo {author} {\bibfnamefont
				{A.}~\bibnamefont {Goebel}}, \bibinfo {author} {\bibfnamefont
				{T.}~\bibnamefont {Yang}},\ and\ \bibinfo {author} {\bibfnamefont {J.-W.}\
				\bibnamefont {Pan}},\ }\bibfield  {title} {\bibinfo {title} {Experimental
				entanglement of six photons in graph states},\ }\href
		{https://doi.org/10.1038/nphys507} {\bibfield  {journal} {\bibinfo  {journal}
				{Nature Physics}\ }\textbf {\bibinfo {volume} {3}},\ \bibinfo {pages} {91}
			(\bibinfo {year} {2007})}\BibitemShut {NoStop}%
		\bibitem [{\citenamefont {Gao}\ \emph {et~al.}(2010)\citenamefont {Gao},
			\citenamefont {Lu}, \citenamefont {Yao}, \citenamefont {Xu}, \citenamefont
			{G{\"u}hne}, \citenamefont {Goebel}, \citenamefont {Chen}, \citenamefont
			{Peng}, \citenamefont {Chen},\ and\ \citenamefont {Pan}}]{Gao2010}%
		\BibitemOpen
		\bibfield  {author} {\bibinfo {author} {\bibfnamefont {W.-B.}\ \bibnamefont
				{Gao}}, \bibinfo {author} {\bibfnamefont {C.-Y.}\ \bibnamefont {Lu}},
			\bibinfo {author} {\bibfnamefont {X.-C.}\ \bibnamefont {Yao}}, \bibinfo
			{author} {\bibfnamefont {P.}~\bibnamefont {Xu}}, \bibinfo {author}
			{\bibfnamefont {O.}~\bibnamefont {G{\"u}hne}}, \bibinfo {author}
			{\bibfnamefont {A.}~\bibnamefont {Goebel}}, \bibinfo {author} {\bibfnamefont
				{Y.-A.}\ \bibnamefont {Chen}}, \bibinfo {author} {\bibfnamefont {C.-Z.}\
				\bibnamefont {Peng}}, \bibinfo {author} {\bibfnamefont {Z.-B.}\ \bibnamefont
				{Chen}},\ and\ \bibinfo {author} {\bibfnamefont {J.-W.}\ \bibnamefont
				{Pan}},\ }\bibfield  {title} {\bibinfo {title} {Experimental demonstration of
				a hyper-entangled ten-qubit schr{\"o}dinger cat state},\ }\href
		{https://doi.org/10.1038/nphys1603} {\bibfield  {journal} {\bibinfo
				{journal} {Nature Physics}\ }\textbf {\bibinfo {volume} {6}},\ \bibinfo
			{pages} {331} (\bibinfo {year} {2010})}\BibitemShut {NoStop}%
		\bibitem [{\citenamefont {Yao}\ \emph {et~al.}(2012)\citenamefont {Yao},
			\citenamefont {Wang}, \citenamefont {Xu}, \citenamefont {Lu}, \citenamefont
			{Pan}, \citenamefont {Bao}, \citenamefont {Peng}, \citenamefont {Lu},
			\citenamefont {Chen},\ and\ \citenamefont {Pan}}]{Yao2012}%
		\BibitemOpen
		\bibfield  {author} {\bibinfo {author} {\bibfnamefont {X.-C.}\ \bibnamefont
				{Yao}}, \bibinfo {author} {\bibfnamefont {T.-X.}\ \bibnamefont {Wang}},
			\bibinfo {author} {\bibfnamefont {P.}~\bibnamefont {Xu}}, \bibinfo {author}
			{\bibfnamefont {H.}~\bibnamefont {Lu}}, \bibinfo {author} {\bibfnamefont
				{G.-S.}\ \bibnamefont {Pan}}, \bibinfo {author} {\bibfnamefont {X.-H.}\
				\bibnamefont {Bao}}, \bibinfo {author} {\bibfnamefont {C.-Z.}\ \bibnamefont
				{Peng}}, \bibinfo {author} {\bibfnamefont {C.-Y.}\ \bibnamefont {Lu}},
			\bibinfo {author} {\bibfnamefont {Y.-A.}\ \bibnamefont {Chen}},\ and\
			\bibinfo {author} {\bibfnamefont {J.-W.}\ \bibnamefont {Pan}},\ }\bibfield
		{title} {\bibinfo {title} {Observation of eight-photon entanglement},\ }\href
		{https://doi.org/10.1038/nphoton.2011.354} {\bibfield  {journal} {\bibinfo
				{journal} {Nature Photonics}\ }\textbf {\bibinfo {volume} {6}},\ \bibinfo
			{pages} {225} (\bibinfo {year} {2012})}\BibitemShut {NoStop}%
		\bibitem [{\citenamefont {Wang}\ \emph {et~al.}(2016)\citenamefont {Wang},
			\citenamefont {Chen}, \citenamefont {Li}, \citenamefont {Huang},
			\citenamefont {Liu}, \citenamefont {Chen}, \citenamefont {Luo}, \citenamefont
			{Su}, \citenamefont {Wu}, \citenamefont {Li}, \citenamefont {Lu},
			\citenamefont {Hu}, \citenamefont {Jiang}, \citenamefont {Peng},
			\citenamefont {Li}, \citenamefont {Liu}, \citenamefont {Chen}, \citenamefont
			{Lu},\ and\ \citenamefont {Pan}}]{Wang2016}%
		\BibitemOpen
		\bibfield  {author} {\bibinfo {author} {\bibfnamefont {X.-L.}\ \bibnamefont
				{Wang}}, \bibinfo {author} {\bibfnamefont {L.-K.}\ \bibnamefont {Chen}},
			\bibinfo {author} {\bibfnamefont {W.}~\bibnamefont {Li}}, \bibinfo {author}
			{\bibfnamefont {H.-L.}\ \bibnamefont {Huang}}, \bibinfo {author}
			{\bibfnamefont {C.}~\bibnamefont {Liu}}, \bibinfo {author} {\bibfnamefont
				{C.}~\bibnamefont {Chen}}, \bibinfo {author} {\bibfnamefont {Y.-H.}\
				\bibnamefont {Luo}}, \bibinfo {author} {\bibfnamefont {Z.-E.}\ \bibnamefont
				{Su}}, \bibinfo {author} {\bibfnamefont {D.}~\bibnamefont {Wu}}, \bibinfo
			{author} {\bibfnamefont {Z.-D.}\ \bibnamefont {Li}}, \bibinfo {author}
			{\bibfnamefont {H.}~\bibnamefont {Lu}}, \bibinfo {author} {\bibfnamefont
				{Y.}~\bibnamefont {Hu}}, \bibinfo {author} {\bibfnamefont {X.}~\bibnamefont
				{Jiang}}, \bibinfo {author} {\bibfnamefont {C.-Z.}\ \bibnamefont {Peng}},
			\bibinfo {author} {\bibfnamefont {L.}~\bibnamefont {Li}}, \bibinfo {author}
			{\bibfnamefont {N.-L.}\ \bibnamefont {Liu}}, \bibinfo {author} {\bibfnamefont
				{Y.-A.}\ \bibnamefont {Chen}}, \bibinfo {author} {\bibfnamefont {C.-Y.}\
				\bibnamefont {Lu}},\ and\ \bibinfo {author} {\bibfnamefont {J.-W.}\
				\bibnamefont {Pan}},\ }\bibfield  {title} {\bibinfo {title} {Experimental
				ten-photon entanglement},\ }\href
		{https://doi.org/10.1103/PhysRevLett.117.210502} {\bibfield  {journal}
			{\bibinfo  {journal} {Phys. Rev. Lett.}\ }\textbf {\bibinfo {volume} {117}},\
			\bibinfo {pages} {210502} (\bibinfo {year} {2016})}\BibitemShut {NoStop}%
		\bibitem [{\citenamefont {Zhong}\ \emph {et~al.}(2018)\citenamefont {Zhong},
			\citenamefont {Li}, \citenamefont {Li}, \citenamefont {Peng}, \citenamefont
			{Su}, \citenamefont {Hu}, \citenamefont {He}, \citenamefont {Ding},
			\citenamefont {Zhang}, \citenamefont {Li}, \citenamefont {Zhang},
			\citenamefont {Wang}, \citenamefont {You}, \citenamefont {Wang},
			\citenamefont {Jiang}, \citenamefont {Li}, \citenamefont {Chen},
			\citenamefont {Liu}, \citenamefont {Lu},\ and\ \citenamefont
			{Pan}}]{Zhong2018}%
		\BibitemOpen
		\bibfield  {author} {\bibinfo {author} {\bibfnamefont {H.-S.}\ \bibnamefont
				{Zhong}}, \bibinfo {author} {\bibfnamefont {Y.}~\bibnamefont {Li}}, \bibinfo
			{author} {\bibfnamefont {W.}~\bibnamefont {Li}}, \bibinfo {author}
			{\bibfnamefont {L.-C.}\ \bibnamefont {Peng}}, \bibinfo {author}
			{\bibfnamefont {Z.-E.}\ \bibnamefont {Su}}, \bibinfo {author} {\bibfnamefont
				{Y.}~\bibnamefont {Hu}}, \bibinfo {author} {\bibfnamefont {Y.-M.}\
				\bibnamefont {He}}, \bibinfo {author} {\bibfnamefont {X.}~\bibnamefont
				{Ding}}, \bibinfo {author} {\bibfnamefont {W.}~\bibnamefont {Zhang}},
			\bibinfo {author} {\bibfnamefont {H.}~\bibnamefont {Li}}, \bibinfo {author}
			{\bibfnamefont {L.}~\bibnamefont {Zhang}}, \bibinfo {author} {\bibfnamefont
				{Z.}~\bibnamefont {Wang}}, \bibinfo {author} {\bibfnamefont {L.}~\bibnamefont
				{You}}, \bibinfo {author} {\bibfnamefont {X.-L.}\ \bibnamefont {Wang}},
			\bibinfo {author} {\bibfnamefont {X.}~\bibnamefont {Jiang}}, \bibinfo
			{author} {\bibfnamefont {L.}~\bibnamefont {Li}}, \bibinfo {author}
			{\bibfnamefont {Y.-A.}\ \bibnamefont {Chen}}, \bibinfo {author}
			{\bibfnamefont {N.-L.}\ \bibnamefont {Liu}}, \bibinfo {author} {\bibfnamefont
				{C.-Y.}\ \bibnamefont {Lu}},\ and\ \bibinfo {author} {\bibfnamefont {J.-W.}\
				\bibnamefont {Pan}},\ }\bibfield  {title} {\bibinfo {title} {12-photon
				entanglement and scalable scattershot boson sampling with optimal
				entangled-photon pairs from parametric down-conversion},\ }\href
		{https://doi.org/10.1103/PhysRevLett.121.250505} {\bibfield  {journal}
			{\bibinfo  {journal} {Phys. Rev. Lett.}\ }\textbf {\bibinfo {volume} {121}},\
			\bibinfo {pages} {250505} (\bibinfo {year} {2018})}\BibitemShut {NoStop}%
		\bibitem [{\citenamefont {Saggio}\ \emph {et~al.}(2019)\citenamefont {Saggio},
			\citenamefont {Dimi{\'{c}}}, \citenamefont {Greganti}, \citenamefont
			{Rozema}, \citenamefont {Walther},\ and\ \citenamefont
			{Daki{\'{c}}}}]{Saggio2019}%
		\BibitemOpen
		\bibfield  {author} {\bibinfo {author} {\bibfnamefont {V.}~\bibnamefont
				{Saggio}}, \bibinfo {author} {\bibfnamefont {A.}~\bibnamefont {Dimi{\'{c}}}},
			\bibinfo {author} {\bibfnamefont {C.}~\bibnamefont {Greganti}}, \bibinfo
			{author} {\bibfnamefont {L.~A.}\ \bibnamefont {Rozema}}, \bibinfo {author}
			{\bibfnamefont {P.}~\bibnamefont {Walther}},\ and\ \bibinfo {author}
			{\bibfnamefont {B.}~\bibnamefont {Daki{\'{c}}}},\ }\bibfield  {title}
		{\bibinfo {title} {Experimental few-copy multipartite entanglement
				detection},\ }\href {https://doi.org/10.1038/s41567-019-0550-4} {\bibfield
			{journal} {\bibinfo  {journal} {Nature Physics}\ }\textbf {\bibinfo {volume}
				{15}},\ \bibinfo {pages} {935} (\bibinfo {year} {2019})}\BibitemShut
		{NoStop}%
		\bibitem [{\citenamefont {Friis}\ \emph {et~al.}(2018)\citenamefont {Friis},
			\citenamefont {Marty}, \citenamefont {Maier}, \citenamefont {Hempel},
			\citenamefont {Holz\"apfel}, \citenamefont {Jurcevic}, \citenamefont
			{Plenio}, \citenamefont {Huber}, \citenamefont {Roos}, \citenamefont
			{Blatt},\ and\ \citenamefont {Lanyon}}]{Friis2018}%
		\BibitemOpen
		\bibfield  {author} {\bibinfo {author} {\bibfnamefont {N.}~\bibnamefont
				{Friis}}, \bibinfo {author} {\bibfnamefont {O.}~\bibnamefont {Marty}},
			\bibinfo {author} {\bibfnamefont {C.}~\bibnamefont {Maier}}, \bibinfo
			{author} {\bibfnamefont {C.}~\bibnamefont {Hempel}}, \bibinfo {author}
			{\bibfnamefont {M.}~\bibnamefont {Holz\"apfel}}, \bibinfo {author}
			{\bibfnamefont {P.}~\bibnamefont {Jurcevic}}, \bibinfo {author}
			{\bibfnamefont {M.~B.}\ \bibnamefont {Plenio}}, \bibinfo {author}
			{\bibfnamefont {M.}~\bibnamefont {Huber}}, \bibinfo {author} {\bibfnamefont
				{C.}~\bibnamefont {Roos}}, \bibinfo {author} {\bibfnamefont {R.}~\bibnamefont
				{Blatt}},\ and\ \bibinfo {author} {\bibfnamefont {B.}~\bibnamefont
				{Lanyon}},\ }\bibfield  {title} {\bibinfo {title} {Observation of entangled
				states of a fully controlled 20-qubit system},\ }\href
		{https://doi.org/10.1103/PhysRevX.8.021012} {\bibfield  {journal} {\bibinfo
				{journal} {Phys. Rev. X}\ }\textbf {\bibinfo {volume} {8}},\ \bibinfo {pages}
			{021012} (\bibinfo {year} {2018})}\BibitemShut {NoStop}%
		\bibitem [{\citenamefont {Thomas}\ \emph {et~al.}(2022)\citenamefont {Thomas},
			\citenamefont {Ruscio}, \citenamefont {Morin},\ and\ \citenamefont
			{Rempe}}]{Thomas2022}%
		\BibitemOpen
		\bibfield  {author} {\bibinfo {author} {\bibfnamefont {P.}~\bibnamefont
				{Thomas}}, \bibinfo {author} {\bibfnamefont {L.}~\bibnamefont {Ruscio}},
			\bibinfo {author} {\bibfnamefont {O.}~\bibnamefont {Morin}},\ and\ \bibinfo
			{author} {\bibfnamefont {G.}~\bibnamefont {Rempe}},\ }\bibfield  {title}
		{\bibinfo {title} {Efficient generation of entangled multiphoton graph states
				from a single atom},\ }\href {https://doi.org/10.1038/s41586-022-04987-5}
		{\bibfield  {journal} {\bibinfo  {journal} {Nature}\ }\textbf {\bibinfo
				{volume} {608}},\ \bibinfo {pages} {677} (\bibinfo {year}
			{2022})}\BibitemShut {NoStop}%
		\bibitem [{\citenamefont {Moses}\ \emph {et~al.}(2023)\citenamefont {Moses},
			\citenamefont {Baldwin}, \citenamefont {Allman}, \citenamefont {Ancona},
			\citenamefont {Ascarrunz}, \citenamefont {Barnes}, \citenamefont
			{Bartolotta}, \citenamefont {Bjork}, \citenamefont {Blanchard}, \citenamefont
			{Bohn}, \citenamefont {Bohnet}, \citenamefont {Brown}, \citenamefont
			{Burdick}, \citenamefont {Burton}, \citenamefont {Campbell}, \citenamefont
			{Campora}, \citenamefont {Carron}, \citenamefont {Chambers}, \citenamefont
			{Chan}, \citenamefont {Chen}, \citenamefont {Chernoguzov}, \citenamefont
			{Chertkov}, \citenamefont {Colina}, \citenamefont {Curtis}, \citenamefont
			{Daniel}, \citenamefont {DeCross}, \citenamefont {Deen}, \citenamefont
			{Delaney}, \citenamefont {Dreiling}, \citenamefont {Ertsgaard}, \citenamefont
			{Esposito}, \citenamefont {Estey}, \citenamefont {Fabrikant}, \citenamefont
			{Figgatt}, \citenamefont {Foltz}, \citenamefont {Foss-Feig}, \citenamefont
			{Francois}, \citenamefont {Gaebler}, \citenamefont {Gatterman}, \citenamefont
			{Gilbreth}, \citenamefont {Giles}, \citenamefont {Glynn}, \citenamefont
			{Hall}, \citenamefont {Hankin}, \citenamefont {Hansen}, \citenamefont
			{Hayes}, \citenamefont {Higashi}, \citenamefont {Hoffman}, \citenamefont
			{Horning}, \citenamefont {Hout}, \citenamefont {Jacobs}, \citenamefont
			{Johansen}, \citenamefont {Jones}, \citenamefont {Karcz}, \citenamefont
			{Klein}, \citenamefont {Lauria}, \citenamefont {Lee}, \citenamefont {Liefer},
			\citenamefont {Lu}, \citenamefont {Lucchetti}, \citenamefont {Lytle},
			\citenamefont {Malm}, \citenamefont {Matheny}, \citenamefont {Mathewson},
			\citenamefont {Mayer}, \citenamefont {Miller}, \citenamefont {Mills},
			\citenamefont {Neyenhuis}, \citenamefont {Nugent}, \citenamefont {Olson},
			\citenamefont {Parks}, \citenamefont {Price}, \citenamefont {Price},
			\citenamefont {Pugh}, \citenamefont {Ransford}, \citenamefont {Reed},
			\citenamefont {Roman}, \citenamefont {Rowe}, \citenamefont {Ryan-Anderson},
			\citenamefont {Sanders}, \citenamefont {Sedlacek}, \citenamefont {Shevchuk},
			\citenamefont {Siegfried}, \citenamefont {Skripka}, \citenamefont {Spaun},
			\citenamefont {Sprenkle}, \citenamefont {Stutz}, \citenamefont {Swallows},
			\citenamefont {Tobey}, \citenamefont {Tran}, \citenamefont {Tran},
			\citenamefont {Vogt}, \citenamefont {Volin}, \citenamefont {Walker},
			\citenamefont {Zolot},\ and\ \citenamefont {Pino}}]{Moses2023}%
		\BibitemOpen
		\bibfield  {author} {\bibinfo {author} {\bibfnamefont {S.~A.}\ \bibnamefont
				{Moses}}, \bibinfo {author} {\bibfnamefont {C.~H.}\ \bibnamefont {Baldwin}},
			\bibinfo {author} {\bibfnamefont {M.~S.}\ \bibnamefont {Allman}}, \bibinfo
			{author} {\bibfnamefont {R.}~\bibnamefont {Ancona}}, \bibinfo {author}
			{\bibfnamefont {L.}~\bibnamefont {Ascarrunz}}, \bibinfo {author}
			{\bibfnamefont {C.}~\bibnamefont {Barnes}}, \bibinfo {author} {\bibfnamefont
				{J.}~\bibnamefont {Bartolotta}}, \bibinfo {author} {\bibfnamefont
				{B.}~\bibnamefont {Bjork}}, \bibinfo {author} {\bibfnamefont
				{P.}~\bibnamefont {Blanchard}}, \bibinfo {author} {\bibfnamefont
				{M.}~\bibnamefont {Bohn}}, \bibinfo {author} {\bibfnamefont {J.~G.}\
				\bibnamefont {Bohnet}}, \bibinfo {author} {\bibfnamefont {N.~C.}\
				\bibnamefont {Brown}}, \bibinfo {author} {\bibfnamefont {N.~Q.}\ \bibnamefont
				{Burdick}}, \bibinfo {author} {\bibfnamefont {W.~C.}\ \bibnamefont {Burton}},
			\bibinfo {author} {\bibfnamefont {S.~L.}\ \bibnamefont {Campbell}}, \bibinfo
			{author} {\bibfnamefont {J.~P.}\ \bibnamefont {Campora}}, \bibinfo {author}
			{\bibfnamefont {C.}~\bibnamefont {Carron}}, \bibinfo {author} {\bibfnamefont
				{J.}~\bibnamefont {Chambers}}, \bibinfo {author} {\bibfnamefont {J.~W.}\
				\bibnamefont {Chan}}, \bibinfo {author} {\bibfnamefont {Y.~H.}\ \bibnamefont
				{Chen}}, \bibinfo {author} {\bibfnamefont {A.}~\bibnamefont {Chernoguzov}},
			\bibinfo {author} {\bibfnamefont {E.}~\bibnamefont {Chertkov}}, \bibinfo
			{author} {\bibfnamefont {J.}~\bibnamefont {Colina}}, \bibinfo {author}
			{\bibfnamefont {J.~P.}\ \bibnamefont {Curtis}}, \bibinfo {author}
			{\bibfnamefont {R.}~\bibnamefont {Daniel}}, \bibinfo {author} {\bibfnamefont
				{M.}~\bibnamefont {DeCross}}, \bibinfo {author} {\bibfnamefont
				{D.}~\bibnamefont {Deen}}, \bibinfo {author} {\bibfnamefont {C.}~\bibnamefont
				{Delaney}}, \bibinfo {author} {\bibfnamefont {J.~M.}\ \bibnamefont
				{Dreiling}}, \bibinfo {author} {\bibfnamefont {C.~T.}\ \bibnamefont
				{Ertsgaard}}, \bibinfo {author} {\bibfnamefont {J.}~\bibnamefont {Esposito}},
			\bibinfo {author} {\bibfnamefont {B.}~\bibnamefont {Estey}}, \bibinfo
			{author} {\bibfnamefont {M.}~\bibnamefont {Fabrikant}}, \bibinfo {author}
			{\bibfnamefont {C.}~\bibnamefont {Figgatt}}, \bibinfo {author} {\bibfnamefont
				{C.}~\bibnamefont {Foltz}}, \bibinfo {author} {\bibfnamefont
				{M.}~\bibnamefont {Foss-Feig}}, \bibinfo {author} {\bibfnamefont
				{D.}~\bibnamefont {Francois}}, \bibinfo {author} {\bibfnamefont {J.~P.}\
				\bibnamefont {Gaebler}}, \bibinfo {author} {\bibfnamefont {T.~M.}\
				\bibnamefont {Gatterman}}, \bibinfo {author} {\bibfnamefont {C.~N.}\
				\bibnamefont {Gilbreth}}, \bibinfo {author} {\bibfnamefont {J.}~\bibnamefont
				{Giles}}, \bibinfo {author} {\bibfnamefont {E.}~\bibnamefont {Glynn}},
			\bibinfo {author} {\bibfnamefont {A.}~\bibnamefont {Hall}}, \bibinfo {author}
			{\bibfnamefont {A.~M.}\ \bibnamefont {Hankin}}, \bibinfo {author}
			{\bibfnamefont {A.}~\bibnamefont {Hansen}}, \bibinfo {author} {\bibfnamefont
				{D.}~\bibnamefont {Hayes}}, \bibinfo {author} {\bibfnamefont
				{B.}~\bibnamefont {Higashi}}, \bibinfo {author} {\bibfnamefont {I.~M.}\
				\bibnamefont {Hoffman}}, \bibinfo {author} {\bibfnamefont {B.}~\bibnamefont
				{Horning}}, \bibinfo {author} {\bibfnamefont {J.~J.}\ \bibnamefont {Hout}},
			\bibinfo {author} {\bibfnamefont {R.}~\bibnamefont {Jacobs}}, \bibinfo
			{author} {\bibfnamefont {J.}~\bibnamefont {Johansen}}, \bibinfo {author}
			{\bibfnamefont {L.}~\bibnamefont {Jones}}, \bibinfo {author} {\bibfnamefont
				{J.}~\bibnamefont {Karcz}}, \bibinfo {author} {\bibfnamefont
				{T.}~\bibnamefont {Klein}}, \bibinfo {author} {\bibfnamefont
				{P.}~\bibnamefont {Lauria}}, \bibinfo {author} {\bibfnamefont
				{P.}~\bibnamefont {Lee}}, \bibinfo {author} {\bibfnamefont {D.}~\bibnamefont
				{Liefer}}, \bibinfo {author} {\bibfnamefont {S.~T.}\ \bibnamefont {Lu}},
			\bibinfo {author} {\bibfnamefont {D.}~\bibnamefont {Lucchetti}}, \bibinfo
			{author} {\bibfnamefont {C.}~\bibnamefont {Lytle}}, \bibinfo {author}
			{\bibfnamefont {A.}~\bibnamefont {Malm}}, \bibinfo {author} {\bibfnamefont
				{M.}~\bibnamefont {Matheny}}, \bibinfo {author} {\bibfnamefont
				{B.}~\bibnamefont {Mathewson}}, \bibinfo {author} {\bibfnamefont
				{K.}~\bibnamefont {Mayer}}, \bibinfo {author} {\bibfnamefont {D.~B.}\
				\bibnamefont {Miller}}, \bibinfo {author} {\bibfnamefont {M.}~\bibnamefont
				{Mills}}, \bibinfo {author} {\bibfnamefont {B.}~\bibnamefont {Neyenhuis}},
			\bibinfo {author} {\bibfnamefont {L.}~\bibnamefont {Nugent}}, \bibinfo
			{author} {\bibfnamefont {S.}~\bibnamefont {Olson}}, \bibinfo {author}
			{\bibfnamefont {J.}~\bibnamefont {Parks}}, \bibinfo {author} {\bibfnamefont
				{G.~N.}\ \bibnamefont {Price}}, \bibinfo {author} {\bibfnamefont
				{Z.}~\bibnamefont {Price}}, \bibinfo {author} {\bibfnamefont
				{M.}~\bibnamefont {Pugh}}, \bibinfo {author} {\bibfnamefont {A.}~\bibnamefont
				{Ransford}}, \bibinfo {author} {\bibfnamefont {A.~P.}\ \bibnamefont {Reed}},
			\bibinfo {author} {\bibfnamefont {C.}~\bibnamefont {Roman}}, \bibinfo
			{author} {\bibfnamefont {M.}~\bibnamefont {Rowe}}, \bibinfo {author}
			{\bibfnamefont {C.}~\bibnamefont {Ryan-Anderson}}, \bibinfo {author}
			{\bibfnamefont {S.}~\bibnamefont {Sanders}}, \bibinfo {author} {\bibfnamefont
				{J.}~\bibnamefont {Sedlacek}}, \bibinfo {author} {\bibfnamefont
				{P.}~\bibnamefont {Shevchuk}}, \bibinfo {author} {\bibfnamefont
				{P.}~\bibnamefont {Siegfried}}, \bibinfo {author} {\bibfnamefont
				{T.}~\bibnamefont {Skripka}}, \bibinfo {author} {\bibfnamefont
				{B.}~\bibnamefont {Spaun}}, \bibinfo {author} {\bibfnamefont {R.~T.}\
				\bibnamefont {Sprenkle}}, \bibinfo {author} {\bibfnamefont {R.~P.}\
				\bibnamefont {Stutz}}, \bibinfo {author} {\bibfnamefont {M.}~\bibnamefont
				{Swallows}}, \bibinfo {author} {\bibfnamefont {R.~I.}\ \bibnamefont {Tobey}},
			\bibinfo {author} {\bibfnamefont {A.}~\bibnamefont {Tran}}, \bibinfo {author}
			{\bibfnamefont {T.}~\bibnamefont {Tran}}, \bibinfo {author} {\bibfnamefont
				{E.}~\bibnamefont {Vogt}}, \bibinfo {author} {\bibfnamefont {C.}~\bibnamefont
				{Volin}}, \bibinfo {author} {\bibfnamefont {J.}~\bibnamefont {Walker}},
			\bibinfo {author} {\bibfnamefont {A.~M.}\ \bibnamefont {Zolot}},\ and\
			\bibinfo {author} {\bibfnamefont {J.~M.}\ \bibnamefont {Pino}},\ }\bibfield
		{title} {\bibinfo {title} {A race-track trapped-ion quantum processor},\
		}\href {https://doi.org/10.1103/PhysRevX.13.041052} {\bibfield  {journal}
			{\bibinfo  {journal} {Phys. Rev. X}\ }\textbf {\bibinfo {volume} {13}},\
			\bibinfo {pages} {041052} (\bibinfo {year} {2023})}\BibitemShut {NoStop}%
		\bibitem [{\citenamefont {Cao}\ \emph {et~al.}(2023)\citenamefont {Cao},
			\citenamefont {Wu}, \citenamefont {Chen}, \citenamefont {Gong}, \citenamefont
			{Wu}, \citenamefont {Ye}, \citenamefont {Zha}, \citenamefont {Qian},
			\citenamefont {Ying}, \citenamefont {Guo}, \citenamefont {Zhu}, \citenamefont
			{Huang}, \citenamefont {Zhao}, \citenamefont {Li}, \citenamefont {Wang},
			\citenamefont {Yu}, \citenamefont {Fan}, \citenamefont {Wu}, \citenamefont
			{Su}, \citenamefont {Deng}, \citenamefont {Rong}, \citenamefont {Li},
			\citenamefont {Zhang}, \citenamefont {Chung}, \citenamefont {Liang},
			\citenamefont {Lin}, \citenamefont {Xu}, \citenamefont {Sun}, \citenamefont
			{Guo}, \citenamefont {Li}, \citenamefont {Huo}, \citenamefont {Peng},
			\citenamefont {Lu}, \citenamefont {Yuan}, \citenamefont {Zhu},\ and\
			\citenamefont {Pan}}]{Cao2023}%
		\BibitemOpen
		\bibfield  {author} {\bibinfo {author} {\bibfnamefont {S.}~\bibnamefont
				{Cao}}, \bibinfo {author} {\bibfnamefont {B.}~\bibnamefont {Wu}}, \bibinfo
			{author} {\bibfnamefont {F.}~\bibnamefont {Chen}}, \bibinfo {author}
			{\bibfnamefont {M.}~\bibnamefont {Gong}}, \bibinfo {author} {\bibfnamefont
				{Y.}~\bibnamefont {Wu}}, \bibinfo {author} {\bibfnamefont {Y.}~\bibnamefont
				{Ye}}, \bibinfo {author} {\bibfnamefont {C.}~\bibnamefont {Zha}}, \bibinfo
			{author} {\bibfnamefont {H.}~\bibnamefont {Qian}}, \bibinfo {author}
			{\bibfnamefont {C.}~\bibnamefont {Ying}}, \bibinfo {author} {\bibfnamefont
				{S.}~\bibnamefont {Guo}}, \bibinfo {author} {\bibfnamefont {Q.}~\bibnamefont
				{Zhu}}, \bibinfo {author} {\bibfnamefont {H.-L.}\ \bibnamefont {Huang}},
			\bibinfo {author} {\bibfnamefont {Y.}~\bibnamefont {Zhao}}, \bibinfo {author}
			{\bibfnamefont {S.}~\bibnamefont {Li}}, \bibinfo {author} {\bibfnamefont
				{S.}~\bibnamefont {Wang}}, \bibinfo {author} {\bibfnamefont {J.}~\bibnamefont
				{Yu}}, \bibinfo {author} {\bibfnamefont {D.}~\bibnamefont {Fan}}, \bibinfo
			{author} {\bibfnamefont {D.}~\bibnamefont {Wu}}, \bibinfo {author}
			{\bibfnamefont {H.}~\bibnamefont {Su}}, \bibinfo {author} {\bibfnamefont
				{H.}~\bibnamefont {Deng}}, \bibinfo {author} {\bibfnamefont {H.}~\bibnamefont
				{Rong}}, \bibinfo {author} {\bibfnamefont {Y.}~\bibnamefont {Li}}, \bibinfo
			{author} {\bibfnamefont {K.}~\bibnamefont {Zhang}}, \bibinfo {author}
			{\bibfnamefont {T.-H.}\ \bibnamefont {Chung}}, \bibinfo {author}
			{\bibfnamefont {F.}~\bibnamefont {Liang}}, \bibinfo {author} {\bibfnamefont
				{J.}~\bibnamefont {Lin}}, \bibinfo {author} {\bibfnamefont {Y.}~\bibnamefont
				{Xu}}, \bibinfo {author} {\bibfnamefont {L.}~\bibnamefont {Sun}}, \bibinfo
			{author} {\bibfnamefont {C.}~\bibnamefont {Guo}}, \bibinfo {author}
			{\bibfnamefont {N.}~\bibnamefont {Li}}, \bibinfo {author} {\bibfnamefont
				{Y.-H.}\ \bibnamefont {Huo}}, \bibinfo {author} {\bibfnamefont {C.-Z.}\
				\bibnamefont {Peng}}, \bibinfo {author} {\bibfnamefont {C.-Y.}\ \bibnamefont
				{Lu}}, \bibinfo {author} {\bibfnamefont {X.}~\bibnamefont {Yuan}}, \bibinfo
			{author} {\bibfnamefont {X.}~\bibnamefont {Zhu}},\ and\ \bibinfo {author}
			{\bibfnamefont {J.-W.}\ \bibnamefont {Pan}},\ }\bibfield  {title} {\bibinfo
			{title} {Generation of genuine entanglement up to 51 superconducting
				qubits},\ }\href {https://doi.org/10.1038/s41586-023-06195-1} {\bibfield
			{journal} {\bibinfo  {journal} {Nature}\ }\textbf {\bibinfo {volume} {619}},\
			\bibinfo {pages} {738} (\bibinfo {year} {2023})}\BibitemShut {NoStop}%
		\bibitem [{\citenamefont {Cerf}\ \emph {et~al.}(2002)\citenamefont {Cerf},
			\citenamefont {Bourennane}, \citenamefont {Karlsson},\ and\ \citenamefont
			{Gisin}}]{Cerf2002}%
		\BibitemOpen
		\bibfield  {author} {\bibinfo {author} {\bibfnamefont {N.~J.}\ \bibnamefont
				{Cerf}}, \bibinfo {author} {\bibfnamefont {M.}~\bibnamefont {Bourennane}},
			\bibinfo {author} {\bibfnamefont {A.}~\bibnamefont {Karlsson}},\ and\
			\bibinfo {author} {\bibfnamefont {N.}~\bibnamefont {Gisin}},\ }\bibfield
		{title} {\bibinfo {title} {Security of quantum key distribution using
				$\mathit{d}$-level systems},\ }\href
		{https://doi.org/10.1103/PhysRevLett.88.127902} {\bibfield  {journal}
			{\bibinfo  {journal} {Phys. Rev. Lett.}\ }\textbf {\bibinfo {volume} {88}},\
			\bibinfo {pages} {127902} (\bibinfo {year} {2002})}\BibitemShut {NoStop}%
		\bibitem [{\citenamefont {Sheridan}\ and\ \citenamefont
			{Scarani}(2010)}]{Sheridan2010}%
		\BibitemOpen
		\bibfield  {author} {\bibinfo {author} {\bibfnamefont {L.}~\bibnamefont
				{Sheridan}}\ and\ \bibinfo {author} {\bibfnamefont {V.}~\bibnamefont
				{Scarani}},\ }\bibfield  {title} {\bibinfo {title} {Security proof for
				quantum key distribution using qudit systems},\ }\href
		{https://doi.org/10.1103/PhysRevA.82.030301} {\bibfield  {journal} {\bibinfo
				{journal} {Phys. Rev. A}\ }\textbf {\bibinfo {volume} {82}},\ \bibinfo
			{pages} {030301} (\bibinfo {year} {2010})}\BibitemShut {NoStop}%
		\bibitem [{\citenamefont {Ecker}\ \emph {et~al.}(2019)\citenamefont {Ecker},
			\citenamefont {Bouchard}, \citenamefont {Bulla}, \citenamefont {Brandt},
			\citenamefont {Kohout}, \citenamefont {Steinlechner}, \citenamefont
			{Fickler}, \citenamefont {Malik}, \citenamefont {Guryanova}, \citenamefont
			{Ursin},\ and\ \citenamefont {Huber}}]{Ecker2019}%
		\BibitemOpen
		\bibfield  {author} {\bibinfo {author} {\bibfnamefont {S.}~\bibnamefont
				{Ecker}}, \bibinfo {author} {\bibfnamefont {F.}~\bibnamefont {Bouchard}},
			\bibinfo {author} {\bibfnamefont {L.}~\bibnamefont {Bulla}}, \bibinfo
			{author} {\bibfnamefont {F.}~\bibnamefont {Brandt}}, \bibinfo {author}
			{\bibfnamefont {O.}~\bibnamefont {Kohout}}, \bibinfo {author} {\bibfnamefont
				{F.}~\bibnamefont {Steinlechner}}, \bibinfo {author} {\bibfnamefont
				{R.}~\bibnamefont {Fickler}}, \bibinfo {author} {\bibfnamefont
				{M.}~\bibnamefont {Malik}}, \bibinfo {author} {\bibfnamefont
				{Y.}~\bibnamefont {Guryanova}}, \bibinfo {author} {\bibfnamefont
				{R.}~\bibnamefont {Ursin}},\ and\ \bibinfo {author} {\bibfnamefont
				{M.}~\bibnamefont {Huber}},\ }\bibfield  {title} {\bibinfo {title}
			{Overcoming noise in entanglement distribution},\ }\href
		{https://doi.org/10.1103/PhysRevX.9.041042} {\bibfield  {journal} {\bibinfo
				{journal} {Phys. Rev. X}\ }\textbf {\bibinfo {volume} {9}},\ \bibinfo {pages}
			{041042} (\bibinfo {year} {2019})}\BibitemShut {NoStop}%
		\bibitem [{\citenamefont {Wang}\ \emph {et~al.}(2015)\citenamefont {Wang},
			\citenamefont {Cai}, \citenamefont {Su}, \citenamefont {Chen}, \citenamefont
			{Wu}, \citenamefont {Li}, \citenamefont {Liu}, \citenamefont {Lu},\ and\
			\citenamefont {Pan}}]{Wang2015}%
		\BibitemOpen
		\bibfield  {author} {\bibinfo {author} {\bibfnamefont {X.-L.}\ \bibnamefont
				{Wang}}, \bibinfo {author} {\bibfnamefont {X.-D.}\ \bibnamefont {Cai}},
			\bibinfo {author} {\bibfnamefont {Z.-E.}\ \bibnamefont {Su}}, \bibinfo
			{author} {\bibfnamefont {M.-C.}\ \bibnamefont {Chen}}, \bibinfo {author}
			{\bibfnamefont {D.}~\bibnamefont {Wu}}, \bibinfo {author} {\bibfnamefont
				{L.}~\bibnamefont {Li}}, \bibinfo {author} {\bibfnamefont {N.-L.}\
				\bibnamefont {Liu}}, \bibinfo {author} {\bibfnamefont {C.-Y.}\ \bibnamefont
				{Lu}},\ and\ \bibinfo {author} {\bibfnamefont {J.-W.}\ \bibnamefont {Pan}},\
		}\bibfield  {title} {\bibinfo {title} {Quantum teleportation of multiple
				degrees of freedom of a single photon},\ }\href
		{https://doi.org/10.1038/nature14246} {\bibfield  {journal} {\bibinfo
				{journal} {Nature}\ }\textbf {\bibinfo {volume} {518}},\ \bibinfo {pages}
			{516} (\bibinfo {year} {2015})}\BibitemShut {NoStop}%
		\bibitem [{\citenamefont {Collins}\ \emph {et~al.}(2002)\citenamefont
			{Collins}, \citenamefont {Gisin}, \citenamefont {Linden}, \citenamefont
			{Massar},\ and\ \citenamefont {Popescu}}]{Collins2002}%
		\BibitemOpen
		\bibfield  {author} {\bibinfo {author} {\bibfnamefont {D.}~\bibnamefont
				{Collins}}, \bibinfo {author} {\bibfnamefont {N.}~\bibnamefont {Gisin}},
			\bibinfo {author} {\bibfnamefont {N.}~\bibnamefont {Linden}}, \bibinfo
			{author} {\bibfnamefont {S.}~\bibnamefont {Massar}},\ and\ \bibinfo {author}
			{\bibfnamefont {S.}~\bibnamefont {Popescu}},\ }\bibfield  {title} {\bibinfo
			{title} {Bell inequalities for arbitrarily high-dimensional systems},\ }\href
		{https://doi.org/10.1103/PhysRevLett.88.040404} {\bibfield  {journal}
			{\bibinfo  {journal} {Phys. Rev. Lett.}\ }\textbf {\bibinfo {volume} {88}},\
			\bibinfo {pages} {040404} (\bibinfo {year} {2002})}\BibitemShut {NoStop}%
		\bibitem [{\citenamefont {Skrzypczyk}\ and\ \citenamefont
			{Cavalcanti}(2018)}]{Skrzypczyk2018}%
		\BibitemOpen
		\bibfield  {author} {\bibinfo {author} {\bibfnamefont {P.}~\bibnamefont
				{Skrzypczyk}}\ and\ \bibinfo {author} {\bibfnamefont {D.}~\bibnamefont
				{Cavalcanti}},\ }\bibfield  {title} {\bibinfo {title} {Maximal randomness
				generation from steering inequality violations using qudits},\ }\href
		{https://doi.org/10.1103/PhysRevLett.120.260401} {\bibfield  {journal}
			{\bibinfo  {journal} {Phys. Rev. Lett.}\ }\textbf {\bibinfo {volume} {120}},\
			\bibinfo {pages} {260401} (\bibinfo {year} {2018})}\BibitemShut {NoStop}%
		\bibitem [{\citenamefont {Tavakoli}\ \emph {et~al.}(2021)\citenamefont
			{Tavakoli}, \citenamefont {Farkas}, \citenamefont {Rosset}, \citenamefont
			{Bancal},\ and\ \citenamefont {Kaniewski}}]{Tavakoli2021}%
		\BibitemOpen
		\bibfield  {author} {\bibinfo {author} {\bibfnamefont {A.}~\bibnamefont
				{Tavakoli}}, \bibinfo {author} {\bibfnamefont {M.}~\bibnamefont {Farkas}},
			\bibinfo {author} {\bibfnamefont {D.}~\bibnamefont {Rosset}}, \bibinfo
			{author} {\bibfnamefont {J.-D.}\ \bibnamefont {Bancal}},\ and\ \bibinfo
			{author} {\bibfnamefont {J.}~\bibnamefont {Kaniewski}},\ }\bibfield  {title}
		{\bibinfo {title} {Mutually unbiased bases and symmetric informationally
				complete measurements in bell experiments},\ }\href
		{https://doi.org/10.1126/sciadv.abc3847} {\bibfield  {journal} {\bibinfo
				{journal} {Science Advances}\ }\textbf {\bibinfo {volume} {7}},\ \bibinfo
			{pages} {eabc3847} (\bibinfo {year} {2021})},\ \Eprint
		{https://arxiv.org/abs/https://www.science.org/doi/pdf/10.1126/sciadv.abc3847}
		{https://www.science.org/doi/pdf/10.1126/sciadv.abc3847} \BibitemShut
		{NoStop}%
		\bibitem [{\citenamefont {Terhal}\ and\ \citenamefont
			{Horodecki}(2000)}]{Terhal_2000}%
		\BibitemOpen
		\bibfield  {author} {\bibinfo {author} {\bibfnamefont {B.~M.}\ \bibnamefont
				{Terhal}}\ and\ \bibinfo {author} {\bibfnamefont {P.}~\bibnamefont
				{Horodecki}},\ }\bibfield  {title} {\bibinfo {title} {Schmidt number for
				density matrices},\ }\bibfield  {journal} {\bibinfo  {journal} {Physical
				Review A}\ }\textbf {\bibinfo {volume} {61}},\ \href
		{https://doi.org/10.1103/physreva.61.040301} {10.1103/physreva.61.040301}
		(\bibinfo {year} {2000})\BibitemShut {NoStop}%
		\bibitem [{\citenamefont {Friis}\ \emph {et~al.}(2019)\citenamefont {Friis},
			\citenamefont {Vitagliano}, \citenamefont {Malik},\ and\ \citenamefont
			{Huber}}]{Friis2019}%
		\BibitemOpen
		\bibfield  {author} {\bibinfo {author} {\bibfnamefont {N.}~\bibnamefont
				{Friis}}, \bibinfo {author} {\bibfnamefont {G.}~\bibnamefont {Vitagliano}},
			\bibinfo {author} {\bibfnamefont {M.}~\bibnamefont {Malik}},\ and\ \bibinfo
			{author} {\bibfnamefont {M.}~\bibnamefont {Huber}},\ }\bibfield  {title}
		{\bibinfo {title} {Entanglement certification from theory to experiment},\
		}\href {https://doi.org/10.1038/s42254-018-0003-5} {\bibfield  {journal}
			{\bibinfo  {journal} {Nature Reviews Physics}\ }\textbf {\bibinfo {volume}
				{1}},\ \bibinfo {pages} {72} (\bibinfo {year} {2019})}\BibitemShut {NoStop}%
		\bibitem [{\citenamefont {Erhard}\ \emph {et~al.}(2020)\citenamefont {Erhard},
			\citenamefont {Krenn},\ and\ \citenamefont {Zeilinger}}]{Erhard2020}%
		\BibitemOpen
		\bibfield  {author} {\bibinfo {author} {\bibfnamefont {M.}~\bibnamefont
				{Erhard}}, \bibinfo {author} {\bibfnamefont {M.}~\bibnamefont {Krenn}},\ and\
			\bibinfo {author} {\bibfnamefont {A.}~\bibnamefont {Zeilinger}},\ }\bibfield
		{title} {\bibinfo {title} {Advances in high-dimensional quantum
				entanglement},\ }\href {https://doi.org/10.1038/s42254-020-0193-5} {\bibfield
			{journal} {\bibinfo  {journal} {Nature Reviews Physics}\ }\textbf {\bibinfo
				{volume} {2}},\ \bibinfo {pages} {365} (\bibinfo {year} {2020})}\BibitemShut
		{NoStop}%
		\bibitem [{\citenamefont {Sanpera}\ \emph {et~al.}(2001)\citenamefont
			{Sanpera}, \citenamefont {Bru\ss{}},\ and\ \citenamefont
			{Lewenstein}}]{Sanpera2001}%
		\BibitemOpen
		\bibfield  {author} {\bibinfo {author} {\bibfnamefont {A.}~\bibnamefont
				{Sanpera}}, \bibinfo {author} {\bibfnamefont {D.}~\bibnamefont {Bru\ss{}}},\
			and\ \bibinfo {author} {\bibfnamefont {M.}~\bibnamefont {Lewenstein}},\
		}\bibfield  {title} {\bibinfo {title} {Schmidt-number witnesses and bound
				entanglement},\ }\href {https://doi.org/10.1103/PhysRevA.63.050301}
		{\bibfield  {journal} {\bibinfo  {journal} {Phys. Rev. A}\ }\textbf {\bibinfo
				{volume} {63}},\ \bibinfo {pages} {050301} (\bibinfo {year}
			{2001})}\BibitemShut {NoStop}%
		\bibitem [{\citenamefont {Hulpke}\ \emph {et~al.}(2004)\citenamefont {Hulpke},
			\citenamefont {Bruss}, \citenamefont {Lewenstein},\ and\ \citenamefont
			{Sanpera}}]{Hulpke2004}%
		\BibitemOpen
		\bibfield  {author} {\bibinfo {author} {\bibfnamefont {F.}~\bibnamefont
				{Hulpke}}, \bibinfo {author} {\bibfnamefont {D.}~\bibnamefont {Bruss}},
			\bibinfo {author} {\bibfnamefont {M.}~\bibnamefont {Lewenstein}},\ and\
			\bibinfo {author} {\bibfnamefont {A.}~\bibnamefont {Sanpera}},\ }\bibfield
		{title} {\bibinfo {title} {Simplifying {Schmidt} number witnesses via
				higher-dimensional embeddings},\ }\href {https://doi.org/10.26421/QIC4.3-6}
		{\bibfield  {journal} {\bibinfo  {journal} {Quantum Inf. Comput.}\ }\textbf
			{\bibinfo {volume} {4}},\ \bibinfo {pages} {207} (\bibinfo {year} {2004})},\
		\Eprint {https://arxiv.org/abs/quant-ph/0401118} {arXiv:quant-ph/0401118}
		\BibitemShut {NoStop}%
		\bibitem [{\citenamefont {Shahandeh}\ \emph {et~al.}(2013)\citenamefont
			{Shahandeh}, \citenamefont {Sperling},\ and\ \citenamefont
			{Vogel}}]{Shahandeh2013}%
		\BibitemOpen
		\bibfield  {author} {\bibinfo {author} {\bibfnamefont {F.}~\bibnamefont
				{Shahandeh}}, \bibinfo {author} {\bibfnamefont {J.}~\bibnamefont
				{Sperling}},\ and\ \bibinfo {author} {\bibfnamefont {W.}~\bibnamefont
				{Vogel}},\ }\bibfield  {title} {\bibinfo {title} {Operational gaussian
				schmidt-number witnesses},\ }\href
		{https://doi.org/10.1103/PhysRevA.88.062323} {\bibfield  {journal} {\bibinfo
				{journal} {Phys. Rev. A}\ }\textbf {\bibinfo {volume} {88}},\ \bibinfo
			{pages} {062323} (\bibinfo {year} {2013})}\BibitemShut {NoStop}%
		\bibitem [{\citenamefont {Weilenmann}\ \emph {et~al.}(2020)\citenamefont
			{Weilenmann}, \citenamefont {Dive}, \citenamefont {Trillo}, \citenamefont
			{Aguilar},\ and\ \citenamefont {Navascu\'es}}]{Weilenmann2020}%
		\BibitemOpen
		\bibfield  {author} {\bibinfo {author} {\bibfnamefont {M.}~\bibnamefont
				{Weilenmann}}, \bibinfo {author} {\bibfnamefont {B.}~\bibnamefont {Dive}},
			\bibinfo {author} {\bibfnamefont {D.}~\bibnamefont {Trillo}}, \bibinfo
			{author} {\bibfnamefont {E.~A.}\ \bibnamefont {Aguilar}},\ and\ \bibinfo
			{author} {\bibfnamefont {M.}~\bibnamefont {Navascu\'es}},\ }\bibfield
		{title} {\bibinfo {title} {Entanglement detection beyond measuring
				fidelities},\ }\href {https://doi.org/10.1103/PhysRevLett.124.200502}
		{\bibfield  {journal} {\bibinfo  {journal} {Phys. Rev. Lett.}\ }\textbf
			{\bibinfo {volume} {124}},\ \bibinfo {pages} {200502} (\bibinfo {year}
			{2020})}\BibitemShut {NoStop}%
		\bibitem [{\citenamefont {Wyderka}\ \emph {et~al.}(2023)\citenamefont
			{Wyderka}, \citenamefont {Chesi}, \citenamefont {Kampermann}, \citenamefont
			{Macchiavello},\ and\ \citenamefont {Bru\ss{}}}]{Wyderka2023}%
		\BibitemOpen
		\bibfield  {author} {\bibinfo {author} {\bibfnamefont {N.}~\bibnamefont
				{Wyderka}}, \bibinfo {author} {\bibfnamefont {G.}~\bibnamefont {Chesi}},
			\bibinfo {author} {\bibfnamefont {H.}~\bibnamefont {Kampermann}}, \bibinfo
			{author} {\bibfnamefont {C.}~\bibnamefont {Macchiavello}},\ and\ \bibinfo
			{author} {\bibfnamefont {D.}~\bibnamefont {Bru\ss{}}},\ }\bibfield  {title}
		{\bibinfo {title} {Construction of efficient schmidt-number witnesses for
				high-dimensional quantum states},\ }\href
		{https://doi.org/10.1103/PhysRevA.107.022431} {\bibfield  {journal} {\bibinfo
				{journal} {Phys. Rev. A}\ }\textbf {\bibinfo {volume} {107}},\ \bibinfo
			{pages} {022431} (\bibinfo {year} {2023})}\BibitemShut {NoStop}%
		\bibitem [{\citenamefont {Morelli}\ \emph {et~al.}(2023)\citenamefont
			{Morelli}, \citenamefont {Huber},\ and\ \citenamefont
			{Tavakoli}}]{Morelli2023}%
		\BibitemOpen
		\bibfield  {author} {\bibinfo {author} {\bibfnamefont {S.}~\bibnamefont
				{Morelli}}, \bibinfo {author} {\bibfnamefont {M.}~\bibnamefont {Huber}},\
			and\ \bibinfo {author} {\bibfnamefont {A.}~\bibnamefont {Tavakoli}},\
		}\bibfield  {title} {\bibinfo {title} {Resource-efficient high-dimensional
				entanglement detection via symmetric projections},\ }\href
		{https://doi.org/10.1103/PhysRevLett.131.170201} {\bibfield  {journal}
			{\bibinfo  {journal} {Phys. Rev. Lett.}\ }\textbf {\bibinfo {volume} {131}},\
			\bibinfo {pages} {170201} (\bibinfo {year} {2023})}\BibitemShut {NoStop}%
		\bibitem [{\citenamefont {Tavakoli}\ and\ \citenamefont
			{Morelli}(2024)}]{Tavakoli2024}%
		\BibitemOpen
		\bibfield  {author} {\bibinfo {author} {\bibfnamefont {A.}~\bibnamefont
				{Tavakoli}}\ and\ \bibinfo {author} {\bibfnamefont {S.}~\bibnamefont
				{Morelli}},\ }\href@noop {} {\bibinfo {title} {Enhanced schmidt number
				criteria based on correlation trace norms}} (\bibinfo {year} {2024}),\
		\Eprint {https://arxiv.org/abs/2402.09972} {arXiv:2402.09972 [quant-ph]}
		\BibitemShut {NoStop}%
		\bibitem [{\citenamefont {Dada}\ \emph {et~al.}(2011)\citenamefont {Dada},
			\citenamefont {Leach}, \citenamefont {Buller}, \citenamefont {Padgett},\ and\
			\citenamefont {Andersson}}]{Dada2011}%
		\BibitemOpen
		\bibfield  {author} {\bibinfo {author} {\bibfnamefont {A.~C.}\ \bibnamefont
				{Dada}}, \bibinfo {author} {\bibfnamefont {J.}~\bibnamefont {Leach}},
			\bibinfo {author} {\bibfnamefont {G.~S.}\ \bibnamefont {Buller}}, \bibinfo
			{author} {\bibfnamefont {M.~J.}\ \bibnamefont {Padgett}},\ and\ \bibinfo
			{author} {\bibfnamefont {E.}~\bibnamefont {Andersson}},\ }\bibfield  {title}
		{\bibinfo {title} {Experimental high-dimensional two-photon entanglement and
				violations of generalized bell inequalities},\ }\href
		{https://doi.org/10.1038/nphys1996} {\bibfield  {journal} {\bibinfo
				{journal} {Nature Physics}\ }\textbf {\bibinfo {volume} {7}},\ \bibinfo
			{pages} {677} (\bibinfo {year} {2011})}\BibitemShut {NoStop}%
		\bibitem [{\citenamefont {Krenn}\ \emph {et~al.}(2014)\citenamefont {Krenn},
			\citenamefont {Huber}, \citenamefont {Fickler}, \citenamefont {Lapkiewicz},
			\citenamefont {Ramelow},\ and\ \citenamefont {Zeilinger}}]{Krenn2014}%
		\BibitemOpen
		\bibfield  {author} {\bibinfo {author} {\bibfnamefont {M.}~\bibnamefont
				{Krenn}}, \bibinfo {author} {\bibfnamefont {M.}~\bibnamefont {Huber}},
			\bibinfo {author} {\bibfnamefont {R.}~\bibnamefont {Fickler}}, \bibinfo
			{author} {\bibfnamefont {R.}~\bibnamefont {Lapkiewicz}}, \bibinfo {author}
			{\bibfnamefont {S.}~\bibnamefont {Ramelow}},\ and\ \bibinfo {author}
			{\bibfnamefont {A.}~\bibnamefont {Zeilinger}},\ }\bibfield  {title} {\bibinfo
			{title} {Generation and confirmation of a (100 × 100)-dimensional entangled
				quantum system},\ }\href {https://doi.org/10.1073/pnas.1402365111} {\bibfield
			{journal} {\bibinfo  {journal} {Proceedings of the National Academy of
					Sciences}\ }\textbf {\bibinfo {volume} {111}},\ \bibinfo {pages} {6243}
			(\bibinfo {year} {2014})},\ \Eprint
		{https://arxiv.org/abs/https://www.pnas.org/doi/pdf/10.1073/pnas.1402365111}
		{https://www.pnas.org/doi/pdf/10.1073/pnas.1402365111} \BibitemShut {NoStop}%
		\bibitem [{\citenamefont {Martin}\ \emph {et~al.}(2017)\citenamefont {Martin},
			\citenamefont {Guerreiro}, \citenamefont {Tiranov}, \citenamefont
			{Designolle}, \citenamefont {Fr\"owis}, \citenamefont {Brunner},
			\citenamefont {Huber},\ and\ \citenamefont {Gisin}}]{PhysRevLett.118.110501}%
		\BibitemOpen
		\bibfield  {author} {\bibinfo {author} {\bibfnamefont {A.}~\bibnamefont
				{Martin}}, \bibinfo {author} {\bibfnamefont {T.}~\bibnamefont {Guerreiro}},
			\bibinfo {author} {\bibfnamefont {A.}~\bibnamefont {Tiranov}}, \bibinfo
			{author} {\bibfnamefont {S.}~\bibnamefont {Designolle}}, \bibinfo {author}
			{\bibfnamefont {F.}~\bibnamefont {Fr\"owis}}, \bibinfo {author}
			{\bibfnamefont {N.}~\bibnamefont {Brunner}}, \bibinfo {author} {\bibfnamefont
				{M.}~\bibnamefont {Huber}},\ and\ \bibinfo {author} {\bibfnamefont
				{N.}~\bibnamefont {Gisin}},\ }\bibfield  {title} {\bibinfo {title}
			{Quantifying photonic high-dimensional entanglement},\ }\href
		{https://doi.org/10.1103/PhysRevLett.118.110501} {\bibfield  {journal}
			{\bibinfo  {journal} {Phys. Rev. Lett.}\ }\textbf {\bibinfo {volume} {118}},\
			\bibinfo {pages} {110501} (\bibinfo {year} {2017})}\BibitemShut {NoStop}%
		\bibitem [{\citenamefont {Bavaresco}\ \emph {et~al.}(2018)\citenamefont
			{Bavaresco}, \citenamefont {Herrera~Valencia}, \citenamefont {Kl{\"o}ckl},
			\citenamefont {Pivoluska}, \citenamefont {Erker}, \citenamefont {Friis},
			\citenamefont {Malik},\ and\ \citenamefont {Huber}}]{Bavaresco2018}%
		\BibitemOpen
		\bibfield  {author} {\bibinfo {author} {\bibfnamefont {J.}~\bibnamefont
				{Bavaresco}}, \bibinfo {author} {\bibfnamefont {N.}~\bibnamefont
				{Herrera~Valencia}}, \bibinfo {author} {\bibfnamefont {C.}~\bibnamefont
				{Kl{\"o}ckl}}, \bibinfo {author} {\bibfnamefont {M.}~\bibnamefont
				{Pivoluska}}, \bibinfo {author} {\bibfnamefont {P.}~\bibnamefont {Erker}},
			\bibinfo {author} {\bibfnamefont {N.}~\bibnamefont {Friis}}, \bibinfo
			{author} {\bibfnamefont {M.}~\bibnamefont {Malik}},\ and\ \bibinfo {author}
			{\bibfnamefont {M.}~\bibnamefont {Huber}},\ }\bibfield  {title} {\bibinfo
			{title} {Measurements in two bases are sufficient for certifying
				high-dimensional entanglement},\ }\href
		{https://doi.org/10.1038/s41567-018-0203-z} {\bibfield  {journal} {\bibinfo
				{journal} {Nature Physics}\ }\textbf {\bibinfo {volume} {14}},\ \bibinfo
			{pages} {1032} (\bibinfo {year} {2018})}\BibitemShut {NoStop}%
		\bibitem [{\citenamefont {Herrera~Valencia}\ \emph {et~al.}(2020)\citenamefont
			{Herrera~Valencia}, \citenamefont {Srivastav}, \citenamefont {Pivoluska},
			\citenamefont {Huber}, \citenamefont {Friis}, \citenamefont {McCutcheon},\
			and\ \citenamefont {Malik}}]{Herrera2020}%
		\BibitemOpen
		\bibfield  {author} {\bibinfo {author} {\bibfnamefont {N.}~\bibnamefont
				{Herrera~Valencia}}, \bibinfo {author} {\bibfnamefont {V.}~\bibnamefont
				{Srivastav}}, \bibinfo {author} {\bibfnamefont {M.}~\bibnamefont
				{Pivoluska}}, \bibinfo {author} {\bibfnamefont {M.}~\bibnamefont {Huber}},
			\bibinfo {author} {\bibfnamefont {N.}~\bibnamefont {Friis}}, \bibinfo
			{author} {\bibfnamefont {W.}~\bibnamefont {McCutcheon}},\ and\ \bibinfo
			{author} {\bibfnamefont {M.}~\bibnamefont {Malik}},\ }\bibfield  {title}
		{\bibinfo {title} {High-{D}imensional {P}ixel {E}ntanglement: {E}fficient
				{G}eneration and {C}ertification},\ }\href
		{https://doi.org/10.22331/q-2020-12-24-376} {\bibfield  {journal} {\bibinfo
				{journal} {{Quantum}}\ }\textbf {\bibinfo {volume} {4}},\ \bibinfo {pages}
			{376} (\bibinfo {year} {2020})}\BibitemShut {NoStop}%
		\bibitem [{\citenamefont {Designolle}\ \emph {et~al.}(2021)\citenamefont
			{Designolle}, \citenamefont {Srivastav}, \citenamefont {Uola}, \citenamefont
			{Valencia}, \citenamefont {McCutcheon}, \citenamefont {Malik},\ and\
			\citenamefont {Brunner}}]{Designolle2021}%
		\BibitemOpen
		\bibfield  {author} {\bibinfo {author} {\bibfnamefont {S.}~\bibnamefont
				{Designolle}}, \bibinfo {author} {\bibfnamefont {V.}~\bibnamefont
				{Srivastav}}, \bibinfo {author} {\bibfnamefont {R.}~\bibnamefont {Uola}},
			\bibinfo {author} {\bibfnamefont {N.~H.}\ \bibnamefont {Valencia}}, \bibinfo
			{author} {\bibfnamefont {W.}~\bibnamefont {McCutcheon}}, \bibinfo {author}
			{\bibfnamefont {M.}~\bibnamefont {Malik}},\ and\ \bibinfo {author}
			{\bibfnamefont {N.}~\bibnamefont {Brunner}},\ }\bibfield  {title} {\bibinfo
			{title} {Genuine high-dimensional quantum steering},\ }\href
		{https://doi.org/10.1103/PhysRevLett.126.200404} {\bibfield  {journal}
			{\bibinfo  {journal} {Phys. Rev. Lett.}\ }\textbf {\bibinfo {volume} {126}},\
			\bibinfo {pages} {200404} (\bibinfo {year} {2021})}\BibitemShut {NoStop}%
		\bibitem [{\citenamefont {Goel}\ \emph {et~al.}(2024)\citenamefont {Goel},
			\citenamefont {Leedumrongwatthanakun}, \citenamefont {Valencia},
			\citenamefont {McCutcheon}, \citenamefont {Tavakoli}, \citenamefont {Conti},
			\citenamefont {Pinkse},\ and\ \citenamefont {Malik}}]{Goel2024}%
		\BibitemOpen
		\bibfield  {author} {\bibinfo {author} {\bibfnamefont {S.}~\bibnamefont
				{Goel}}, \bibinfo {author} {\bibfnamefont {S.}~\bibnamefont
				{Leedumrongwatthanakun}}, \bibinfo {author} {\bibfnamefont {N.~H.}\
				\bibnamefont {Valencia}}, \bibinfo {author} {\bibfnamefont {W.}~\bibnamefont
				{McCutcheon}}, \bibinfo {author} {\bibfnamefont {A.}~\bibnamefont
				{Tavakoli}}, \bibinfo {author} {\bibfnamefont {C.}~\bibnamefont {Conti}},
			\bibinfo {author} {\bibfnamefont {P.~W.~H.}\ \bibnamefont {Pinkse}},\ and\
			\bibinfo {author} {\bibfnamefont {M.}~\bibnamefont {Malik}},\ }\bibfield
		{title} {\bibinfo {title} {Inverse design of high-dimensional quantum optical
				circuits in a complex medium},\ }\href
		{https://doi.org/10.1038/s41567-023-02319-6} {\bibfield  {journal} {\bibinfo
				{journal} {Nature Physics}\ }\textbf {\bibinfo {volume} {20}},\ \bibinfo
			{pages} {232} (\bibinfo {year} {2024})}\BibitemShut {NoStop}%
		\bibitem [{\citenamefont {Huber}\ and\ \citenamefont
			{de~Vicente}(2013)}]{Huber_2013}%
		\BibitemOpen
		\bibfield  {author} {\bibinfo {author} {\bibfnamefont {M.}~\bibnamefont
				{Huber}}\ and\ \bibinfo {author} {\bibfnamefont {J.~I.}\ \bibnamefont
				{de~Vicente}},\ }\bibfield  {title} {\bibinfo {title} {Structure of
				multidimensional entanglement in multipartite systems},\ }\href
		{https://doi.org/10.1103/PhysRevLett.110.030501} {\bibfield  {journal}
			{\bibinfo  {journal} {Phys. Rev. Lett.}\ }\textbf {\bibinfo {volume} {110}},\
			\bibinfo {pages} {030501} (\bibinfo {year} {2013})}\BibitemShut {NoStop}%
		\bibitem [{\citenamefont {Chi}\ \emph {et~al.}(2022)\citenamefont {Chi},
			\citenamefont {Huang}, \citenamefont {Zhang}, \citenamefont {Mao},
			\citenamefont {Zhou}, \citenamefont {Chen}, \citenamefont {Zhai},
			\citenamefont {Bao}, \citenamefont {Dai}, \citenamefont {Yuan}, \citenamefont
			{Zhang}, \citenamefont {Dai}, \citenamefont {Tang}, \citenamefont {Yang},
			\citenamefont {Li}, \citenamefont {Ding}, \citenamefont {Oxenl{\o}we},
			\citenamefont {Thompson}, \citenamefont {O'Brien}, \citenamefont {Li},
			\citenamefont {Gong},\ and\ \citenamefont {Wang}}]{Chi2022}%
		\BibitemOpen
		\bibfield  {author} {\bibinfo {author} {\bibfnamefont {Y.}~\bibnamefont
				{Chi}}, \bibinfo {author} {\bibfnamefont {J.}~\bibnamefont {Huang}}, \bibinfo
			{author} {\bibfnamefont {Z.}~\bibnamefont {Zhang}}, \bibinfo {author}
			{\bibfnamefont {J.}~\bibnamefont {Mao}}, \bibinfo {author} {\bibfnamefont
				{Z.}~\bibnamefont {Zhou}}, \bibinfo {author} {\bibfnamefont {X.}~\bibnamefont
				{Chen}}, \bibinfo {author} {\bibfnamefont {C.}~\bibnamefont {Zhai}}, \bibinfo
			{author} {\bibfnamefont {J.}~\bibnamefont {Bao}}, \bibinfo {author}
			{\bibfnamefont {T.}~\bibnamefont {Dai}}, \bibinfo {author} {\bibfnamefont
				{H.}~\bibnamefont {Yuan}}, \bibinfo {author} {\bibfnamefont {M.}~\bibnamefont
				{Zhang}}, \bibinfo {author} {\bibfnamefont {D.}~\bibnamefont {Dai}}, \bibinfo
			{author} {\bibfnamefont {B.}~\bibnamefont {Tang}}, \bibinfo {author}
			{\bibfnamefont {Y.}~\bibnamefont {Yang}}, \bibinfo {author} {\bibfnamefont
				{Z.}~\bibnamefont {Li}}, \bibinfo {author} {\bibfnamefont {Y.}~\bibnamefont
				{Ding}}, \bibinfo {author} {\bibfnamefont {L.~K.}\ \bibnamefont
				{Oxenl{\o}we}}, \bibinfo {author} {\bibfnamefont {M.~G.}\ \bibnamefont
				{Thompson}}, \bibinfo {author} {\bibfnamefont {J.~L.}\ \bibnamefont
				{O'Brien}}, \bibinfo {author} {\bibfnamefont {Y.}~\bibnamefont {Li}},
			\bibinfo {author} {\bibfnamefont {Q.}~\bibnamefont {Gong}},\ and\ \bibinfo
			{author} {\bibfnamefont {J.}~\bibnamefont {Wang}},\ }\bibfield  {title}
		{\bibinfo {title} {A programmable qudit-based quantum processor},\ }\href
		{https://doi.org/10.1038/s41467-022-28767-x} {\bibfield  {journal} {\bibinfo
				{journal} {Nature Communications}\ }\textbf {\bibinfo {volume} {13}},\
			\bibinfo {pages} {1166} (\bibinfo {year} {2022})}\BibitemShut {NoStop}%
		\bibitem [{\citenamefont {Ringbauer}\ \emph {et~al.}(2022)\citenamefont
			{Ringbauer}, \citenamefont {Meth}, \citenamefont {Postler}, \citenamefont
			{Stricker}, \citenamefont {Blatt}, \citenamefont {Schindler},\ and\
			\citenamefont {Monz}}]{Ringbauer2022}%
		\BibitemOpen
		\bibfield  {author} {\bibinfo {author} {\bibfnamefont {M.}~\bibnamefont
				{Ringbauer}}, \bibinfo {author} {\bibfnamefont {M.}~\bibnamefont {Meth}},
			\bibinfo {author} {\bibfnamefont {L.}~\bibnamefont {Postler}}, \bibinfo
			{author} {\bibfnamefont {R.}~\bibnamefont {Stricker}}, \bibinfo {author}
			{\bibfnamefont {R.}~\bibnamefont {Blatt}}, \bibinfo {author} {\bibfnamefont
				{P.}~\bibnamefont {Schindler}},\ and\ \bibinfo {author} {\bibfnamefont
				{T.}~\bibnamefont {Monz}},\ }\bibfield  {title} {\bibinfo {title} {A
				universal qudit quantum processor with trapped ions},\ }\href
		{https://doi.org/10.1038/s41567-022-01658-0} {\bibfield  {journal} {\bibinfo
				{journal} {Nature Physics}\ }\textbf {\bibinfo {volume} {18}},\ \bibinfo
			{pages} {1053} (\bibinfo {year} {2022})}\BibitemShut {NoStop}%
		\bibitem [{\citenamefont {Tang}\ \emph {et~al.}(2013)\citenamefont {Tang},
			\citenamefont {Yu},\ and\ \citenamefont {Oh}}]{Tang2013}%
		\BibitemOpen
		\bibfield  {author} {\bibinfo {author} {\bibfnamefont {W.}~\bibnamefont
				{Tang}}, \bibinfo {author} {\bibfnamefont {S.}~\bibnamefont {Yu}},\ and\
			\bibinfo {author} {\bibfnamefont {C.~H.}\ \bibnamefont {Oh}},\ }\bibfield
		{title} {\bibinfo {title} {Greenberger-horne-zeilinger paradoxes from qudit
				graph states},\ }\href {https://doi.org/10.1103/PhysRevLett.110.100403}
		{\bibfield  {journal} {\bibinfo  {journal} {Phys. Rev. Lett.}\ }\textbf
			{\bibinfo {volume} {110}},\ \bibinfo {pages} {100403} (\bibinfo {year}
			{2013})}\BibitemShut {NoStop}%
		\bibitem [{\citenamefont {Augusiak}\ \emph {et~al.}(2019)\citenamefont
			{Augusiak}, \citenamefont {Salavrakos}, \citenamefont {Tura},\ and\
			\citenamefont {Acín}}]{Augusiak2019}%
		\BibitemOpen
		\bibfield  {author} {\bibinfo {author} {\bibfnamefont {R.}~\bibnamefont
				{Augusiak}}, \bibinfo {author} {\bibfnamefont {A.}~\bibnamefont
				{Salavrakos}}, \bibinfo {author} {\bibfnamefont {J.}~\bibnamefont {Tura}},\
			and\ \bibinfo {author} {\bibfnamefont {A.}~\bibnamefont {Acín}},\ }\bibfield
		{title} {\bibinfo {title} {Bell inequalities tailored to the
				greenberger–horne–zeilinger states of arbitrary local dimension},\ }\href
		{https://doi.org/10.1088/1367-2630/ab4d9f} {\bibfield  {journal} {\bibinfo
				{journal} {New Journal of Physics}\ }\textbf {\bibinfo {volume} {21}},\
			\bibinfo {pages} {113001} (\bibinfo {year} {2019})}\BibitemShut {NoStop}%
		\bibitem [{\citenamefont {Fitzi}\ \emph {et~al.}(2001)\citenamefont {Fitzi},
			\citenamefont {Gisin},\ and\ \citenamefont {Maurer}}]{Fitzi2001}%
		\BibitemOpen
		\bibfield  {author} {\bibinfo {author} {\bibfnamefont {M.}~\bibnamefont
				{Fitzi}}, \bibinfo {author} {\bibfnamefont {N.}~\bibnamefont {Gisin}},\ and\
			\bibinfo {author} {\bibfnamefont {U.}~\bibnamefont {Maurer}},\ }\bibfield
		{title} {\bibinfo {title} {Quantum solution to the byzantine agreement
				problem},\ }\href {https://doi.org/10.1103/PhysRevLett.87.217901} {\bibfield
			{journal} {\bibinfo  {journal} {Phys. Rev. Lett.}\ }\textbf {\bibinfo
				{volume} {87}},\ \bibinfo {pages} {217901} (\bibinfo {year}
			{2001})}\BibitemShut {NoStop}%
		\bibitem [{\citenamefont {Cabello}(2002)}]{Cabello2002}%
		\BibitemOpen
		\bibfield  {author} {\bibinfo {author} {\bibfnamefont {A.}~\bibnamefont
				{Cabello}},\ }\bibfield  {title} {\bibinfo {title} {$n$-particle $n$-level
				singlet states: Some properties and applications},\ }\href
		{https://doi.org/10.1103/PhysRevLett.89.100402} {\bibfield  {journal}
			{\bibinfo  {journal} {Phys. Rev. Lett.}\ }\textbf {\bibinfo {volume} {89}},\
			\bibinfo {pages} {100402} (\bibinfo {year} {2002})}\BibitemShut {NoStop}%
		\bibitem [{\citenamefont {Malik}\ \emph {et~al.}(2016)\citenamefont {Malik},
			\citenamefont {Erhard}, \citenamefont {Huber}, \citenamefont {Krenn},
			\citenamefont {Fickler},\ and\ \citenamefont {Zeilinger}}]{Malik2016}%
		\BibitemOpen
		\bibfield  {author} {\bibinfo {author} {\bibfnamefont {M.}~\bibnamefont
				{Malik}}, \bibinfo {author} {\bibfnamefont {M.}~\bibnamefont {Erhard}},
			\bibinfo {author} {\bibfnamefont {M.}~\bibnamefont {Huber}}, \bibinfo
			{author} {\bibfnamefont {M.}~\bibnamefont {Krenn}}, \bibinfo {author}
			{\bibfnamefont {R.}~\bibnamefont {Fickler}},\ and\ \bibinfo {author}
			{\bibfnamefont {A.}~\bibnamefont {Zeilinger}},\ }\bibfield  {title} {\bibinfo
			{title} {Multi-photon entanglement in high dimensions},\ }\href
		{https://doi.org/10.1038/nphoton.2016.12} {\bibfield  {journal} {\bibinfo
				{journal} {Nature Photonics}\ }\textbf {\bibinfo {volume} {10}},\ \bibinfo
			{pages} {248} (\bibinfo {year} {2016})}\BibitemShut {NoStop}%
		\bibitem [{\citenamefont {Erhard}\ \emph {et~al.}(2018)\citenamefont {Erhard},
			\citenamefont {Malik}, \citenamefont {Krenn},\ and\ \citenamefont
			{Zeilinger}}]{Erhard2018}%
		\BibitemOpen
		\bibfield  {author} {\bibinfo {author} {\bibfnamefont {M.}~\bibnamefont
				{Erhard}}, \bibinfo {author} {\bibfnamefont {M.}~\bibnamefont {Malik}},
			\bibinfo {author} {\bibfnamefont {M.}~\bibnamefont {Krenn}},\ and\ \bibinfo
			{author} {\bibfnamefont {A.}~\bibnamefont {Zeilinger}},\ }\bibfield  {title}
		{\bibinfo {title} {Experimental greenberger--horne--zeilinger entanglement
				beyond qubits},\ }\href {https://doi.org/10.1038/s41566-018-0257-6}
		{\bibfield  {journal} {\bibinfo  {journal} {Nature Photonics}\ }\textbf
			{\bibinfo {volume} {12}},\ \bibinfo {pages} {759} (\bibinfo {year}
			{2018})}\BibitemShut {NoStop}%
		\bibitem [{\citenamefont {Imany}\ \emph {et~al.}(2019)\citenamefont {Imany},
			\citenamefont {Jaramillo-Villegas}, \citenamefont {Alshaykh}, \citenamefont
			{Lukens}, \citenamefont {Odele}, \citenamefont {Moore}, \citenamefont
			{Leaird}, \citenamefont {Qi},\ and\ \citenamefont {Weiner}}]{Imany2019}%
		\BibitemOpen
		\bibfield  {author} {\bibinfo {author} {\bibfnamefont {P.}~\bibnamefont
				{Imany}}, \bibinfo {author} {\bibfnamefont {J.~A.}\ \bibnamefont
				{Jaramillo-Villegas}}, \bibinfo {author} {\bibfnamefont {M.~S.}\ \bibnamefont
				{Alshaykh}}, \bibinfo {author} {\bibfnamefont {J.~M.}\ \bibnamefont
				{Lukens}}, \bibinfo {author} {\bibfnamefont {O.~D.}\ \bibnamefont {Odele}},
			\bibinfo {author} {\bibfnamefont {A.~J.}\ \bibnamefont {Moore}}, \bibinfo
			{author} {\bibfnamefont {D.~E.}\ \bibnamefont {Leaird}}, \bibinfo {author}
			{\bibfnamefont {M.}~\bibnamefont {Qi}},\ and\ \bibinfo {author}
			{\bibfnamefont {A.~M.}\ \bibnamefont {Weiner}},\ }\bibfield  {title}
		{\bibinfo {title} {High-dimensional optical quantum logic in large
				operational spaces},\ }\href {https://doi.org/10.1038/s41534-019-0173-8}
		{\bibfield  {journal} {\bibinfo  {journal} {npj Quantum Information}\
			}\textbf {\bibinfo {volume} {5}},\ \bibinfo {pages} {59} (\bibinfo {year}
			{2019})}\BibitemShut {NoStop}%
		\bibitem [{\citenamefont {Xing}\ \emph {et~al.}(2023)\citenamefont {Xing},
			\citenamefont {Hu}, \citenamefont {Guo}, \citenamefont {Liu}, \citenamefont
			{Li},\ and\ \citenamefont {Guo}}]{Xing2023}%
		\BibitemOpen
		\bibfield  {author} {\bibinfo {author} {\bibfnamefont {W.-B.}\ \bibnamefont
				{Xing}}, \bibinfo {author} {\bibfnamefont {X.-M.}\ \bibnamefont {Hu}},
			\bibinfo {author} {\bibfnamefont {Y.}~\bibnamefont {Guo}}, \bibinfo {author}
			{\bibfnamefont {B.-H.}\ \bibnamefont {Liu}}, \bibinfo {author} {\bibfnamefont
				{C.-F.}\ \bibnamefont {Li}},\ and\ \bibinfo {author} {\bibfnamefont {G.-C.}\
				\bibnamefont {Guo}},\ }\bibfield  {title} {\bibinfo {title} {Preparation of
				multiphoton high-dimensional ghz states},\ }\href
		{https://doi.org/10.1364/OE.494850} {\bibfield  {journal} {\bibinfo
				{journal} {Opt. Express}\ }\textbf {\bibinfo {volume} {31}},\ \bibinfo
			{pages} {24887} (\bibinfo {year} {2023})}\BibitemShut {NoStop}%
		\bibitem [{\citenamefont {Reimer}\ \emph {et~al.}(2019)\citenamefont {Reimer},
			\citenamefont {Sciara}, \citenamefont {Roztocki}, \citenamefont {Islam},
			\citenamefont {Romero~Cort{\'e}s}, \citenamefont {Zhang}, \citenamefont
			{Fischer}, \citenamefont {Loranger}, \citenamefont {Kashyap}, \citenamefont
			{Cino}, \citenamefont {Chu}, \citenamefont {Little}, \citenamefont {Moss},
			\citenamefont {Caspani}, \citenamefont {Munro}, \citenamefont {Aza{\~{n}}a},
			\citenamefont {Kues},\ and\ \citenamefont {Morandotti}}]{Reimer2019}%
		\BibitemOpen
		\bibfield  {author} {\bibinfo {author} {\bibfnamefont {C.}~\bibnamefont
				{Reimer}}, \bibinfo {author} {\bibfnamefont {S.}~\bibnamefont {Sciara}},
			\bibinfo {author} {\bibfnamefont {P.}~\bibnamefont {Roztocki}}, \bibinfo
			{author} {\bibfnamefont {M.}~\bibnamefont {Islam}}, \bibinfo {author}
			{\bibfnamefont {L.}~\bibnamefont {Romero~Cort{\'e}s}}, \bibinfo {author}
			{\bibfnamefont {Y.}~\bibnamefont {Zhang}}, \bibinfo {author} {\bibfnamefont
				{B.}~\bibnamefont {Fischer}}, \bibinfo {author} {\bibfnamefont
				{S.}~\bibnamefont {Loranger}}, \bibinfo {author} {\bibfnamefont
				{R.}~\bibnamefont {Kashyap}}, \bibinfo {author} {\bibfnamefont
				{A.}~\bibnamefont {Cino}}, \bibinfo {author} {\bibfnamefont {S.~T.}\
				\bibnamefont {Chu}}, \bibinfo {author} {\bibfnamefont {B.~E.}\ \bibnamefont
				{Little}}, \bibinfo {author} {\bibfnamefont {D.~J.}\ \bibnamefont {Moss}},
			\bibinfo {author} {\bibfnamefont {L.}~\bibnamefont {Caspani}}, \bibinfo
			{author} {\bibfnamefont {W.~J.}\ \bibnamefont {Munro}}, \bibinfo {author}
			{\bibfnamefont {J.}~\bibnamefont {Aza{\~{n}}a}}, \bibinfo {author}
			{\bibfnamefont {M.}~\bibnamefont {Kues}},\ and\ \bibinfo {author}
			{\bibfnamefont {R.}~\bibnamefont {Morandotti}},\ }\bibfield  {title}
		{\bibinfo {title} {High-dimensional one-way quantum processing implemented on
				d-level cluster states},\ }\href {https://doi.org/10.1038/s41567-018-0347-x}
		{\bibfield  {journal} {\bibinfo  {journal} {Nature Physics}\ }\textbf
			{\bibinfo {volume} {15}},\ \bibinfo {pages} {148} (\bibinfo {year}
			{2019})}\BibitemShut {NoStop}%
		\bibitem [{\citenamefont {Bao}\ \emph {et~al.}(2023)\citenamefont {Bao},
			\citenamefont {Fu}, \citenamefont {Pramanik}, \citenamefont {Mao},
			\citenamefont {Chi}, \citenamefont {Cao}, \citenamefont {Zhai}, \citenamefont
			{Mao}, \citenamefont {Dai}, \citenamefont {Chen}, \citenamefont {Jia},
			\citenamefont {Zhao}, \citenamefont {Zheng}, \citenamefont {Tang},
			\citenamefont {Li}, \citenamefont {Luo}, \citenamefont {Wang}, \citenamefont
			{Yang}, \citenamefont {Peng}, \citenamefont {Liu}, \citenamefont {Dai},
			\citenamefont {He}, \citenamefont {Muthali}, \citenamefont {Oxenl{\o}we},
			\citenamefont {Vigliar}, \citenamefont {Paesani}, \citenamefont {Hou},
			\citenamefont {Santagati}, \citenamefont {Silverstone}, \citenamefont
			{Laing}, \citenamefont {Thompson}, \citenamefont {O'Brien}, \citenamefont
			{Ding}, \citenamefont {Gong},\ and\ \citenamefont {Wang}}]{Bao2023}%
		\BibitemOpen
		\bibfield  {author} {\bibinfo {author} {\bibfnamefont {J.}~\bibnamefont
				{Bao}}, \bibinfo {author} {\bibfnamefont {Z.}~\bibnamefont {Fu}}, \bibinfo
			{author} {\bibfnamefont {T.}~\bibnamefont {Pramanik}}, \bibinfo {author}
			{\bibfnamefont {J.}~\bibnamefont {Mao}}, \bibinfo {author} {\bibfnamefont
				{Y.}~\bibnamefont {Chi}}, \bibinfo {author} {\bibfnamefont {Y.}~\bibnamefont
				{Cao}}, \bibinfo {author} {\bibfnamefont {C.}~\bibnamefont {Zhai}}, \bibinfo
			{author} {\bibfnamefont {Y.}~\bibnamefont {Mao}}, \bibinfo {author}
			{\bibfnamefont {T.}~\bibnamefont {Dai}}, \bibinfo {author} {\bibfnamefont
				{X.}~\bibnamefont {Chen}}, \bibinfo {author} {\bibfnamefont {X.}~\bibnamefont
				{Jia}}, \bibinfo {author} {\bibfnamefont {L.}~\bibnamefont {Zhao}}, \bibinfo
			{author} {\bibfnamefont {Y.}~\bibnamefont {Zheng}}, \bibinfo {author}
			{\bibfnamefont {B.}~\bibnamefont {Tang}}, \bibinfo {author} {\bibfnamefont
				{Z.}~\bibnamefont {Li}}, \bibinfo {author} {\bibfnamefont {J.}~\bibnamefont
				{Luo}}, \bibinfo {author} {\bibfnamefont {W.}~\bibnamefont {Wang}}, \bibinfo
			{author} {\bibfnamefont {Y.}~\bibnamefont {Yang}}, \bibinfo {author}
			{\bibfnamefont {Y.}~\bibnamefont {Peng}}, \bibinfo {author} {\bibfnamefont
				{D.}~\bibnamefont {Liu}}, \bibinfo {author} {\bibfnamefont {D.}~\bibnamefont
				{Dai}}, \bibinfo {author} {\bibfnamefont {Q.}~\bibnamefont {He}}, \bibinfo
			{author} {\bibfnamefont {A.~L.}\ \bibnamefont {Muthali}}, \bibinfo {author}
			{\bibfnamefont {L.~K.}\ \bibnamefont {Oxenl{\o}we}}, \bibinfo {author}
			{\bibfnamefont {C.}~\bibnamefont {Vigliar}}, \bibinfo {author} {\bibfnamefont
				{S.}~\bibnamefont {Paesani}}, \bibinfo {author} {\bibfnamefont
				{H.}~\bibnamefont {Hou}}, \bibinfo {author} {\bibfnamefont {R.}~\bibnamefont
				{Santagati}}, \bibinfo {author} {\bibfnamefont {J.~W.}\ \bibnamefont
				{Silverstone}}, \bibinfo {author} {\bibfnamefont {A.}~\bibnamefont {Laing}},
			\bibinfo {author} {\bibfnamefont {M.~G.}\ \bibnamefont {Thompson}}, \bibinfo
			{author} {\bibfnamefont {J.~L.}\ \bibnamefont {O'Brien}}, \bibinfo {author}
			{\bibfnamefont {Y.}~\bibnamefont {Ding}}, \bibinfo {author} {\bibfnamefont
				{Q.}~\bibnamefont {Gong}},\ and\ \bibinfo {author} {\bibfnamefont
				{J.}~\bibnamefont {Wang}},\ }\bibfield  {title} {\bibinfo {title}
			{Very-large-scale integrated quantum graph photonics},\ }\href
		{https://doi.org/10.1038/s41566-023-01187-z} {\bibfield  {journal} {\bibinfo
				{journal} {Nature Photonics}\ }\textbf {\bibinfo {volume} {17}},\ \bibinfo
			{pages} {573} (\bibinfo {year} {2023})}\BibitemShut {NoStop}%
		\bibitem [{\citenamefont {Cervera-Lierta}\ \emph {et~al.}(2022)\citenamefont
			{Cervera-Lierta}, \citenamefont {Krenn}, \citenamefont {Aspuru-Guzik},\ and\
			\citenamefont {Galda}}]{Cervera2022}%
		\BibitemOpen
		\bibfield  {author} {\bibinfo {author} {\bibfnamefont {A.}~\bibnamefont
				{Cervera-Lierta}}, \bibinfo {author} {\bibfnamefont {M.}~\bibnamefont
				{Krenn}}, \bibinfo {author} {\bibfnamefont {A.}~\bibnamefont
				{Aspuru-Guzik}},\ and\ \bibinfo {author} {\bibfnamefont {A.}~\bibnamefont
				{Galda}},\ }\bibfield  {title} {\bibinfo {title} {Experimental
				high-dimensional greenberger-horne-zeilinger entanglement with
				superconducting transmon qutrits},\ }\href
		{https://doi.org/10.1103/PhysRevApplied.17.024062} {\bibfield  {journal}
			{\bibinfo  {journal} {Phys. Rev. Appl.}\ }\textbf {\bibinfo {volume} {17}},\
			\bibinfo {pages} {024062} (\bibinfo {year} {2022})}\BibitemShut {NoStop}%
		\bibitem [{\citenamefont {Peres}(1995)}]{Peres1995}%
		\BibitemOpen
		\bibfield  {author} {\bibinfo {author} {\bibfnamefont {A.}~\bibnamefont
				{Peres}},\ }\bibfield  {title} {\bibinfo {title} {Higher order schmidt
				decompositions},\ }\href
		{https://doi.org/https://doi.org/10.1016/0375-9601(95)00315-T} {\bibfield
			{journal} {\bibinfo  {journal} {Physics Letters A}\ }\textbf {\bibinfo
				{volume} {202}},\ \bibinfo {pages} {16} (\bibinfo {year} {1995})}\BibitemShut
		{NoStop}%
		\bibitem [{\citenamefont {Spengler}\ \emph {et~al.}(2013)\citenamefont
			{Spengler}, \citenamefont {Huber}, \citenamefont {Gabriel},\ and\
			\citenamefont {Hiesmayr}}]{Spengler2013}%
		\BibitemOpen
		\bibfield  {author} {\bibinfo {author} {\bibfnamefont {C.}~\bibnamefont
				{Spengler}}, \bibinfo {author} {\bibfnamefont {M.}~\bibnamefont {Huber}},
			\bibinfo {author} {\bibfnamefont {A.}~\bibnamefont {Gabriel}},\ and\ \bibinfo
			{author} {\bibfnamefont {B.~C.}\ \bibnamefont {Hiesmayr}},\ }\bibfield
		{title} {\bibinfo {title} {Examining the dimensionality of genuine
				multipartite entanglement},\ }\href
		{https://doi.org/10.1007/s11128-012-0369-8} {\bibfield  {journal} {\bibinfo
				{journal} {Quantum Information Processing}\ }\textbf {\bibinfo {volume}
				{12}},\ \bibinfo {pages} {269} (\bibinfo {year} {2013})}\BibitemShut
		{NoStop}%
		\bibitem [{\citenamefont {Fickler}\ \emph {et~al.}(2014)\citenamefont
			{Fickler}, \citenamefont {Lapkiewicz}, \citenamefont {Huber}, \citenamefont
			{Lavery}, \citenamefont {Padgett},\ and\ \citenamefont
			{Zeilinger}}]{Fickler_2014}%
		\BibitemOpen
		\bibfield  {author} {\bibinfo {author} {\bibfnamefont {R.}~\bibnamefont
				{Fickler}}, \bibinfo {author} {\bibfnamefont {R.}~\bibnamefont {Lapkiewicz}},
			\bibinfo {author} {\bibfnamefont {M.}~\bibnamefont {Huber}}, \bibinfo
			{author} {\bibfnamefont {M.~P.}\ \bibnamefont {Lavery}}, \bibinfo {author}
			{\bibfnamefont {M.~J.}\ \bibnamefont {Padgett}},\ and\ \bibinfo {author}
			{\bibfnamefont {A.}~\bibnamefont {Zeilinger}},\ }\bibfield  {title} {\bibinfo
			{title} {Interface between path and orbital angular momentum entanglement for
				high-dimensional photonic quantum information},\ }\bibfield  {journal}
		{\bibinfo  {journal} {Nature Communications}\ }\textbf {\bibinfo {volume}
			{5}},\ \href {https://doi.org/10.1038/ncomms5502} {10.1038/ncomms5502}
		(\bibinfo {year} {2014})\BibitemShut {NoStop}%
		\bibitem [{\citenamefont {Huber}\ \emph {et~al.}(2018)\citenamefont {Huber},
			\citenamefont {Eltschka}, \citenamefont {Siewert},\ and\ \citenamefont
			{Gühne}}]{Huber_2018}%
		\BibitemOpen
		\bibfield  {author} {\bibinfo {author} {\bibfnamefont {F.}~\bibnamefont
				{Huber}}, \bibinfo {author} {\bibfnamefont {C.}~\bibnamefont {Eltschka}},
			\bibinfo {author} {\bibfnamefont {J.}~\bibnamefont {Siewert}},\ and\ \bibinfo
			{author} {\bibfnamefont {O.}~\bibnamefont {Gühne}},\ }\bibfield  {title}
		{\bibinfo {title} {Bounds on absolutely maximally entangled states from
				shadow inequalities, and the quantum macwilliams identity},\ }\href
		{https://doi.org/10.1088/1751-8121/aaade5} {\bibfield  {journal} {\bibinfo
				{journal} {Journal of Physics A: Mathematical and Theoretical}\ }\textbf
			{\bibinfo {volume} {51}},\ \bibinfo {pages} {175301} (\bibinfo {year}
			{2018})}\BibitemShut {NoStop}%
		\bibitem [{\citenamefont {Huber}\ and\ \citenamefont
			{Wyderka}()}]{Huber_table}%
		\BibitemOpen
		\bibfield  {author} {\bibinfo {author} {\bibfnamefont {F.}~\bibnamefont
				{Huber}}\ and\ \bibinfo {author} {\bibfnamefont {N.}~\bibnamefont
				{Wyderka}},\ }\href@noop {} {\bibinfo {title} {Table of ame states}},\
		\bibinfo {note} {\url{http://www.tp.nt.uni-siegen.de/+fhuber/ame.html}
			[Accessed: (Use the date of access)]}\BibitemShut {NoStop}%
		\bibitem [{\citenamefont {Raussendorf}\ and\ \citenamefont
			{Briegel}(2001)}]{Raussendorf2001}%
		\BibitemOpen
		\bibfield  {author} {\bibinfo {author} {\bibfnamefont {R.}~\bibnamefont
				{Raussendorf}}\ and\ \bibinfo {author} {\bibfnamefont {H.~J.}\ \bibnamefont
				{Briegel}},\ }\bibfield  {title} {\bibinfo {title} {A one-way quantum
				computer},\ }\href {https://doi.org/10.1103/PhysRevLett.86.5188} {\bibfield
			{journal} {\bibinfo  {journal} {Phys. Rev. Lett.}\ }\textbf {\bibinfo
				{volume} {86}},\ \bibinfo {pages} {5188} (\bibinfo {year}
			{2001})}\BibitemShut {NoStop}%
		\bibitem [{\citenamefont {Raussendorf}\ \emph {et~al.}(2003)\citenamefont
			{Raussendorf}, \citenamefont {Browne},\ and\ \citenamefont
			{Briegel}}]{Raussendorf2003}%
		\BibitemOpen
		\bibfield  {author} {\bibinfo {author} {\bibfnamefont {R.}~\bibnamefont
				{Raussendorf}}, \bibinfo {author} {\bibfnamefont {D.~E.}\ \bibnamefont
				{Browne}},\ and\ \bibinfo {author} {\bibfnamefont {H.~J.}\ \bibnamefont
				{Briegel}},\ }\bibfield  {title} {\bibinfo {title} {Measurement-based quantum
				computation on cluster states},\ }\href
		{https://doi.org/10.1103/PhysRevA.68.022312} {\bibfield  {journal} {\bibinfo
				{journal} {Phys. Rev. A}\ }\textbf {\bibinfo {volume} {68}},\ \bibinfo
			{pages} {022312} (\bibinfo {year} {2003})}\BibitemShut {NoStop}%
		\bibitem [{\citenamefont {Zhou}\ \emph {et~al.}(2003)\citenamefont {Zhou},
			\citenamefont {Zeng}, \citenamefont {Xu},\ and\ \citenamefont
			{Sun}}]{PhysRevA.68.062303}%
		\BibitemOpen
		\bibfield  {author} {\bibinfo {author} {\bibfnamefont {D.~L.}\ \bibnamefont
				{Zhou}}, \bibinfo {author} {\bibfnamefont {B.}~\bibnamefont {Zeng}}, \bibinfo
			{author} {\bibfnamefont {Z.}~\bibnamefont {Xu}},\ and\ \bibinfo {author}
			{\bibfnamefont {C.~P.}\ \bibnamefont {Sun}},\ }\bibfield  {title} {\bibinfo
			{title} {Quantum computation based on d-level cluster state},\ }\href
		{https://doi.org/10.1103/PhysRevA.68.062303} {\bibfield  {journal} {\bibinfo
				{journal} {Phys. Rev. A}\ }\textbf {\bibinfo {volume} {68}},\ \bibinfo
			{pages} {062303} (\bibinfo {year} {2003})}\BibitemShut {NoStop}%
		\bibitem [{\citenamefont {Tavakoli}\ \emph {et~al.}(2023)\citenamefont
			{Tavakoli}, \citenamefont {Pozas-Kerstjens}, \citenamefont {Brown},\ and\
			\citenamefont {Araújo}}]{tavakoli2023semidefinite}%
		\BibitemOpen
		\bibfield  {author} {\bibinfo {author} {\bibfnamefont {A.}~\bibnamefont
				{Tavakoli}}, \bibinfo {author} {\bibfnamefont {A.}~\bibnamefont
				{Pozas-Kerstjens}}, \bibinfo {author} {\bibfnamefont {P.}~\bibnamefont
				{Brown}},\ and\ \bibinfo {author} {\bibfnamefont {M.}~\bibnamefont
				{Araújo}},\ }\href@noop {} {\bibinfo {title} {Semidefinite programming
				relaxations for quantum correlations}} (\bibinfo {year} {2023}),\ \Eprint
		{https://arxiv.org/abs/2307.02551} {arXiv:2307.02551 [quant-ph]} \BibitemShut
		{NoStop}%
		\bibitem [{\citenamefont {Jungnitsch}\ \emph
			{et~al.}(2011{\natexlab{a}})\citenamefont {Jungnitsch}, \citenamefont
			{Moroder},\ and\ \citenamefont {G\"uhne}}]{Jungnitsch2011}%
		\BibitemOpen
		\bibfield  {author} {\bibinfo {author} {\bibfnamefont {B.}~\bibnamefont
				{Jungnitsch}}, \bibinfo {author} {\bibfnamefont {T.}~\bibnamefont
				{Moroder}},\ and\ \bibinfo {author} {\bibfnamefont {O.}~\bibnamefont
				{G\"uhne}},\ }\bibfield  {title} {\bibinfo {title} {Taming multiparticle
				entanglement},\ }\href {https://doi.org/10.1103/PhysRevLett.106.190502}
		{\bibfield  {journal} {\bibinfo  {journal} {Phys. Rev. Lett.}\ }\textbf
			{\bibinfo {volume} {106}},\ \bibinfo {pages} {190502} (\bibinfo {year}
			{2011}{\natexlab{a}})}\BibitemShut {NoStop}%
		\bibitem [{\citenamefont {Tomiyama}(1985)}]{Tomiyama1985}%
		\BibitemOpen
		\bibfield  {author} {\bibinfo {author} {\bibfnamefont {J.}~\bibnamefont
				{Tomiyama}},\ }\bibfield  {title} {\bibinfo {title} {On the geometry of
				positive maps in matrix algebras. ii},\ }\href
		{https://www.sciencedirect.com/science/article/pii/0024379585900746}
		{\bibfield  {journal} {\bibinfo  {journal} {Linear Algebra and its
					Applications}\ }\textbf {\bibinfo {volume} {69}},\ \bibinfo {pages} {169}
			(\bibinfo {year} {1985})}\BibitemShut {NoStop}%
		\bibitem [{\citenamefont {Jungnitsch}\ \emph
			{et~al.}(2011{\natexlab{b}})\citenamefont {Jungnitsch}, \citenamefont
			{Moroder},\ and\ \citenamefont {Gühne}}]{Jungnitsch_2011}%
		\BibitemOpen
		\bibfield  {author} {\bibinfo {author} {\bibfnamefont {B.}~\bibnamefont
				{Jungnitsch}}, \bibinfo {author} {\bibfnamefont {T.}~\bibnamefont
				{Moroder}},\ and\ \bibinfo {author} {\bibfnamefont {O.}~\bibnamefont
				{Gühne}},\ }\bibfield  {title} {\bibinfo {title} {Entanglement witnesses for
				graph states: General theory and examples},\ }\bibfield  {journal} {\bibinfo
			{journal} {Physical Review A}\ }\textbf {\bibinfo {volume} {84}},\ \href
		{https://doi.org/10.1103/physreva.84.032310} {10.1103/physreva.84.032310}
		(\bibinfo {year} {2011}{\natexlab{b}})\BibitemShut {NoStop}%
		\bibitem [{git()}]{github-code}%
		\BibitemOpen
		\bibfield  {title} {\bibinfo {title} {Convex programming method for
				{GME}-dimension}\ }\href {https://doi.org/10.5281/zenodo.13123607}
		{10.5281/zenodo.13123607}\BibitemShut {NoStop}%
		\bibitem [{\citenamefont {Durt}\ \emph {et~al.}(2010)\citenamefont {Durt},
			\citenamefont {Englert}, \citenamefont {Bengtsson},\ and\ \citenamefont
			{Życzkowski}}]{DURT_2010}%
		\BibitemOpen
		\bibfield  {author} {\bibinfo {author} {\bibfnamefont {T.}~\bibnamefont
				{Durt}}, \bibinfo {author} {\bibfnamefont {B.-G.}\ \bibnamefont {Englert}},
			\bibinfo {author} {\bibfnamefont {I.}~\bibnamefont {Bengtsson}},\ and\
			\bibinfo {author} {\bibfnamefont {K.}~\bibnamefont {Życzkowski}},\
		}\bibfield  {title} {\bibinfo {title} {On mutually unbiased bases},\ }\href
		{https://doi.org/10.1142/s0219749910006502} {\bibfield  {journal} {\bibinfo
				{journal} {International Journal of Quantum Information}\ }\textbf {\bibinfo
				{volume} {08}},\ \bibinfo {pages} {535–640} (\bibinfo {year}
			{2010})}\BibitemShut {NoStop}%
		\bibitem [{\citenamefont {Huber}\ \emph {et~al.}(2010)\citenamefont {Huber},
			\citenamefont {Mintert}, \citenamefont {Gabriel},\ and\ \citenamefont
			{Hiesmayr}}]{Huber_2010}%
		\BibitemOpen
		\bibfield  {author} {\bibinfo {author} {\bibfnamefont {M.}~\bibnamefont
				{Huber}}, \bibinfo {author} {\bibfnamefont {F.}~\bibnamefont {Mintert}},
			\bibinfo {author} {\bibfnamefont {A.}~\bibnamefont {Gabriel}},\ and\ \bibinfo
			{author} {\bibfnamefont {B.~C.}\ \bibnamefont {Hiesmayr}},\ }\bibfield
		{title} {\bibinfo {title} {Detection of high-dimensional genuine multipartite
				entanglement of mixed states},\ }\bibfield  {journal} {\bibinfo  {journal}
			{Physical Review Letters}\ }\textbf {\bibinfo {volume} {104}},\ \href
		{https://doi.org/10.1103/physrevlett.104.210501}
		{10.1103/physrevlett.104.210501} (\bibinfo {year} {2010})\BibitemShut
		{NoStop}%
		\bibitem [{\citenamefont {Gao}\ and\ \citenamefont {Hong}(2011)}]{Gao2011}%
		\BibitemOpen
		\bibfield  {author} {\bibinfo {author} {\bibfnamefont {T.}~\bibnamefont
				{Gao}}\ and\ \bibinfo {author} {\bibfnamefont {Y.}~\bibnamefont {Hong}},\
		}\bibfield  {title} {\bibinfo {title} {Separability criteria for several
				classes of n-partite quantum states},\ }\href
		{https://doi.org/10.1140/epjd/e2010-10432-4} {\bibfield  {journal} {\bibinfo
				{journal} {The European Physical Journal D}\ }\textbf {\bibinfo {volume}
				{61}},\ \bibinfo {pages} {765} (\bibinfo {year} {2011})}\BibitemShut
		{NoStop}%
		\bibitem [{\citenamefont {Ananth}\ \emph {et~al.}(2015)\citenamefont {Ananth},
			\citenamefont {Chandrasekar},\ and\ \citenamefont
			{Senthilvelan}}]{Ananth2015}%
		\BibitemOpen
		\bibfield  {author} {\bibinfo {author} {\bibfnamefont {N.}~\bibnamefont
				{Ananth}}, \bibinfo {author} {\bibfnamefont {V.~K.}\ \bibnamefont
				{Chandrasekar}},\ and\ \bibinfo {author} {\bibfnamefont {M.}~\bibnamefont
				{Senthilvelan}},\ }\bibfield  {title} {\bibinfo {title} {Criteria for
				non-k-separability of n-partite quantum states},\ }\href
		{https://doi.org/10.1140/epjd/e2015-50538-5} {\bibfield  {journal} {\bibinfo
				{journal} {The European Physical Journal D}\ }\textbf {\bibinfo {volume}
				{69}},\ \bibinfo {pages} {56} (\bibinfo {year} {2015})}\BibitemShut {NoStop}%
			
			
			
			\bibitem [{\citenamefont {Bengtsson}\ and\ \citenamefont
				{Zyczkowski}(2006)}]{bengtsson_zyczkowski_2006}%
			\BibitemOpen
			\bibfield  {author} {\bibinfo {author} {\bibfnamefont {I.}~\bibnamefont
					{Bengtsson}}\ and\ \bibinfo {author} {\bibfnamefont {K.}~\bibnamefont
					{Zyczkowski}},\ }\href {https://doi.org/10.1017/CBO9780511535048} {\emph
				{\bibinfo {title} {Geometry of Quantum States: An Introduction to Quantum
						Entanglement}}}\ (\bibinfo  {publisher} {Cambridge University Press},\
			\bibinfo {year} {2006})\BibitemShut {NoStop}%
			\bibitem [{\citenamefont {Barnum}\ \emph {et~al.}(1996)\citenamefont {Barnum},
				\citenamefont {Caves}, \citenamefont {Fuchs}, \citenamefont {Jozsa},\ and\
				\citenamefont {Schumacher}}]{Barnum_1996}%
			\BibitemOpen
			\bibfield  {author} {\bibinfo {author} {\bibfnamefont {H.}~\bibnamefont
					{Barnum}}, \bibinfo {author} {\bibfnamefont {C.~M.}\ \bibnamefont {Caves}},
				\bibinfo {author} {\bibfnamefont {C.~A.}\ \bibnamefont {Fuchs}}, \bibinfo
				{author} {\bibfnamefont {R.}~\bibnamefont {Jozsa}},\ and\ \bibinfo {author}
				{\bibfnamefont {B.}~\bibnamefont {Schumacher}},\ }\bibfield  {title}
			{\bibinfo {title} {Noncommuting mixed states cannot be broadcast},\ }\href
			{https://doi.org/10.1103/physrevlett.76.2818} {\bibfield  {journal} {\bibinfo
					{journal} {Physical Review Letters}\ }\textbf {\bibinfo {volume} {76}},\
				\bibinfo {pages} {2818–2821} (\bibinfo {year} {1996})}\BibitemShut
			{NoStop}%
			\bibitem [{\citenamefont {Preskill}(1998)}]{Preskill1998}%
			\BibitemOpen
			\bibfield  {author} {\bibinfo {author} {\bibfnamefont {J.}~\bibnamefont
					{Preskill}},\ }\href@noop {} {\bibinfo {title} {Lecture notes for physics
					229: Quantum information and computation}} (\bibinfo {year}
			{1998})\BibitemShut {NoStop}%
			\bibitem [{\citenamefont {Bertsekas}(1996)}]{Lagrange_multiplier}%
			\BibitemOpen
			\bibfield  {author} {\bibinfo {author} {\bibfnamefont {D.~P.}\ \bibnamefont
					{Bertsekas}},\ }\href
			{http://www.amazon.com/Constrained-Optimization-Lagrange-Multiplier-Computation/dp/1886529043%3FSubscriptionId%3D192BW6DQ43CK9FN0ZGG2%26tag%3Dws%26linkCode%3Dxm2%26camp%3D2025%26creative%3D165953%26creativeASIN%3D1886529043}
			{\emph {\bibinfo {title} {Constrained Optimization and Lagrange Multiplier
						Methods (Optimization and Neural Computation Series)}}},\ \bibinfo {edition}
			{1st}\ ed.\ (\bibinfo  {publisher} {Athena Scientific},\ \bibinfo {year}
			{1996})\BibitemShut {NoStop}%
			\bibitem [{\citenamefont {Matouek}\ and\ \citenamefont
				{G\"{a}rtner}(2006)}]{Linear_Programming}%
			\BibitemOpen
			\bibfield  {author} {\bibinfo {author} {\bibfnamefont {J.}~\bibnamefont
					{Matouek}}\ and\ \bibinfo {author} {\bibfnamefont {B.}~\bibnamefont
					{G\"{a}rtner}},\ }\href@noop {} {\emph {\bibinfo {title} {Understanding and
						Using Linear Programming (Universitext)}}}\ (\bibinfo  {publisher}
			{Springer-Verlag},\ \bibinfo {address} {Berlin, Heidelberg},\ \bibinfo {year}
			{2006})\BibitemShut {NoStop}%
			\bibitem [{\citenamefont {Skrzypczyk}\ and\ \citenamefont
				{Cavalcanti}(2023)}]{Cavalcanti-Skrzypczyk}%
			\BibitemOpen
			\bibfield  {author} {\bibinfo {author} {\bibfnamefont {P.}~\bibnamefont
					{Skrzypczyk}}\ and\ \bibinfo {author} {\bibfnamefont {D.}~\bibnamefont
					{Cavalcanti}},\ }\href {https://doi.org/10.1088/978-0-7503-3343-6} {\emph
				{\bibinfo {title} {Semidefinite Programming in Quantum Information
						Science}}},\ 2053-2563\ (\bibinfo  {publisher} {IOP Publishing},\ \bibinfo
			{year} {2023})\BibitemShut {NoStop}%
			\bibitem [{\citenamefont {Bavaresco}\ \emph {et~al.}(2018)\citenamefont
				{Bavaresco}, \citenamefont {Herrera~Valencia}, \citenamefont {Kl{\"o}ckl},
				\citenamefont {Pivoluska}, \citenamefont {Erker}, \citenamefont {Friis},
				\citenamefont {Malik},\ and\ \citenamefont {Huber}}]{Bavaresco2018}%
			\BibitemOpen
			\bibfield  {author} {\bibinfo {author} {\bibfnamefont {J.}~\bibnamefont
					{Bavaresco}}, \bibinfo {author} {\bibfnamefont {N.}~\bibnamefont
					{Herrera~Valencia}}, \bibinfo {author} {\bibfnamefont {C.}~\bibnamefont
					{Kl{\"o}ckl}}, \bibinfo {author} {\bibfnamefont {M.}~\bibnamefont
					{Pivoluska}}, \bibinfo {author} {\bibfnamefont {P.}~\bibnamefont {Erker}},
				\bibinfo {author} {\bibfnamefont {N.}~\bibnamefont {Friis}}, \bibinfo
				{author} {\bibfnamefont {M.}~\bibnamefont {Malik}},\ and\ \bibinfo {author}
				{\bibfnamefont {M.}~\bibnamefont {Huber}},\ }\bibfield  {title} {\bibinfo
				{title} {Measurements in two bases are sufficient for certifying
					high-dimensional entanglement},\ }\href
			{https://doi.org/10.1038/s41567-018-0203-z} {\bibfield  {journal} {\bibinfo
					{journal} {Nature Physics}\ }\textbf {\bibinfo {volume} {14}},\ \bibinfo
				{pages} {1032} (\bibinfo {year} {2018})}\BibitemShut {NoStop}%
			
			
	\end{thebibliography}
\end{document}